\RequirePackage{fix-cm}
\documentclass[smallextended]{svjour3}       % onecolumn (second format)
\smartqed  % flush right qed marks, e.g. at end of proof
\usepackage{amsmath,amsfonts,amssymb,booktabs,siunitx,bm,xfrac,nicefrac}
\usepackage{algorithm,algpseudocode}
\usepackage{xcolor}
\usepackage{graphicx}
\usepackage[hidelinks]{hyperref}
\hypersetup{
    colorlinks,
    linkcolor={green!50!black},
    citecolor={blue!50!black},
    urlcolor={blue!80!black}
}  
\usepackage{relsize}
\usepackage{tikz}
\usetikzlibrary{shapes.geometric, arrows}
\tikzstyle{startstop} = [nodes = {draw,align=center},rectangle, rounded corners, minimum width=3cm, minimum height=1cm,text centered, draw=black, fill=red!30]
\tikzstyle{io} = [trapezium, trapezium left angle=70, trapezium right angle=110, minimum width=3cm, minimum height=1cm, text centered, draw=black, fill=blue!30]
\tikzstyle{process} = [rectangle, minimum width=3cm, minimum height=1cm, text centered, draw=black, fill=orange!30]
\tikzstyle{decision} = [diamond, minimum width=3cm, minimum height=1cm, text centered, draw=black, fill=green!30]
\tikzstyle{arrow} = [thick,->,>=stealth]
\usepackage{subfigure}

\usepackage[margin=0.5in]{geometry}

\begin{document}

\title{High order hybridizable discontinuous Galerkin method for three-phase flow in porous media}
\titlerunning{HDG for three-phase flow}   

\author{
	Maurice S. Fabien
}

%\authorrunning{Short form of author list} % if too long for running head

\institute{Maurice S. Fabien \at
              Department of Mathematics, University of Wisconsin, Madison, WI 53706, USA \\
              %Tel.: +123-45-678910\\
             % Fax: +123-45-678910\\
              \email{mfabien@wisc.edu}           %  \\
%             \emph{Present address:} of F. Author  %  if needed
%           \and
%           S. Author \at
%              second address
}

\date{Received: date / Accepted: date}
% The correct dates will be entered by the editor

\maketitle

\begin{abstract} 
We present a high-order hybridizable discontinuous Galerkin method for the numerical solution of time-dependent three-phase flow in heterogeneous porous media.  The underlying algorithm is a semi-implicit operator splitting approach that relaxes the nonlinearity present in the governing equations.  By treating the subsequent equations implicitly, we obtain solution that remain stable for large time steps.  The hybridizable discontinuous Galerkin method allows for static condensation, which significantly reduces the total number of degrees of freedom, especially when compared to classical discontinuous Galerkin methods. Several numerical tests are given, for example, we verify analytic convergence rates for the method, as well as examine its robustness in both homogeneous and heterogeneous porous media.
\end{abstract}

\section*{Introduction} % The \section*{} command stops section numbering
Simulation of multiphase flows has been vital to the understanding, prediction, and analysis of flows through porous media.  This is especially true when slow subterranean flows are considered, as large scale space-time observable data is limited.  Multiphase flow problems are able to successfully model a variety of physical phenomena; for instance, hydrology applications (e.g. sediment and pollution transport and other air--water problems \cite{eslinger2005discontinuous}), chemical \cite{class2009benchmark}, industrial and environmental processes (e.g. oil and gas recovery \cite{chen2006computational}), and biomedical applications \cite{ricardo2012single}.

In this work we study incompressible immiscible three-phase flows in porous media.  A hybridizable discontinuous Galerkin (HDG) method is employed to discretize the spatial dimensions.  Discontinuous Galerkin (DG) methods pervade finite element literature, and are especially popular for PDEs that model complicated transport behavior.  Originally DG methods were applied to hyperbolic problems that could be resolved in an explicit time stepping context \cite{reed1973triangularmesh}.  Later, DG methods intended for elliptic and parabolic PDEs were developed separately from the explicit Runge-Kutta DG (see \cite{cockburn1998runge}) type methods.  In \cite{ArnoldBCM02}, the authors put all DG methods for elliptic equations known at the time under a unified framework.  Even though the equations governing Darcy flow are of elliptic type, continuous Galerkin methods are typically not a suitable discretization as it does not locally conserve mass.  The lack of mass conservation may result in instability and spurious oscillations, which renders approximations ineffective. Various stablization and post-processing techniques can render continuous Galerkin type methods viable for the applications under consideration \cite{flux_cg,TEZDUYAR19911,flux_eg}.

	Discretizations that are locally conservative for these applications enjoy robust stability properties \cite{dawson2004compatible}; although local conservation for numerical schemes can be argued to not be strictly necessary; it does not change the fact that the governing equations obey a local and global conservation statement.  Many DG methods satisfy a local mass conservation property.  However, one draw back of DG methods is that they give rise to a large number of degrees of freedom compared to their continuous counterparts.  For problems that can be resolved using explicit time stepping this disadvantage is minimal as no large sparse linear systems need to be formed or solved.  This is not the case for elliptic and parabolic problems.

In an effort to curtail the growth of DG degrees of freedom, reduced stencil approaches were examined, for instance see \cite{peraire2008compact}.  The HDG method \cite{CockburnDG08} was developed to establish enhanced DG methods that have the ability to statically condense interior degrees of freedom.  That is, the hybridization of DG methods results in a globally coupled linear system whose degrees of freedom are defined only on the mesh skeleton.  Thus, for higher orders, the HDG method significantly reduces the number of degrees of freedom compared to classical DG methods.
 
We point out an important but subtle observation for locally conservative discretizations of flow-transport systems.  Local conservation is often found to numerically to improve stability \cite{dawson2004compatible}, but it is not sufficient.  For instance many authors have utilized IPDG-IPDG discretizations for flow-transport systems (e.g. \cite{ern2009accurate,jamei2016novel,ern2010discontinuous,arbogast2013discontinuous,li2015high}), but require a $H(\text{div})$ reconstruction of velocity in order to stabilize the method.  The underlying issue is that of \textit{compatible} discretizations \cite{dawson2004compatible}.  The compatibility property is a more stringent statement of local and global conservation.  The HDG-HDG method (of the LDG variety) is a compatible discretization, which further motivates its use for flow-transport systems.

Although high-order methods are often met with skepticism for problems in porous media, a number of works suggest their applicability. DG methods have been applied to single-(\cite{dawson2004compatible,li2015high}), two-(\cite{eslinger2005discontinuous,arbogast2013discontinuous,ern2010discontinuous}), and three-phase flows (\cite{moortgat2013higher,dong2014high,dong2016semi}). Moreover, it is well-accepted that $hp$-adaptivity \cite{babuvska1992h} is the best way to obtain additional accuracy for finite element approximations. Furthermore, for many applications, there are large regions in the computational domain were the solution is ``smooth''; as such, higher-order approximations are ideal candidates. To the best of our knowledge, HDG methods have not been explored in the context of three phase-flows in porous media. Moreover, it is generally accepted that any state-of-the-art finite element method will employ $hp$-adaptivity \cite{babuvska1988p}, where both the elements in the mesh and polynomial degree vary non-uniformly throughout the computational domain.

HDG methods have garnered the interest of researchers for applications in porous media. For instance, the first works using HDG for porous media applications are \cite{fabien1} (two-phase flow) and \cite{fabien2}  (one-phase flow). Since then, HDG has been applied to various porous media problems in one and two-phase flows, e.g., \cite{drad054,costa2021high1,moon2022,troescher2023fully,costa2021high2,costa2021one}.

\addcontentsline{toc}{section}{Introduction} % Adds this section to the table of contents
\section{Model problem}

We focus on incompressible immiscible three-phase flow in porous media.  The phases are liquid (heavy oil), vapor (light oil) and aqueous (water) and the components are oil (light or heavy), and water.  The variable $p_o$ denotes pressure of the liquid phase, and the variables $s_w$ and $s_g$ denote saturation of the aqueous and vapor phases respectively.  As the flow is immisible we assume that no mass transfer occurs between the phases.  The equations that govern this flow are a nonlinear system of partial differential equations.
%Pressure-saturation formulations have a structure similar to that of single-phase flow, which allows the nonlinearities present in the system of differential equations to be handled in several different ways \cite{lacroix2000iterative}, \cite{lu2008iteratively}, \cite{chen2006computational}. 

We adopt a pressure-saturation formulation which takes as primary variables $p_o$, $s_w$, and $s_g$.  Other formulations for three-phase flow can be found in \cite{chen1997comparison}.  The resulting equations are as follows:
\begin{align}
{\bm u}_t &= 
- {\bm K}\lambda_t \nabla p_o  
- {\bm K}\lambda_w  p_{cwo}'\nabla s_w 
- {\bm K}\lambda_g  p_{cgo}'\nabla s_g ,
\label{pressure_velocity_eq0}
%\tag{1a}
\\
\nabla \cdot {\bm u}_t  &= 0,
\label{pressure_velocity_eq1}
%\tag{1b}
\end{align}
\begin{align}
\frac{\partial (\phi s_w) }{\partial t}
&+
\nabla \cdot 
\bigg(
\frac{\lambda_w}{\lambda_t} 
{\bm u}_t 
\bigg)
+
\nabla \cdot 
\bigg(
{\bm K} \frac{\lambda_g \lambda_w}{\lambda_t} p_{cgo}' \nabla s_w
\bigg)
\notag
\\
&-
\nabla \cdot 
\bigg(
{\bm K} \frac{\lambda_w(\lambda_o+\lambda_w)}{\lambda_t}
p_{cwo}' \nabla s_w
\bigg)
\notag
\\
&=0,
\label{wet_saturation_eq0} 
\end{align}
\begin{align}
\frac{\partial (\phi s_g) }{\partial t}
&+
\nabla \cdot 
\bigg(
\frac{\lambda_g}{\lambda_t} 
{\bm u}_t 
\bigg)
+
\nabla \cdot 
\bigg(
{\bm K} \frac{\lambda_g \lambda_w}{\lambda_t} p_{cwo}' \nabla s_g
\bigg)
\notag
\\
&-
\nabla \cdot 
\bigg(
{\bm K} \frac{\lambda_g(\lambda_o+\lambda_w)}{\lambda_t}
p_{cgo}' \nabla s_g
\bigg)
\notag
\\
&=0.
\label{light_saturation_eq0}
\end{align}
Subscripts $\alpha \in \{o,w,g\}$ denote heavy oil, wetting, and light oil phases, respectively.  The velocity of phase $\alpha$ is defined by ${\bm u}_\alpha=f_\alpha {\bm u}_t + {\bm K} f_\alpha \sum_{\beta \in \{w,o,g\}} (\lambda_\beta \nabla (p_{c\beta o} - p_{c\alpha o}))$.  A subscript of $t$ indicates a `total' quantity.  That is, ${\bm u}_t$ denotes the total velocity (as defined by equation~(\ref{pressure_velocity_eq0}), and $\lambda_t=\lambda_w+\lambda_o+\lambda_g$ denotes the total mobility.  The permeability of the porous medium $\Omega \subset \mathbb{R}^2$ is given by $\bm K$, and the symbol $\phi$ represents its porosity.  We define fractional flow for the phase $\alpha$ to be $f_\alpha=\lambda_\alpha/\lambda_t$.  We make specific use of the fractional flow for the wetting phase as $f_w=\lambda_w/\lambda_t$ and for the vapor phase $f_g=\lambda_g/\lambda_t$.  The following phase mobilities are used (see \cite{chen2006computational}), unless stated otherwise:
\begin{align*}
\lambda_w(s_w) &= s_w^2 \mu_w^{-1},
\quad
\lambda_g(s_g) = s_g^2\mu_g^{-1},
\\
\quad
\lambda_o(s_w,s_g) &= (1-s_w)(1-s_g)(1-s_w-s_g) \mu_o^{-1}.
\end{align*}
Capillary pressure forces are included in our model.  For the wetting phase, $p_{cwo}$ is its associated capillary pressure.  Similarly, for the vapor phase, $p_{cgo}$ is its associated capillary pressure.  For notation purposes we set $p_{c o o}\equiv 0$.  The derivatives of capillary pressure with respect to their corresponding phases is needed:
\begin{align*}
p_{cwo}' = \frac{\partial p_{cwo}}{\partial s_w},
\quad
p_{cgo}' = \frac{\partial p_{cgo}}{\partial s_g}
.
\end{align*}
The specific capillary pressures we use, unless stated otherwise, are taken from \cite{chen2006computational}:
\begin{align*}
p_{cwo}(s_w) &= \frac{6.3 \text{ psi}}{ \log{\frac{\epsilon}{1-s_{wr} } }} \log{ \bigg( \frac{s_w-s_{rw}+\epsilon}{1-s_{rw}}\bigg)},
\\
p_{cgo}(s_g) &= \frac{3.9 \text{ psi}}{ \log{\frac{\epsilon}{1-s_{wr}-s_{or}} }}  \log{ \bigg( \frac{1-s_g-s_{or}-s_{rw}+\epsilon}{1-s_{or}-s_{rw}}\bigg)},
\end{align*}
where $\epsilon=0.01$, and $s_{wr}$, $s_{or}$, are residual saturations for the wetting and vapor phases, respectively.  In this paper we do consider other models of relative permeability and capillary pressure.  If no specificity is given, we assume the capillary pressures and phase mobilities as defined above.

 No source terms are included in our model, so we require the prescription of boundary conditions to drive the simulation.  The boundary of $\Omega$ is partitioned as 
\begin{align*}
\partial \Omega &= \Gamma_{P_D} \cup \Gamma_{P_N} = \Gamma_{s_{w,D}} \cup \Gamma_{s_{w,N}}= \Gamma_{s_{g,D}} \cup \Gamma_{s_{g,N}},
\\
\emptyset &= \Gamma_{P_D} \cap \Gamma_{P_N} = \Gamma_{s_{w,D}} \cap \Gamma_{s_{w,N}}= \Gamma_{s_{g,D}} \cap \Gamma_{s_{g,N}}.
\end{align*}
For the pressure equation we have
\begin{alignat*}{3}
p_o &= \bar{p}_{D}, \quad \quad &\text{on  }\Gamma_{P_D},
\\
{\bm u}_t \cdot {\bm n} & = 0
 \quad \quad &\text{on  }\Gamma_{P_N},
\end{alignat*} 
and for the saturation equations we have
\begin{alignat*}{3}
s_w &= \bar{s}_{w,D}, \quad \quad &\text{on  }\Gamma_{s_{w,D}},
\\
%{\bm K}  \frac{\lambda_w(\lambda_o + \lambda_g)}{\lambda_t} p_{cwo}'\nabla s_w \cdot {\bm n} & = 0
% \quad \quad &\text{on  }\Gamma_{s_{w,N}},
{\bm u}_w \cdot {\bm n} & = 0
 \quad \quad &\text{on  }\Gamma_{s_{w,N}},
\\
s_g &= \bar{s}_{g,D}, \quad \quad &\text{on  }\Gamma_{s_{g,D}},
\\
%{\bm K}  \frac{\lambda_g(\lambda_o + \lambda_w)}{\lambda_t} p_{cgo}'\nabla s_g \cdot {\bm n} & = 0
% \quad \quad &\text{on  }\Gamma_{s_{g,N}}.
{\bm u}_g \cdot {\bm n} & = 0
 \quad \quad &\text{on  }\Gamma_{s_{g,N}}.
\end{alignat*} 
To complete the system, we impose initial conditions for $s_w$ and $s_g$.
%  In Fig.~\ref{boundary_condition} we clarify how the boundary of $\Omega$ is partitioned into inflow, outflow, and no flow boundaries.
%\begin{figure}[!ht]
%\centering
% 
%\begin{tikzpicture} 
%%\draw (0,0) -- (5,0) -- (5,3) -- (0,3) -- (0,0);
%\draw [dashed] (5,0) -- (5,3) ;
%\draw [dashed] (0,3) -- (0,0) ;
%\draw (0,0) -- (5,0);
%\draw (5,3) -- (0,3) ;
%\draw (-1.8,2.8) node {$\Gamma_{P_D}=\Gamma_{s_{w,D}}=\Gamma_{s_{g,D}}$};
%\draw (-1.8,2.3) node {(dashed lines)};
%
%\draw (-1.8,0.8) node {$\Gamma_{P_N}=\Gamma_{s_{w,N}}=\Gamma_{s_{g,N}}$};
%\draw (-1.8,0.3) node {(solid lines)};
%\draw (1.0,1.0) node {$\Omega$};
%\end{tikzpicture}
%\vspace*{1ex}
%\caption{Sample Dirichlet--Neumann boundary conditions.}
%\label{boundary_condition} 
%\end{figure} 
\section{Semi-implicit algorithm}
Equations (\ref{pressure_velocity_eq0}), (\ref{pressure_velocity_eq1}), (\ref{wet_saturation_eq0}), and (\ref{light_saturation_eq0}) are coupled and highly nonlinear.  To relax the nonlinearity we employ a splitting algorithm.  There are multiple techniques to handle the nonlinearities of multiphase flow problems, e.g. implicit explicit, semi-implicit, fully implicit \cite{lacroix2000iterative}, \cite{lu2008iteratively}, \cite{chen2006computational}.  In this paper we use a semi-implicit splitting algorithm.   That is, we solve the pressure-velocity system, wetting phase system, and vapor phase system sequentially, while keeping each system implicit.  We first solve equations (\ref{pressure_velocity_eq0}) and (\ref{pressure_velocity_eq1}) using saturation values from the previous time step.  Then, using the updated total velocity, we solve equation (\ref{wet_saturation_eq0}) using the vapor phase saturation from the previous time step.  Finally, we solve equation (\ref{light_saturation_eq0}) using the updated total velocity and updated wetting phase saturation.

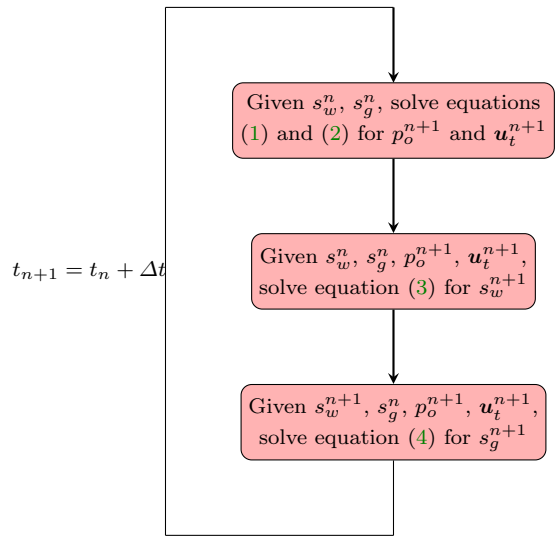
\begin{figure}[h!]
\centering
 \begin{tikzpicture}[node distance=2cm]
 \node (start)  [ align=center ,rectangle, rounded corners, minimum width=3cm, minimum height=1cm,text centered, draw=black, fill=red!30] {Given $s_w^n$, $s_g^n$, solve equations\\ (\ref{pressure_velocity_eq0}) and (\ref{pressure_velocity_eq1}) for $p_o^{n+1}$ and ${\bm u}_t^{n+1}$};
\node (in1) [io, below of=start,align=center ,rectangle, rounded corners, minimum width=3cm, minimum height=1cm,text centered, draw=black, fill=red!30] {Given $s_w^n$, $s_g^n$, $p_o^{n+1}$, ${\bm u}_t^{n+1}$, \\ solve equation (\ref{wet_saturation_eq0}) for  $s_w^{n+1}$};

\node (pro1) [process, below of=in1,align=center ,rectangle, rounded corners, minimum width=3cm, minimum height=1cm,text centered, draw=black, fill=red!30] {Given $s_w^{n+1}$, $s_g^n$, $p_o^{n+1}$, ${\bm u}_t^{n+1}$, \\solve equation (\ref{light_saturation_eq0})  for  $s_g^{n+1}$};

\draw [arrow] (start) -- (in1);
\draw [arrow] (in1) -- (pro1);

\draw (pro1) -- (0,-5.5);

\draw (0,-5.5) -- (-3,-5.5);
\draw (-3,-5.5) -- (-3,1.5);
\draw (-3,1.5) -- (0,1.5);
\draw [arrow] (0,1.5) -- (0,.5);

\draw (-4,-2.0) node {$t_{n+1} = t_n +\Delta t$};
%\draw [arrow,bend left=70] (pro1)--(start);
%\draw [arrow] (pro1) -- (start);
%\draw [arrow] (pro1) -- (dec1);
%\draw [arrow] (dec1) -- (pro2a);
%\draw [arrow] (dec1) -- (pro2b);
\end{tikzpicture}
\caption{Flow chart of semi-implicit method for three-phase flow.  Subscript of $n$ refers to the $n$th time step.}
\label{flow_chart}
\end{figure}

We note that equation (\ref{wet_saturation_eq0}) and (\ref{light_saturation_eq0}) are nonlinear, and as such, at each time step a Newton-Raphson method is utilized to solve these systems.  Fig.~\ref{flow_chart} has a flow chart illustrating the general steps in the semi-implicit algorithm.
\section{The hybridizable DG method}
 
We use meshes of straight-sided triangular elements.  As such, to ensure a well conditioned approximations we use a high-order nodal basis based on the hierarchical orthogonal basis \cite{proriol1957famille}, \cite{koornwinder1975two}, \cite{dubiner1991spectral}, \cite{Owens857}.
 
\begin{figure}[!ht]
\centering
%\hspace*{-2ex}
\begin{tikzpicture}[scale = 0.91]
\begin{scope} 
\foreach \Point in {
(4.000000000000,-0.000000000000),
(2.000000000000,3.464101615138),
(-0.000000000000,0.000000000000),
(3.830229155400,0.294051728491),
(1.830229155400,3.170049886647),
(0.339541689200,0.000000000000),
(3.468869719600,0.919944631091),
(1.468869719600,2.544156984047),
(1.062260560800,0.000000000000),
(3.000000000000,1.732050807569),
(1.000000000000,1.732050807569),
(2.000000000000,0.000000000000),
(2.531130280400,2.544156984047),
(0.531130280400,0.919944631091),
(2.937739439200,-0.000000000000),
(2.169770844600,3.170049886647),
(0.169770844600,0.294051728491),
(3.660458310800,-0.000000000000),
(3.361987189600,0.368356867831),
(2.000000000000,2.727387879476),
(0.638012810400,0.368356867831),
(1.601822242400,0.346816776607),
(2.500558982200,0.405926604189),
(2.898736739800,1.213810365962),
(2.398177757600,1.962584300152),
(1.499441017800,1.903474472569),
(1.101263260200,1.095590710796),
(2.000000000000,1.154700538379)
}{
    %\node at \Point {$\circ$};
    \node at \Point {$\bullet$};
}  
\end{scope}
\begin{scope} [shift={(-.5,0)},rotate=60]
\foreach \Point in {
(4.000000000000,-0.000000000000),
(2.000000000000,3.464101615138),
(-0.000000000000,0.000000000000),
(3.830229155400,0.294051728491),
(1.830229155400,3.170049886647),
(0.339541689200,0.000000000000),
(3.468869719600,0.919944631091),
(1.468869719600,2.544156984047),
(1.062260560800,0.000000000000),
(3.000000000000,1.732050807569),
(1.000000000000,1.732050807569),
(2.000000000000,0.000000000000),
(2.531130280400,2.544156984047),
(0.531130280400,0.919944631091),
(2.937739439200,-0.000000000000),
(2.169770844600,3.170049886647),
(0.169770844600,0.294051728491),
(3.660458310800,-0.000000000000),
(3.361987189600,0.368356867831),
(2.000000000000,2.727387879476),
(0.638012810400,0.368356867831),
(1.601822242400,0.346816776607),
(2.500558982200,0.405926604189),
(2.898736739800,1.213810365962),
(2.398177757600,1.962584300152),
(1.499441017800,1.903474472569),
(1.101263260200,1.095590710796),
(2.000000000000,1.154700538379)
}{
    %\node at \Point {$\circ$};
    \node  at \Point {$\bullet$};
}  
\end{scope}

\begin{scope} [shift={(.5,0)},rotate around={-60:(4,0)},black]
\foreach \Point in {
(4.000000000000,-0.000000000000),
(2.000000000000,3.464101615138),
(-0.000000000000,0.000000000000),
(3.830229155400,0.294051728491),
(1.830229155400,3.170049886647),
(0.339541689200,0.000000000000),
(3.468869719600,0.919944631091),
(1.468869719600,2.544156984047),
(1.062260560800,0.000000000000),
(3.000000000000,1.732050807569),
(1.000000000000,1.732050807569),
(2.000000000000,0.000000000000),
(2.531130280400,2.544156984047),
(0.531130280400,0.919944631091),
(2.937739439200,-0.000000000000),
(2.169770844600,3.170049886647),
(0.169770844600,0.294051728491),
(3.660458310800,-0.000000000000),
(3.361987189600,0.368356867831),
(2.000000000000,2.727387879476),
(0.638012810400,0.368356867831),
(1.601822242400,0.346816776607),
(2.500558982200,0.405926604189),
(2.898736739800,1.213810365962),
(2.398177757600,1.962584300152),
(1.499441017800,1.903474472569),
(1.101263260200,1.095590710796),
(2.000000000000,1.154700538379)
}{
    %\node at \Point {$\circ$};
    \node at \Point {$\bullet$};
}  
\end{scope}
 
\begin{scope} [shift={(0,-.5)},rotate around={300:(0,0)},black]
\foreach \Point in {
(4.000000000000,-0.000000000000),
(2.000000000000,3.464101615138),
(-0.000000000000,0.000000000000),
(3.830229155400,0.294051728491),
(1.830229155400,3.170049886647),
(0.339541689200,0.000000000000),
(3.468869719600,0.919944631091),
(1.468869719600,2.544156984047),
(1.062260560800,0.000000000000),
(3.000000000000,1.732050807569),
(1.000000000000,1.732050807569),
(2.000000000000,0.000000000000),
(2.531130280400,2.544156984047),
(0.531130280400,0.919944631091),
(2.937739439200,-0.000000000000),
(2.169770844600,3.170049886647),
(0.169770844600,0.294051728491),
(3.660458310800,-0.000000000000),
(3.361987189600,0.368356867831),
(2.000000000000,2.727387879476),
(0.638012810400,0.368356867831),
(1.601822242400,0.346816776607),
(2.500558982200,0.405926604189),
(2.898736739800,1.213810365962),
(2.398177757600,1.962584300152),
(1.499441017800,1.903474472569),
(1.101263260200,1.095590710796),
(2.000000000000,1.154700538379)
}{
    %\node at \Point {$\circ$};
    \node at \Point {$\bullet$};
}  
\end{scope}
\draw[line width=0.25mm] (0,0) node[anchor=north]{ }
  --  (4,0) node[anchor=north]{ }
  --  ( 2 ,{sqrt(12)}) node[anchor=south]{  }
  -- cycle;

\draw[line width=0.25mm,shift={(-.5,0)},rotate=60] (0,0) node[anchor=north]{ }
  --  (4,0) node[anchor=north]{ }
  --  ( 2 ,{sqrt(12)}) node[anchor=south]{ }
  -- cycle;
 
\draw[line width=0.25mm,shift={(.5,0)},rotate around={-60:(4,0)},black]  (0,0) node[anchor=north]{  }
  --  (4,0) node[anchor=north]{  }
  --  ( 2 ,{sqrt(12)}) node[anchor=south]{ }
  -- cycle;
  
\draw[line width=0.25mm,shift={(0,-.5)},rotate around={300:(0,0)},black]  (0,0) node[anchor=north]{  }
  --  (4,0) node[anchor=north]{ }
  --  ( 2 ,{sqrt(12)}) node[anchor=south]{ }
  -- cycle; 

\draw[line width=0.25mm]  (0,-0.25)   -- (4,-0.25);
\draw[line width=0.25mm]  (-.25,0)    -- ( {2-.25} ,{sqrt(12)});
\draw[line width=0.25mm]  ({4+.25},0) -- ( {2+.25} ,{sqrt(12)});

\begin{scope} [shift={(-4.99,0)}]
\draw[ line width=0.25mm ]  ({4+.25},0) -- ( {2+.25} ,{sqrt(12)});
\end{scope}

\begin{scope} [shift={(-2.5,4)}]
\draw[line width=0.25mm]  (0,-0.25)   -- (4,-0.25);
\end{scope}

\begin{scope} [shift={(2.5,4)}]
\draw[line width=0.25mm]  (0,-0.25)   -- (4,-0.25);
\end{scope}

\begin{scope} [shift={(2.5,-4)}]
\draw[line width=0.25mm]  (-.25,0)    -- ( {2-.25} ,{sqrt(12)});
\end{scope}

\begin{scope} [shift={(5,0)}]
\draw[line width=0.25mm]  (-.25,0)    -- ( {2-.25} ,{sqrt(12)});
\end{scope}

\begin{scope} [shift={(-2.5,-4)}]
\draw[line width=0.25mm]  ({4+.25},0) -- ( {2+.25} ,{sqrt(12)});
\end{scope}

\begin{scope} [shift={(0,-0.26)},black]
\foreach \Point in {
(4.000000000000,-0.000000000000),
(-0.000000000000,0.000000000000),
(0.339541689200,0.000000000000),
(1.062260560800,0.000000000000),
(2.000000000000,0.000000000000),
(2.937739439200,-0.000000000000),
(3.660458310800,-0.000000000000)
}{
    \node[scale=0.7,blue] at \Point {$\blacksquare$};
    %\node  at \Point {$\bullet$};
} 
\end{scope}

\begin{scope} [shift={(.25,0)},rotate around={-60:(4,0)},black]
\foreach \Point in {
(4.000000000000,-0.000000000000),
(-0.000000000000,0.000000000000),
(0.339541689200,0.000000000000),
(1.062260560800,0.000000000000),
(2.000000000000,0.000000000000),
(2.937739439200,-0.000000000000),
(3.660458310800,-0.000000000000)
}{
    \node[scale=0.7,blue] at \Point {$\blacksquare$};
    %\node  at \Point {$\bullet$};
} 
\end{scope}

\begin{scope} [shift={(-.24,0)},rotate=60]
\foreach \Point in {
(4.000000000000,-0.000000000000),
(-0.000000000000,0.000000000000),
(0.339541689200,0.000000000000),
(1.062260560800,0.000000000000),
(2.000000000000,0.000000000000),
(2.937739439200,-0.000000000000),
(3.660458310800,-0.000000000000)
}{
    \node[scale=0.7,blue] at \Point {$\blacksquare$};
    %\node  at \Point {$\bullet$};
} 
\end{scope}

\begin{scope} [shift={(-.74,0)},rotate=120]
\foreach \Point in {
(4.000000000000,-0.000000000000),
(-0.000000000000,0.000000000000),
(0.339541689200,0.000000000000),
(1.062260560800,0.000000000000),
(2.000000000000,0.000000000000),
(2.937739439200,-0.000000000000),
(3.660458310800,-0.000000000000)
}{
    \node[scale=0.7,blue] at \Point {$\blacksquare$};
    %\node  at \Point {$\bullet$};
} 
\end{scope}
 
\begin{scope} [shift={(2.5,3.75)},rotate=0]
\foreach \Point in {
(4.000000000000,-0.000000000000),
(-0.000000000000,0.000000000000),
(0.339541689200,0.000000000000),
(1.062260560800,0.000000000000),
(2.000000000000,0.000000000000),
(2.937739439200,-0.000000000000),
(3.660458310800,-0.000000000000)
}{
    \node[scale=0.7,blue] at \Point {$\blacksquare$};
    %\node  at \Point {$\bullet$};
} 
\end{scope}

\begin{scope} [shift={(-2.5,3.75)},rotate=0]
\foreach \Point in {
(4.000000000000,-0.000000000000),
(-0.000000000000,0.000000000000),
(0.339541689200,0.000000000000),
(1.062260560800,0.000000000000),
(2.000000000000,0.000000000000),
(2.937739439200,-0.000000000000),
(3.660458310800,-0.000000000000)
}{
    \node[scale=0.7,blue] at \Point {$\blacksquare$};
    %\node  at \Point {$\bullet$};
} 
\end{scope}

\begin{scope} [shift={(6.8,3.5)},rotate=240]
\foreach \Point in {
(4.000000000000,-0.000000000000),
(-0.000000000000,0.000000000000),
(0.339541689200,0.000000000000),
(1.062260560800,0.000000000000),
(2.000000000000,0.000000000000),
(2.937739439200,-0.000000000000),
(3.660458310800,-0.000000000000)
}{
    \node[scale=0.7,blue] at \Point {$\blacksquare$};
    %\node  at \Point {$\bullet$};
} 
\end{scope}

\begin{scope} [shift={(4.3,-.5)},rotate=240]
\foreach \Point in {
(4.000000000000,-0.000000000000),
(-0.000000000000,0.000000000000),
(0.339541689200,0.000000000000),
(1.062260560800,0.000000000000),
(2.000000000000,0.000000000000),
(2.937739439200,-0.000000000000),
(3.660458310800,-0.000000000000)
}{
    \node[scale=0.7,blue] at \Point {$\blacksquare$};
    %\node  at \Point {$\bullet$};
} 
\end{scope}

\begin{scope} [shift={(1.74,-4)},rotate=120]
\foreach \Point in {
(4.000000000000,-0.000000000000),
(-0.000000000000,0.000000000000),
(0.339541689200,0.000000000000),
(1.062260560800,0.000000000000),
(2.000000000000,0.000000000000),
(2.937739439200,-0.000000000000),
(3.660458310800,-0.000000000000)
}{
    \node[scale=0.7,blue] at \Point {$\blacksquare$};
    %\node  at \Point {$\bullet$};
} 
\end{scope}

\node at (5.5,-1) {$M_h$};
\node[scale=0.7,blue] at (6,-1) {$\blacksquare$};

\node at (5.5,-1.5) {$W_h$};
\node  at (6,-1.5) {$\bullet$};
\end{tikzpicture} 

\caption{HDG nodes (of the Fekete type) for a mesh with four elements and polynomial degree $k=6$.}
\label{hdg_doodle}
\end{figure}
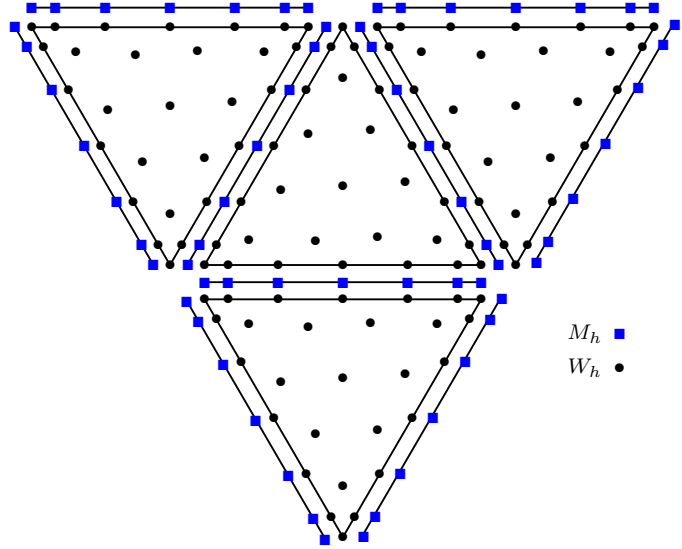

For suitable interpolation nodes, this orthonormal basis has a well-conditioned Vandermonde matrix which makes it suitable for high-order approximations.  Monomial basis functions and equispaced interpolation nodes are well known to give rise to the so-called Runge phenomena \cite{trefethen2013approximation} which can cause unreliable approximations.  We instead use Fekete points on the triangle to obtain a set of energy-minimum configuration interpolation nodes \cite{blyth2006comparison} (Fig.~\ref{hdg_doodle} plots Fekete points for $k=6$).
 
We suppose that the porous medium $\Omega$ has been partitioned into a set of triangular elements, labeled $\mathcal{T}_h=\mathcal{E} _h$.  Hybridized methods introduce additional unknowns on the mesh skeleton, $\Gamma_h,$ which consists of all unique edges of the mesh.  The set $\partial\mathcal{T}_h$ is the collection of all boundaries of mesh elements, and as such, interior edges are duplicated in this collection.  The HDG method we use requires the following approximation spaces:
\begin{equation}
\begin{split}
W_h&= \{ w\in L^2(\Omega): w|_E \in \mathbb{P}_{k}(E),~~~\forall E \in \mathcal{E}_h \},
\\
{\bm V}_h &= W_h \times W_h ,
\\
M_h &= \{ \zeta \in L^2(\Gamma_h): \zeta|_e \in \mathbb{P}_k(e),~~~\forall e \in \Gamma_h \},
\\
M_h(0) &= \{ \zeta \in M_h: \mu=0 ~~~\text{on } \Gamma_{D}\},
\end{split}
\label{hdg_spaces}
\end{equation}
where $\mathbb{P}_k$ is the set of all polynomials of degree $k$ or less, and $\Gamma_{D}$ refers to any Dirichlet boundary.  Fig.~\ref{hdg_doodle} has a depiction of the HDG degrees of freedom for $W_h$ and $M_h.$  The blue squares are the introduced trace space unknowns, which act as primary variables we solve for.  Once the trace space solution is known, we can recover the volume space unknowns (if needed).
 
\subsection{Pressure and total velocity discretization}
\label{sub_pre}
We discretize the Darcy system~(\ref{pressure_velocity_eq0}) and~(\ref{pressure_velocity_eq1}) using a HDG method.  The HDG method seeks $({\bm u}_t, p_o,\lambda)\in {\bm V}_h \times W_h\times M_h(0)$ such that
\begin{align*}
(\lambda^{-1}_t {\bf K}^{-1} {\bf u}_t, {\bf v})_{\mathcal{T}_h}
&-
(   p_w, \nabla\cdot {\bf v} )_{\mathcal{T}_h}
\\
&+
\langle \lambda, {\bf v}\cdot {\bf n} \rangle_{\partial\mathcal{T}_h}
+
\langle P_h \bar{p}_D, {\bf v}\cdot {\bf n} \rangle_{\Gamma_{p_D}}
\\
&+
( f_w |p_{cwo}'(s_w)|\nabla s_w, {\bf v} )_{\mathcal{T}_h}
\\
&+
( f_g |p_{cgo}'(s_g)|\nabla s_g, {\bf v} )_{\mathcal{T}_h}
\\
&=0,
\\
-( {\bf u}_t, \nabla w)_{\mathcal{T}_h}
+
\langle
\widehat{{\bf u} }_t
,
w
\rangle_{\partial\mathcal{T}_h}
 &= 0,
\\
\langle  
 [\![ 
  \widehat{\bf u}_t \cdot {\bm n}
 ]\!]
  , \mu  \rangle_{\Gamma_h } 
 &= \langle  g_{p_N}, \mu  \rangle_{  \Gamma_{p_N}}.
\end{align*}
for all $({\bm v}, w,\mu)\in {\bm V}_h \times W_h\times M_h(0)$, where $P_h$ denotes the $L^2$-projection onto $M_h$.  It is assumed that $\bm K$ is symmetric and uniformly positive definite, and $\lambda_t>0$.  The numerical trace for $p_o$ is an unknown to be solved for in the HDG method, where as the numerical trace for ${\bm u}_t$ is prescribed.  In more detail:
\begin{align*}
\widehat{p}_o
&=
\begin{cases}
\lambda,~~~~~~~\text{on } 
\Gamma_h \backslash \Gamma_{p_D}
\\
P_h\bar{p}_D
,~~~\text{on } 
\Gamma_{p_D},
\end{cases}
\\
\widehat{\bf u}_t
&=
{\bf u}_t
+
\tau ( p_o - \widehat{p}_o){\bm n},
\end{align*}
where $\tau>0$ is a stabilization term that is piecewise constant defined on element boundaries.  As long as $\tau =\mathcal{O}(1)$, optimal convergence rates of $k+1$ for all approximate variables are guaranteed for the HDG method \cite{CockburnDG08}.  The suggested stabilization for diffusion type problems is of the form $\tau = \lambda_t {\bm K}/\ell$, where $\ell$ is a diffusive length of scale.  We note this stabilization does not depend on the polynomial order nor the mesh spacing, and adapts according to the heterogeneous permeability without problem dependent tuning.  Some DG methods are very sensitive to the choice of penalization/stabilization in the presence of heterogeneous permeability \cite{klieber2006adaptive}.

Initially, the addition of a new unknown $\lambda$ may seem strange.  However, since the normal component of the total velocity is continuous, we can generate a linear system that has $\lambda$ as its only variable.  This is achieved by locally eliminating the volume unknowns ${\bm u}_t$ and $p_o$.  Let $(U,P,\Lambda)$ be discrete analogs of $({\bm u}_t, p_o,\lambda)$.  At the matrix level we have:
\begin{equation}
\def\arraystretch{1.5}
\begin{bmatrix}
{\bm A} & -{\bm B}^T & {\bm C}^T
\\
{\bm B} & {\bm D} & {\bm E}
\\
{\bm C} & {\bm G} & {\bm H}
\end{bmatrix}
\begin{bmatrix}
U
\\
P
\\
\Lambda
\end{bmatrix}
=
\begin{bmatrix}
R_u
\\
R_p
\\
R_\lambda
\end{bmatrix}
 \notag
.
\end{equation}
As the HDG approximation is totally discontinuous, we can locally express the volume unknowns in terms of the the surface unknowns, revealing:
\begin{equation}
\def\arraystretch{1.5}
\begin{bmatrix}
U
\\
P
\end{bmatrix}
=
\begin{bmatrix}
{\bm A} & -{\bm B}^T
\\
{\bm B}  & {\bm D }
\end{bmatrix}^{-1}
\Bigg(
\begin{bmatrix}
R_u
\\
R_p
\end{bmatrix}
-
\begin{bmatrix}
{\bm C}^T
\\
{\bm E}
\end{bmatrix}
\Lambda
\Bigg)
 \label{hdg_invert}
 .
\end{equation}
We mention that the inverted matrix in equation~(\ref{hdg_invert}) its dimensions are local to an element.  Specifically, it is of size $(3\binom{k+2}{2}\times 3\binom{k+2}{2})$.  The numerical trace of the total velocity is single valued, and this is enforced by the following equation:
\begin{equation}
{\bm C} U + {\bm G}P + {\bm H} \Lambda = R_\lambda.
\label{eq_single_valued}
\end{equation}
 
The Schur complement applied to equations~(\ref{eq_single_valued}) and (\ref{hdg_invert})  locally condenses the interior unknowns.  As a result, a globally coupled system only defined in terms of $\Lambda$, the heavy oil pressure on the mesh skeleton is obtained: $\mathbb{H}\Lambda = \mathbb{F}$.  Explicitly,
\begin{equation}
\begin{split}
\mathbb{H}
& =
{\bm H} - [{\bm C}~{\bm G}]
\begin{bmatrix}
{\bm A} & -{\bm B}^T
\\
{\bm B}  & {\bm D }
\end{bmatrix}^{-1}
\begin{bmatrix}
{\bm C}^T
\\
{\bm E}
\end{bmatrix},
\\
\mathbb{F} &=
R_\lambda
-
[{\bm C}~{\bm G}]
\begin{bmatrix}
{\bm A} & -{\bm B} ^T
\\
{\bm B}  & {\bm D }
\end{bmatrix}^{-1}
\begin{bmatrix}
R_u
\\
R_p
\end{bmatrix}.
\end{split}
\label{eq_press}
\end{equation}
Once $\Lambda$ is obtained, the recovery of $U$ and $P$ may be performed locally through equation~(\ref{hdg_invert}). 
%\onecolumn
%\newpage
%\onecolumn
\subsection{Saturation discretization}
Equations~(\ref{wet_saturation_eq0}) and~(\ref{light_saturation_eq0}) have nonlinear diffusion terms as well as nonlinear convective terms.  Explicit time stepping is not always appropriate for convection-diffusion problems.  Different authors have demonstrated independently that the use of explicit time stepping for multiphase flows requires slope limiters (for instance \cite{ern2010discontinuous}, \cite{arbogast2013discontinuous}, \cite{jamei2016novel}, \cite{klieber2006adaptive}, \cite{lu2008iteratively}, \cite{lacroix2000iterative}).  The design and analysis of slope limiters is a challenge in and of itself, particularly for 2D and 3D problems.  It turns out that for multiphase flows, implicit time stepping may sufficently stabilize approximations without slope limiters (see \cite{bastian2014fully}, \cite{epshteyn2007fully}, \cite{epshteyn2006solution}, \cite{dong2016semi}).  Implicit problems are more expensive, because at each time step multiple linear system are needed. 

 The HDG method is a good candidate for the discretization of the saturation equations.  This is because it is high-order, locally conservative, enables static condensation, and compatible with the HDG discretization of system~(\ref{pressure_velocity_eq0}) and~(\ref{pressure_velocity_eq1}).

%In the context multiphase flows in mind, the use of explicit time stepping requires slope limiters.
\subsubsection{Wetting saturation}
\label{sub_wet}
HDG for equation~(\ref{wet_saturation_eq0}) approximates the variables $(\widehat{s}_w,s_w,\nabla s_w)$, where $\widehat{s}_w$ is the wetting phase saturation restricted to the mesh skeleton.  In order for this to occur, we rewrite equation~(\ref{wet_saturation_eq0}) in first order form
\begin{align*}
{\bf q } - \nabla s_w &= 0
\\ %F_d(\widehat{u},u,{\bf q})
\frac{\partial \phi s_w}{\partial t} 
 + \nabla \cdot ( F_c(s_w) + F_v( s_w,{\bf q}) ) &= 0,
\end{align*}
where
$$
 F_c(s_w) = \frac{\lambda_w}{\lambda_t} {\bf u}_t  ,
\quad
\text{ and }
\quad
 F_v( s_w,{\bf q})  =\frac{\lambda_w \lambda_g}{\lambda_t} {\bf K} \nabla   p_{cgo} -\frac{ \lambda_w(\lambda_o+\lambda_g)}{\lambda_t} {\bf K} \nabla   p_{cwo}. 
$$
To be precise, $F_v( s_w,{\bf q})=F_v(s_w,{\bm q},s_g,\nabla s_g)$, but we time lag the light oil approximation.  Similarly, $F_c(s_w)=F_c(s_w,s_g,{\bm u}_t)$, and we write $F_c(s_w)$ for brevity as $s_g$ and ${\bm u}_t$ have been computed from the previous time step.  Then, the HDG method finds $(\widehat{s}_w,s_w,\nabla s_w)\in M_h\times W_h \times {\bm V}_h$ such that
 \begin{align*}
 ({\bf q},{\bf v})_E 
 + 
 (s_w,\nabla \cdot{\bf v})_E 
 - \langle\widehat{s_w},{\bf v}\cdot {\bf n} \rangle_{\Gamma_h  \setminus \Gamma_{s_w,D}}  
 &= \langle P s_{wD},{\bf v}\cdot {\bf n} \rangle_{\Gamma_{s_w,D}},
 \\
 (\frac{\partial \phi s_w}{\partial t} ,w)_E
-
(F_c(s_w) + F_v(s_w,{\bf q}) ,\nabla w)_E 
+ 
\langle (  {\hat{F_c}} (\widehat{s_w}) + \hat{F_v}(\widehat{s_w},{\bf q}))\cdot {\bf n}  ,w \rangle_{\Gamma_h } 
& = 0,
\\
\langle
[\![ (\hat{F_c}(\widehat{s_w}) + \hat{F_v}(\widehat{s_w},{\bf q}))
\cdot {\bm n}
]\!]   
,
\mu 
\rangle_{\Gamma_h  \setminus \Gamma_{s_w,N}}
&=\langle g_{s_w,N},{\bf v}\cdot {\bf n} \rangle_{\Gamma_{s_w,N}}
,
 \end{align*}
 for all $(\mu,w,{\bf v})\in  M_h\times W_h \times {\bm V}_h$, where $\hat{F_c}$ denotes a convective numerical flux and $\hat{F_v}$ a diffusive numerical flux.  A Lax-Friedrich and LDG flux are used for the convective and diffusive fluxes, respectively:
 \begin{align*}
 \hat{F_c}(\widehat{s_w},s_w)
 &\stackrel{def}{=} F_c(\widehat{s_w}) + \tau_c (s_w - \widehat{s_w}){\bf n},
 \notag
 \\
  \hat{F_v}(\widehat{s_w},s_w,{\bf q} )
 &\stackrel{def}{=} F_v(\widehat{s_w},s_w,{\bf q} ) + \tau_v (s_w - \widehat{s_w}) {\bf n}.
 \notag
\end{align*}
Adding these fluxes together we have
$$
 \hat{F_c}
 +
   \hat{F_v}
   =
    F_c(\widehat{s_w}) 
+
 F_v(\widehat{s_w},s_w,{\bf q} )    
    + \tau  (s_w - \widehat{s_w}){\bf n},
$$
with $0<\tau_c,\tau_v$, and $\tau = \tau_c+\tau_v$.  For higher order approximations the exact form of the fluxes and stabilization parameters are less critical.  Thorough analysis on the selection and analysis of HDG fluxes and stabilization can be found in~\cite{bui2015godunov}.  %The HDG stabilization factor does not depend on the polynomial order or mesh size, where as other DG methods may have a more delicate stabilization and penalization requirement.

To obtain a fully discrete scheme, let $t_{m+1} = t_m+ m \Delta t$, for $m\ge 0$, where $m$ denotes the $m$th time step.  We assume that ${\bm u}_t^{m+1},{\bm s}_g^{m },$ and $\nabla s_g^{m}$ have been computed. 
 \begin{align*}  
 ({\bf q}^{m+1},{\bf v})_E 
 + 
 (s_w^{m+1},\nabla \cdot{\bf v})_E 
 - \langle\widehat{s_w}^{m+1},{\bf v}\cdot {\bf n} \rangle_{\Gamma_h \setminus \Gamma_{s_w,D}}  
 &= \langle P s_{wD},{\bf v}\cdot {\bf n} \rangle_{\Gamma_{s_w,D}},
 \\
 \bigg(\frac{ \phi (s_w^{m+1}-s_w^{m})}{\Delta t} ,w\bigg)_E
-
(F_c(s_w^{m+1}) + F_v(s_w^{m+1},{\bf q}) ,\nabla w)_E 
+ 
\langle (  {\hat{F_c}} (\widehat{s_w}^{m+1}) + \hat{F_v}(\widehat{s_w}^{m+1},{\bf q}^{m+1},))\cdot {\bf n}  ,w \rangle_{\Gamma_h } 
& = 0,
\\
\langle
[\![ 
(\hat{F_c}(\widehat{s_w}^{m+1}) + \hat{F_v}(\widehat{s_w}^{m+1},{\bf q}^{m+1}))
\cdot {\bm n}
 ]\!]   
,
\mu 
\rangle_{\Gamma_h  \setminus \Gamma_{s_w,N}}
&=\langle g_{s_w,N},{\bf v}\cdot {\bf n} \rangle_{\Gamma_{s_w,N}}
.
 \end{align*}  
 %An exhaustive presentation of the discretization can be found in [MSF 2017 HDG TWO PHASE].
The above system is nonlinear in $s_w^{m+1}$, and we utilize a Picard-Newton-Raphson method to linearize it.  After linearization, static condensation can be employed to the Jacobian matrix.  Linearization leads to the following Jacobian system:
\begin{equation}
\def\arraystretch{1.5}
\begin{bmatrix}
{\bm J}_{qq} & {\bm J}_{qs}  & {\bm J}_{q\widehat{s}}
\\
{\bm J}_{sq}& {\bm J}_{ss} & {\bm J}_{s\widehat{s}}
\\
{\bm J}_{ \widehat{s}q} & {\bm J}_{  \widehat{s}s} &{\bm J}_{ \widehat{s}\widehat{s}}
\end{bmatrix}
\begin{bmatrix}
\delta Q
\\
\delta S
\\
\delta \widehat{S}
\end{bmatrix}
=
\begin{bmatrix}
R_{Q}
\\
R_{S}
\\
R_{\widehat{S}}
\end{bmatrix}
 \notag
,
\end{equation}
where $(\delta Q,\delta S,\delta \widehat{S})$ are Newton increments of $({\bm q},s_w,\widehat{s}_w)$.  In other words, $({\bm q}^{m+1},s_w^{m+1},\widehat{s}_w^{m+1})=({\bm q}^{m},s_w^{m},\widehat{s}_w^{m})+(\delta Q,\delta S,\delta \widehat{S})$.  Static condensation can be performed just as done in section~\ref{sub_pre}.

We discuss our linearization strategy.  Newton's method can be fickle and convergence may stagnate if the initial guess is not in the basin of attraction.  At the first time step, we use an Anderson accelerated Picard iteration to obtain an initial guess for Newton's method \cite{anderson1965iterative}.  For all subsequent time steps we use the solution at the previous time step as an initial guess.  A strict convergence tolerance of $10^{-12}$ is fixed.  We say this tolerance is strict, because we require $\|\delta Q\|_\infty \le \text{ tol},$ $\|\delta S\|_\infty \le \text{ tol},$ and $\|\delta \widehat{S}\|_\infty \le \text{ tol}.$  In \cite{woopen2014hybridized} the authors improve convergence of Newton's method by developing a damped iteration based on a CFL condition.
 
\subsubsection{Light oil saturation}
The HDG discretization for equation~(\ref{light_saturation_eq0}) follows closely to subsection~\ref{sub_wet}.  To this end, the HDG unknowns for equation~(\ref{light_saturation_eq0}) are $(\widehat{s}_g,s_g,\nabla s_g)$, where it is understood that $\widehat{s}_g$ is the light oil phase saturation restricted to the mesh skeleton.  In order for this to occur, we rewrite equation~(\ref{wet_saturation_eq0}) in first order form
\begin{align*}
{\bf q } - \nabla s_g &= 0
\\
\frac{\partial \phi s_g}{\partial t} 
 + \nabla \cdot ( F_c(s_g) + F_v( s_g,{\bf q}) ) &= 0,
\end{align*}
where
$$
 F_c(s_g) = \frac{\lambda_g}{\lambda_t} {\bf u}_t  ,
\quad
\text{ and }
\quad
 F_v( s_g,{\bf q})  =\frac{\lambda_w \lambda_g}{\lambda_t} {\bf K} \nabla   p_{cwo} -\frac{ \lambda_g(\lambda_o+\lambda_w)}{\lambda_t} {\bf K} \nabla   p_{cgo}. 
$$
We assume that ${\bm u}_t^{m+1},{\bm s}_w^{m+1},$ and $\nabla s_w^{m+1}$ have been computed.  In an analogous manner one obtains:
 \begin{align*}  
 ({\bf q}^{m+1},{\bf v})_E 
 + 
 (s_g^{m+1},\nabla \cdot{\bf v})_E 
 - \langle\widehat{s_g}^{m+1},{\bf v}\cdot {\bf n} \rangle_{\Gamma_h  \setminus \Gamma_{s_g,D}}  
 &= \langle P s_{gD},{\bf v}\cdot {\bf n} \rangle_{\Gamma_{s_g,D}},
 \\
 \bigg(\frac{ \phi (s_g^{m+1}-s_g^{m})}{\Delta t} ,w\bigg)_E
-
(F_c(s_g^{m+1}) + F_v(s_g^{m+1},{\bf q}) ,\nabla w)_E 
+ 
\langle (  {\hat{F_c}} (\widehat{s_g}^{m+1}) + \hat{F_v}(\widehat{s_g}^{m+1},{\bf q}^{m+1},))\cdot {\bf n}  ,w \rangle_{\Gamma_h } 
& = 0,
\\
\langle
 [\![ 
 (\hat{F_c}(\widehat{s_g}^{m+1}) + \hat{F_v}(\widehat{s_g}^{m+1},{\bf q}^{m+1}))
 \cdot {\bm n}
  ]\!]
,
\mu 
\rangle_{\Gamma_h \setminus \Gamma_{s_g,N}}
&=\langle g_{s_g,N},{\bf v}\cdot {\bf n} \rangle_{\Gamma_{s_g,N}}
.
 \end{align*}
Linearization and static condensation for the light oil saturation follows the discussion presented in section~\ref{sub_wet}.

\subsection{Postprocessing the saturation variable}
\label{sec:postproc}
%The postprocessing procedure used in this work is explained here, and is inspired by the one established in~\cite{nguyen2009implicit}.  The element-by-element postprocessing of the saturation $s_{wh} $ is denoted by $s_{wh}^* $; and results in a new piecewise discontinuous polynomial approximation of degree $k+1$ such that

In this section we describe a simply postprocessing strategy which can enhance the accuracy of our saturation variables even further \cite{stenberg1991postprocessing,CockburnDG08}.  The postprocessing is performed element-by-element, for the saturation $s_{\alpha,h}\in W_h$ for $\alpha\in\{w,g,o\}$. We denote the postprocessed saturation by $s_{\alpha,h}^*$, which is a piecewise discontinuous polynomial approximation of degree $k+1$ that satisfies:
\begin{align}
( \nabla s_{\alpha,h}^*, \nabla w)_{ E } &= (  {\bm q}_{\alpha,h}, \nabla w)_{ E }, \quad \forall w\in \mathbb{P}_{k+1}(E),
\\
(  s_{\alpha,h}^*,1)_{ E } &= (s_{ \alpha,h},1)_{ E },
\label{eq:post_proc}
\end{align}
for $\alpha\in\{w,g,o\}$, and for all $E \in \mathcal{E}_h$. It is understood from context that ${\bm q}_{\alpha,h}=\nabla s_{\alpha,h}$ and $s_{ \alpha,h}$ are given.

In \cite{fabien1,fabien2} a similar postprocessing was adopted, with convergence rates for smooth solutions. There, it was observed that the postprocessed solution with polynomial degree $k>0$ converged at the rate of $k+2$ in the $L^2$-norm, and $k+1$ in $H^1$-norm.
We note that the postprocessing is element local, and thus, easily parallelizable. Moreover, the postprocessing can be applied only when addition accuracy is required. 
% Numerical evidence (see section~\ref{sec:ex_Manu}) shows that this postprocessing converges for $k>0$, at the rate of $k+2$ in the $L^2$-norm, and $k+1$ in $H^1$-norm.  We mention that various postprocessings exist, some of which have their resulting approximation not satisfying the original PDE in any sense \cite{cockburn2009hybridizable_sacco,cockburn2009superconvergent,stenberg1991postprocessing,CockburnDG08}.  Moreover, to guarantee superconvergence on Cartesian meshes, it is necessary to use a slightly larger finite element space than the standard tensor product space \cite{cockburn2012conditions}.  The postprocessing does not need to occur at every time step, it can be activated at whenever an enhanced solution is desired.  Superconvergence of this postprocessing (steady-state or otherwise) is reliant on both $s_{wh}$ and $\nabla s_{wh}$ converging optimally at the rate of $k+1$ in the $L^2$-norm~\cite{nguyen2009implicit}.  The system in equation~\eqref{eq:post_proc} is element local, and as such, is completely data parallel.  Furthermore, it is cheaper to compute than a fully coupled linear system \cite{fabien2}.

%\newpage
\section{Numerical examples}
This section presents a number of numerical examples to test the robustness and accuracy of the proposed approach. In Section~\ref{sec:manu} we perform a convergence analysis utilizing a known smooth manufactured solution.
Sections~\ref{subsub_hom}, \ref{section_het1}, \ref{sec:het2} consider more complicated material parameters. In these sections we always perform post-processing on the scalar variable. In all numerical experiments we leverage a Crank–Nicolson time integration \cite{wanner1996solving}. Higher-order time stepping that does not significantly impose a time step restriction is the subject of further research. All saturation quantities are postprocessed (as described in Section \ref{eq:post_proc}) except for the manufactured solution from Section \ref{sec:manu}.
 
\subsubsection{Manufactured solution}
\label{sec:manu}
Verification of the convergence of our algorithm is performed in this section.  It is essential to ensure that our method is working as intended, and due to the complexity of the three-phase flow model, exact solutions are not available.  As such, in the verification examples we provide in the following sections, no numerical convergence rates can be provided.  Only qualitative data is provided for those numerical examples.

The time step size $\Delta t$ is taken to be small enough to ensure the spatial accuracy is not polluted. We select analytic functions for $s_w$, $s_g$, $p_o$, $\lambda_\alpha$, $p_{cwo}$ and $p_{cgo}$, which our numerical approximations can be compared against.  The domain $\Omega$ is the unit square, and the medium is homogeneous with ${\bf K}=10^{-4}$, and $\phi\equiv 0.2$.  For simplicity, linear functions for capillary pressure and mobilities are used:
\begin{align*}
\lambda_w &= s_w, \quad
\lambda_g = s_g, \quad
\lambda_o = (1-s_w-s_g),
\\
p_{cwo}&=s_w-1, \quad
p_{cgo}=1-s_g.
\end{align*}
It should be noted that even though these coefficients are linear, equations~(\ref{wet_saturation_eq0}), and (\ref{light_saturation_eq0}) are still nonlinear in their respective phase variable.  The primary unknowns $s_w$, $s_g$, $p_o$ are given by the following expressions:
\begin{align*}
p_o&=  \cos{(0.125 \pi  (x+y+t))},\quad
s_w= 0.125(1-\sin{( 0.125\pi  (x+y+t))}),\quad
\\
s_g&= 0.125(1 + xy(1-x)(1-y)e^{-x^2-y^2}).
\end{align*}
The saturations are selected specifically so that $0<s_w,s_g<1,$ and $0<s_o= 1-s_w-s_g<1.$  We enforce Dirichlet boundary conditions for pressure, wetting saturation, and light oil saturation.  The above manufactured solutions create source terms that must be accounted for.  

The unit square is divided into $N\times N$ squares, and from this we form a uniform mesh of $2N\times 2N$ triangles.  Table~\ref{manu_tab} displays the $L^2$ norm error at $t = 0.5$ for the primary variables $s_w,$ $s_g,$ $\nabla s_w,$ $\nabla s_g$, ${\bm u}_t$ and $p_o$.  Optimal convergence rates of $k+1$ are attained, which showcases that the HDG method is highly accurate, even in a nonlinear setting. 

 The Anderson accelerated Picard-Newton linearization is very effective.  For the first time step on average it requires no more than eight iterations.  Subsequent time steps are substantially reduce the number of Newton iterations.  We observe that reducing the time step also improves the Newton convergence rate; however, the semi-implicit algorithm does allow us to take larger time steps at the cost of a few more Newton iterations.
\begin{table}[htb!]
\centering
\begin{tabular}{cccccccc} \toprule
   {$k$} &  {$N$} & {$\| s_w - s_w^h\|_2$} & {$\| \nabla s_w - \nabla s_w^h\|_2$} & {$\| s_g - s_g^h\|_2$} & {$\| \nabla s_g - \nabla s_g^h\|_2$} & {$\|  p_o -  p_o^h\|_2$} & {$\| {\bm u}_t-{\bm u}_h^h\|_2$}\\ \midrule
    1& 8   & 4.38e-02  & 3.08e-02 & 7.09e-02  & 4.84e-02 & 3.04e-02 & 4.47e-02  \\
     & 16  & 1.27e-02  & 9.03e-03 & 2.05e-02  & 1.41e-02 & 9.08e-03 & 1.31e-02   \\
     & 32  & 3.47e-03  & 2.46e-03 & 5.58e-03  & 3.86e-03 & 2.52e-03 & 3.60e-03   \\
     & 64  & 9.08e-04  & 6.45e-04 & 1.45e-03  & 1.01e-03 & 6.71e-04 & 9.46e-04     \\ \midrule
    
2   &8   & 5.23e-03  & 1.52e-03 & 1.08e-03  & 3.36e-03 & 2.32e-03 & 4.09e-03 \\
    &16  & 7.60e-04  & 2.12e-04 & 1.51e-04  & 4.67e-04 & 3.24e-04 & 6.15e-04    \\
    &32  & 1.03e-04  & 2.81e-05 & 2.00e-05  & 6.18e-05 & 4.30e-05 & 8.62e-05   \\
    &64  & 1.34e-05  & 3.62e-06 & 2.58e-06  & 7.95e-06 & 5.55e-06 & 1.15e-05      \\ \midrule
    
3   &8  & 3.43e-05  & 3.27e-05 & 4.60e-05  & 7.75e-05 & 1.05e-04 & 1.51e-04 \\
    &16 & 2.25e-06  & 2.21e-06 & 3.10e-06  & 5.05e-06 & 7.11e-06 & 1.01e-05   \\
    &32 & 1.45e-07  & 1.43e-07 & 2.02e-07  & 3.26e-07 & 4.64e-07 & 6.61e-07   \\ 
    &64 & 9.28e-09  & 9.19e-09 & 1.29e-08  & 2.08e-08 & 2.96e-08 & 4.21e-08    \\ 
    \midrule
    
4   &8  & 6.78e-07  & 6.75e-07 & 9.36e-07  & 2.21e-06 & 4.91e-06 & 3.43e-06 \\
    &16 & 2.30e-08  & 2.27e-08 & 3.13e-08  & 7.50e-08 & 1.64e-07 & 1.15e-07   \\
    &32 & 7.49e-10  & 7.37e-10 & 1.01e-09  & 2.44e-09 & 5.30e-09 & 3.75e-09    \\ 
    &64 & 2.39e-11  & 2.36e-11 & 3.23e-11  & 7.83e-11 & 1.68e-10 & 1.19e-10    \\ 
    \bottomrule
\end{tabular}
\caption{Convergence rates in $L^2$ norm on a uniform triangular mesh with $2N^2$ elements.  Optimal rates of $k+1$ are obtained for all approximate variables.}
\label{manu_tab}
\end{table}
\newpage
\subsubsection{Homogeneous porous media}
\label{subsub_hom}

For the first validation test we assume a homogeneous domain with dimensions $\Omega=[0,1000]\times[0,1000]$ (meters).  The permeability and porosity are fixed throughout the domain, with ${\bf K}\equiv 10^{-10}$ and $\phi\equiv 0.2$.  Phase viscosities are taken to be $\mu_w= 0.5$ cp, $\mu_o=1$ cp, and $\mu_g = 0.3 $ cp.  The left boundary has $s_w=0.82$, $s_g=0.11$, and $p_o= 19 \cdot 10^6$ Pa.  On the right boundary we set $s_w=0.3$, $s_g=0.54$, and $p_o= 15 \cdot 10^6$ Pa.  No flow conditions are imposed on the remaining boundaries.  Equipped with these boundary conditions, we expect the saturation data to propagate from the left boundary to the right.

The Leverett model \cite{leverett1941steady} is invoked for the capillary pressures and generalized Brooks-Corey model for the  mobilities:
\begin{align*}
\lambda_w &= s_w^2/\mu_w, \quad
\lambda_g = s_g^2/\mu_g, \quad
\lambda_o = (1-a_g)(1-s_w-s_g)^2/\mu_o+a_g(1-s_w-s_g)/\mu_o,
\\
p_{cwo}&=5\epsilon (2-s_w)(1-s_w), \quad
p_{cgo}=\epsilon(2-s_g)(1-s_g),
\end{align*}
where $\epsilon = 10^{-3}$ and $0\le a_g\le 1$.

\begin{figure}[htb!]
\hspace*{-8ex}
\subfigure[Wetting saturation]{\includegraphics[trim = 10mm 80mm 20mm 85mm, clip, scale = 0.4]{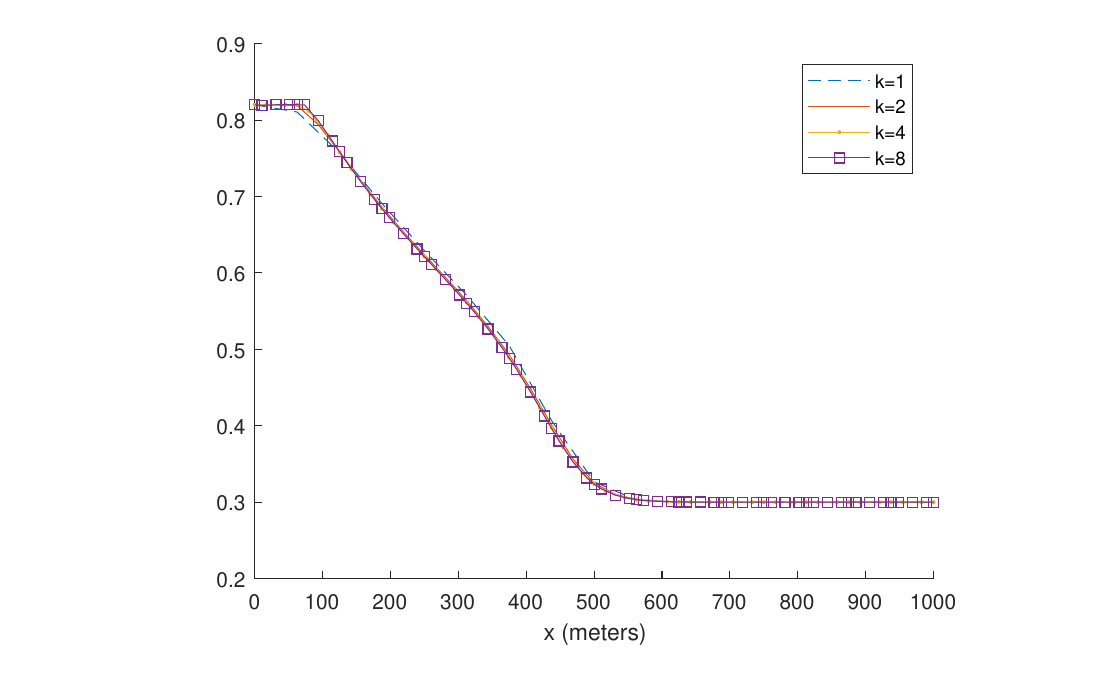}}
\hspace*{-8ex}
\subfigure[Wetting saturation (near $x=100$)]{\includegraphics[trim = 10mm 80mm 20mm 85mm, clip, scale = 0.4]{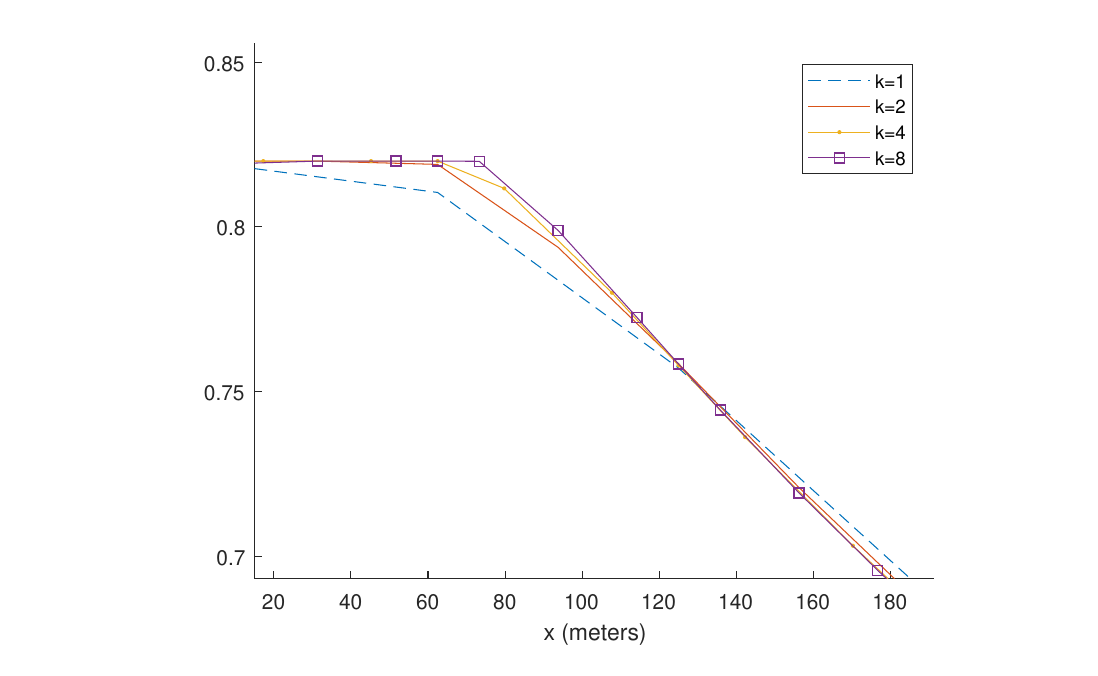}}
\caption{Wetting phase saturation profile along the line $y=500$ for various polynomial orders.  Simulation is terminated at $t=100$ days.}
\label{fig_hom1}
\end{figure}

To qualitatively assess the convergence of the numerical scheme, we examine the profile of the saturations $s_w$ and $s_g$ along various lines.  We do this by using the high-order trace space approximation that results from the HDG method.  A coarse uniform mesh of 512 triangles is used, and we vary the polynomial order from one to eight.  Termination of the simulation occurs at $t=100$ days, and the time step is set at $\Delta t = 1$ day.  The results are given in Fig.~\ref{fig_hom1}.  As the polynomial order increases the approximations converge.

\begin{figure}[htb!]
\hspace*{-8ex}
\subfigure[Light oil saturation]{\includegraphics[trim = 10mm 80mm 20mm 85mm, clip, scale = 0.4]{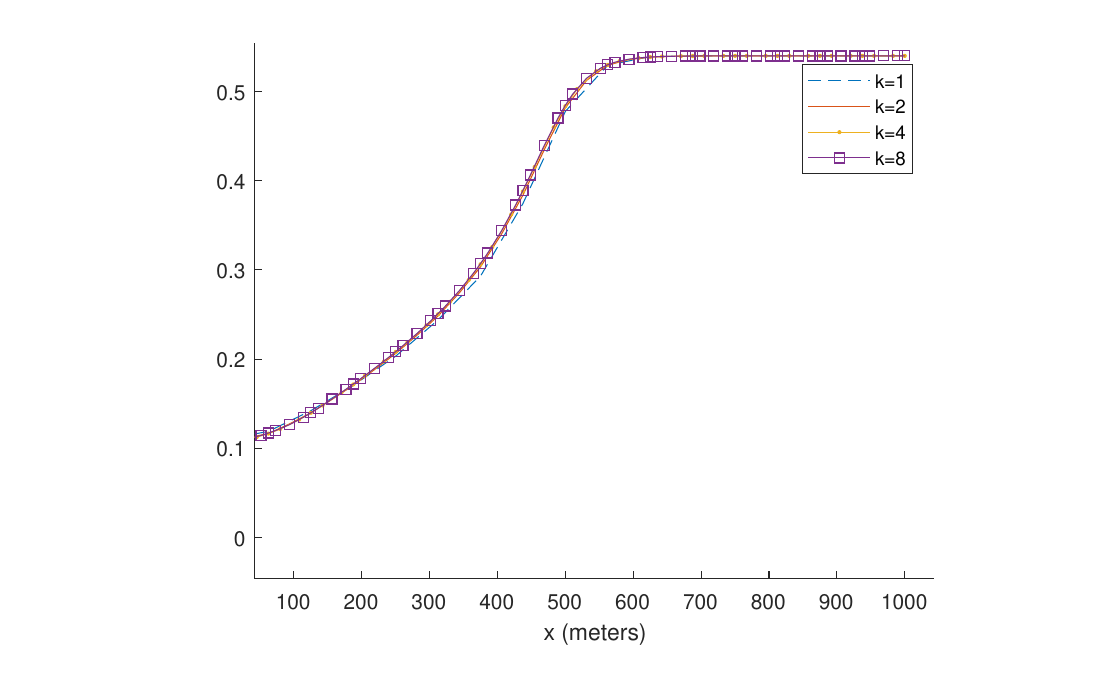}}
\hspace*{-8ex}
\subfigure[Light oil saturation (near $x=600$)]{\includegraphics[trim = 10mm 80mm 20mm 85mm, clip, scale = 0.4]{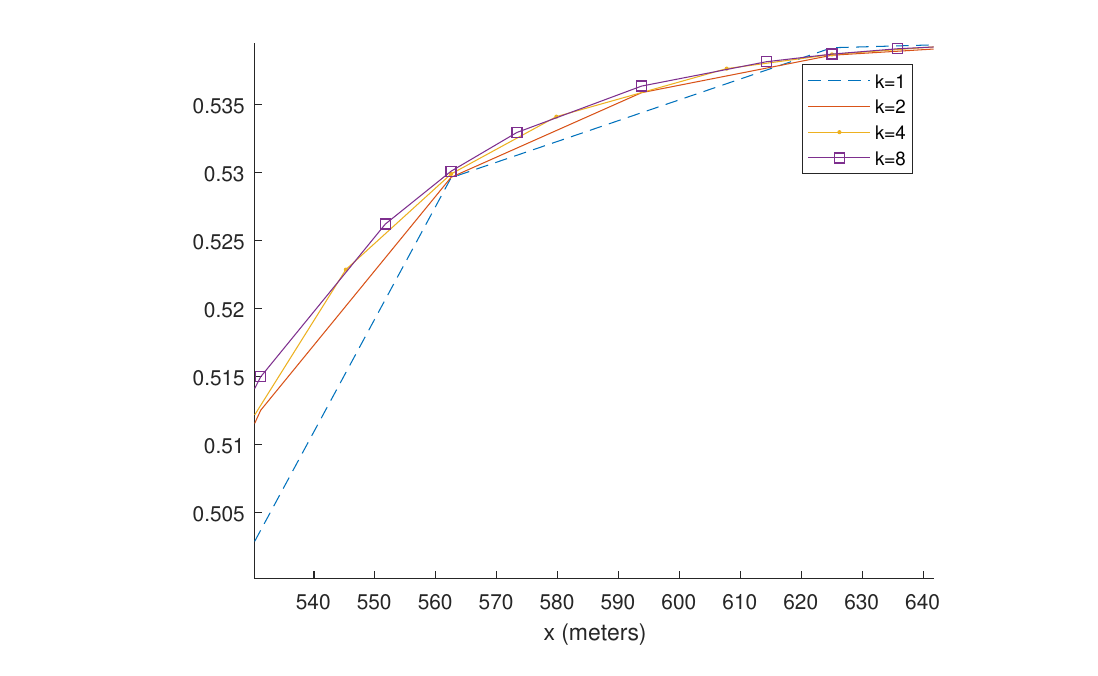}}
\caption{Light oil phase saturation profile along the line $y=500$ for various polynomial orders.  Simulation is terminated at $t=100$ days.}
\label{fig_hom2}
\end{figure}
\newpage
\subsubsection{Heterogeneous porous media}
\label{section_het1}

The following numerical experiments examines how our method performs in a heterogeneous medium.  In particular we explore different sets of relative permeabilities and capillary pressures, as well as discontinuous absolute permeability.

  Example 1 considers three-phase flow in a heterogeneous medium with a square permeability lens. Here the Leverett model capillary pressures and generalized Brooks-Corey model mobilities are used.  Example 2 has a rectangular medium with a permeability disk and logarithmic capillary pressure.  Examples 3 and 4 demonstrates the robustness of our method by simulating three-phase flow in highly heterogeneous permeability.
\paragraph{Example 1}
All parameters and boundary conditions are the same as subsection~\ref{subsub_hom} besides the permeability.  We impose a discontinuous permeability: $\bm K$ is equal to $10^{-13}$ in the lens $[250,500]\times [250,500]$ and $10^{-10}$ elsewhere. 

\begin{figure}[ht!]
\hspace*{-10ex}
\subfigure[Wetting saturation, $k=1$]{\includegraphics[trim = 10mm 80mm 20mm 85mm, clip, scale = 0.4]{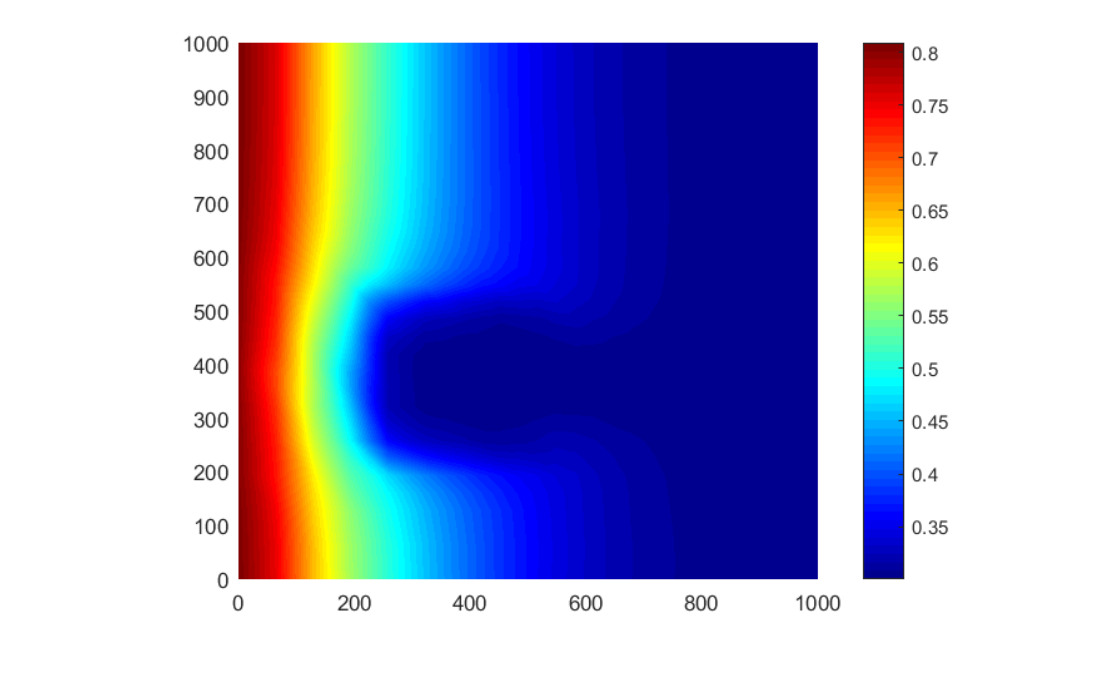}}
\hspace*{-15ex}
\subfigure[Light oil saturation, $k=1$]{\includegraphics[trim = 10mm 80mm 20mm 85mm, clip, scale = 0.4]{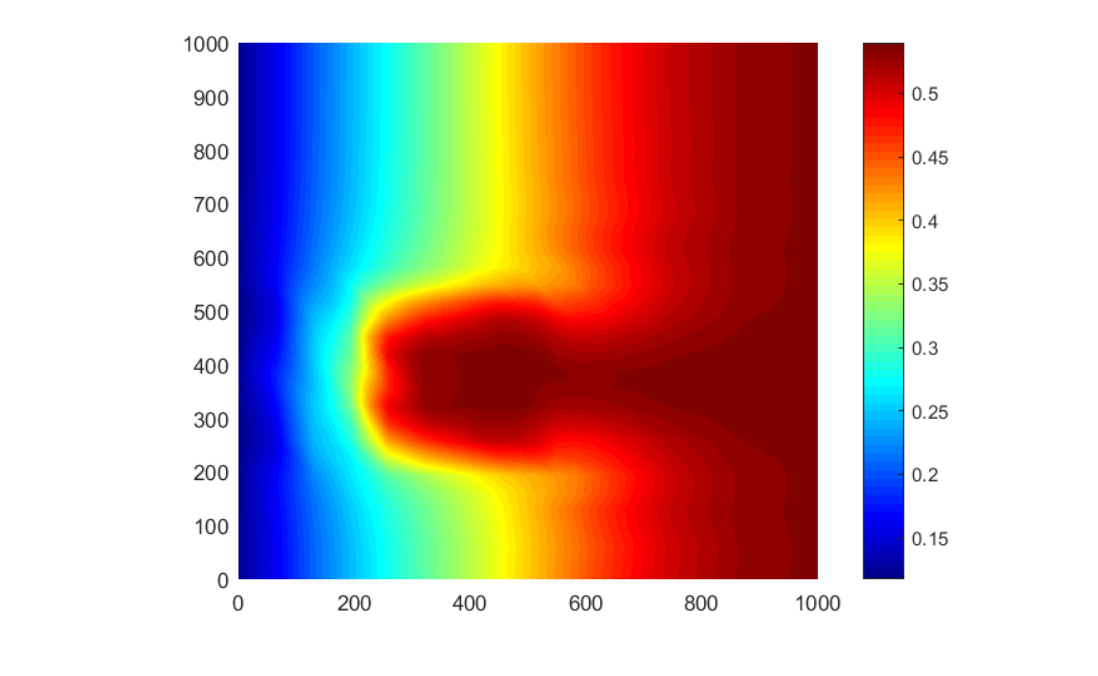}}
\newline
\hspace*{-10ex}
\subfigure[Wetting saturation, $k=2$]{\includegraphics[trim = 10mm 80mm 20mm 85mm, clip, scale = 0.4]{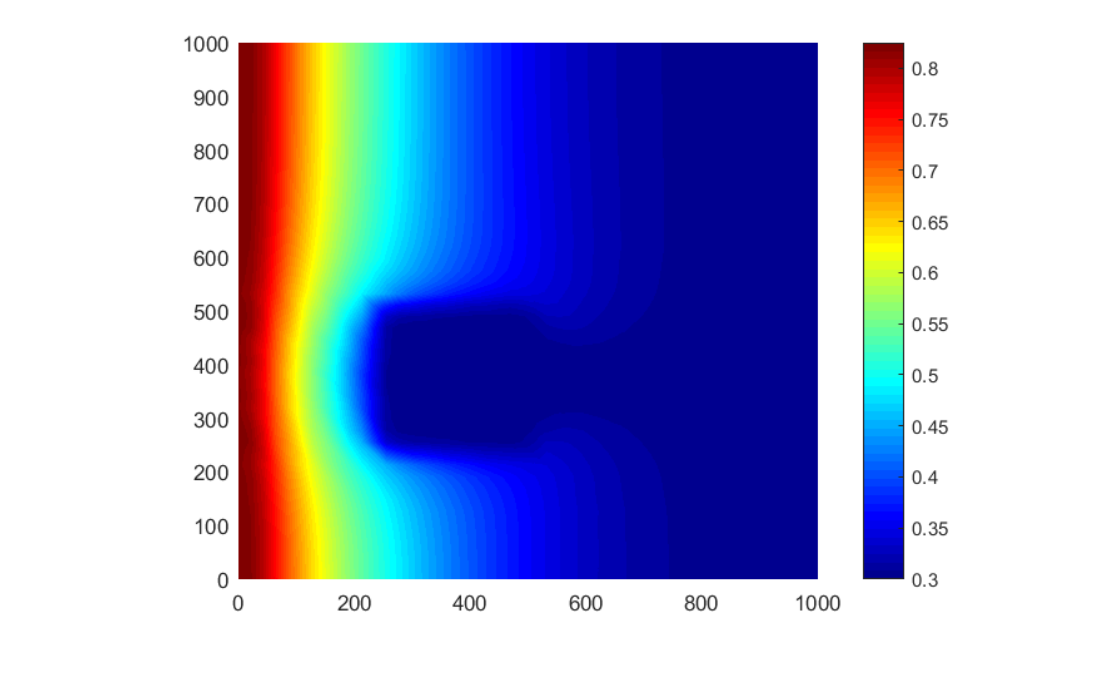}}
\hspace*{-15ex}
\subfigure[Light oil saturation, $k=2$]{\includegraphics[trim = 10mm 80mm 20mm 85mm, clip, scale = 0.4]{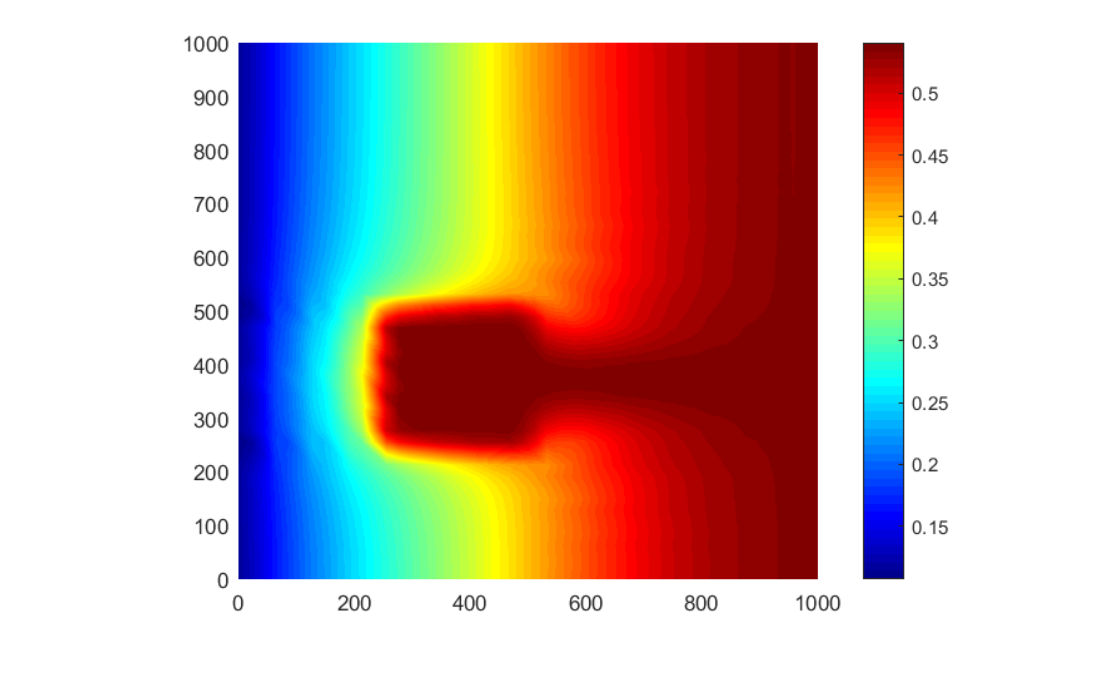}}
\caption{Wetting and light oil phase saturations for various polynomial orders.  Simulation is terminated at $t=100$ days.}
\label{fig_het000}
\end{figure}

\begin{figure}[htb!]
\hspace*{-10ex}
\subfigure[Wetting saturation, $k=4$]{\includegraphics[trim = 10mm 80mm 20mm 85mm, clip, scale = 0.4]{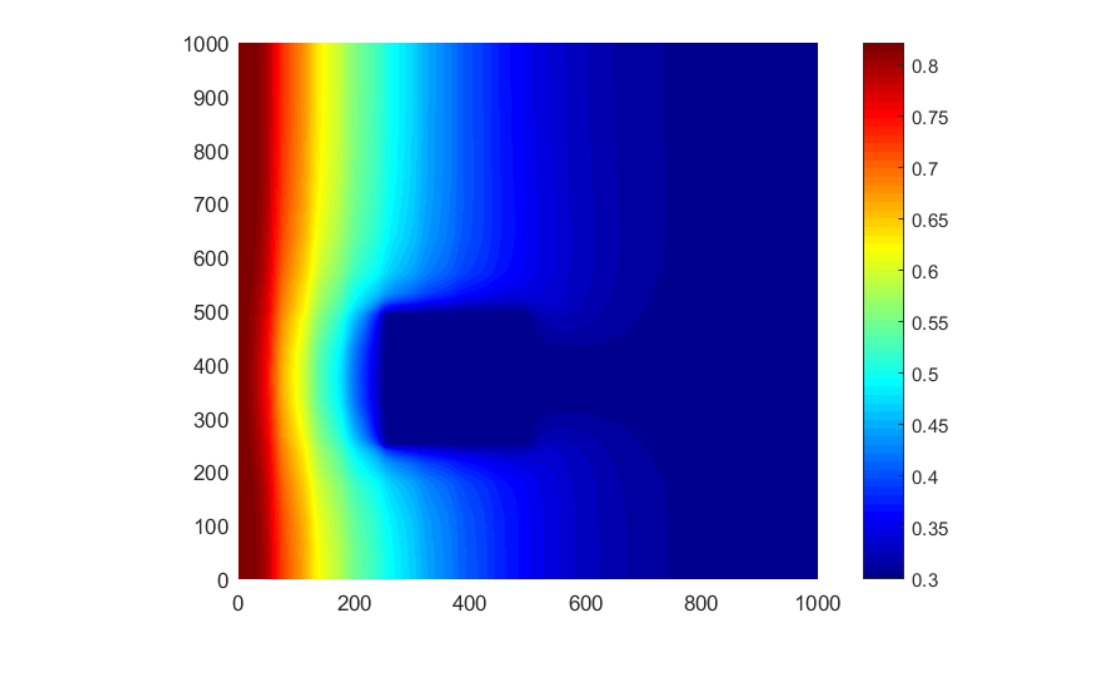}}
\hspace*{-15ex}
\subfigure[Light oil saturation, $k=4$]{\includegraphics[trim = 10mm 80mm 20mm 85mm, clip, scale = 0.4]{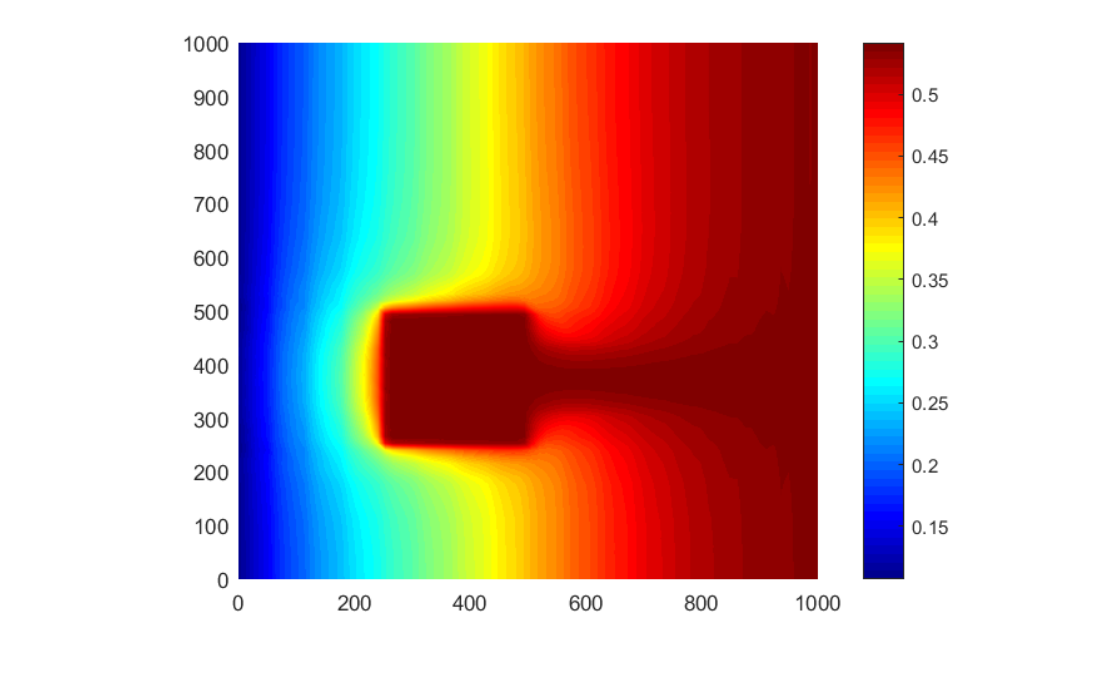}}
\newline
\hspace*{-10ex}
\subfigure[Wetting saturation, $k=8$]{\includegraphics[trim = 10mm 80mm 20mm 85mm, clip, scale = 0.4]{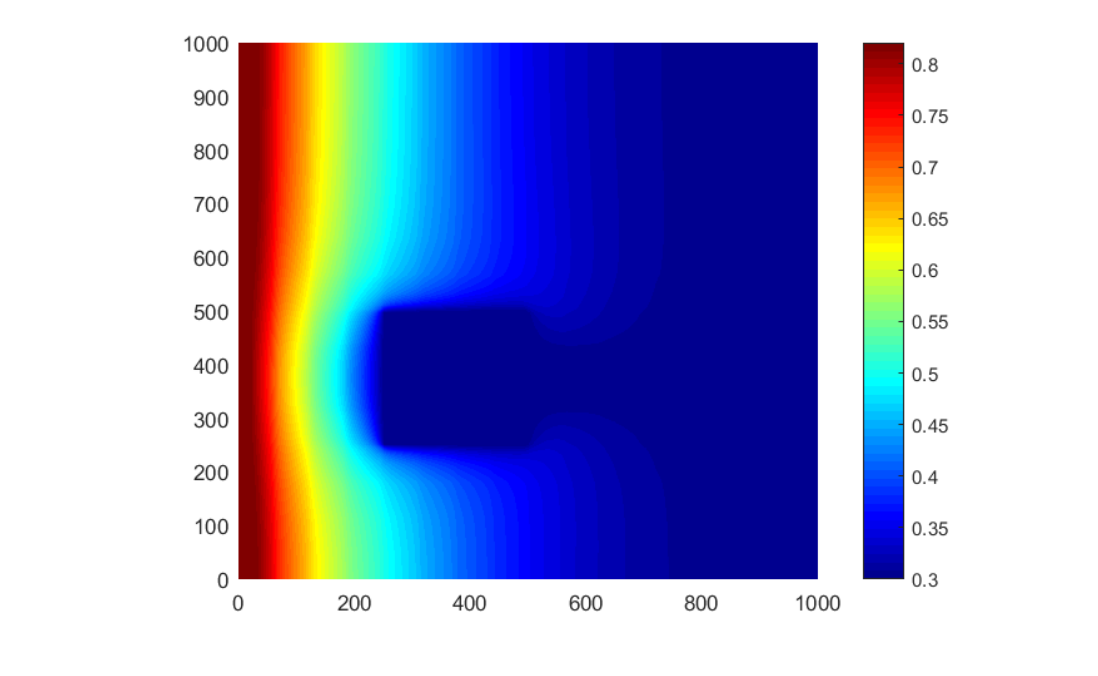}}
\hspace*{-15ex}
\subfigure[Light oil saturation, $k=8$]{\includegraphics[trim = 10mm 80mm 20mm 85mm, clip, scale = 0.4]{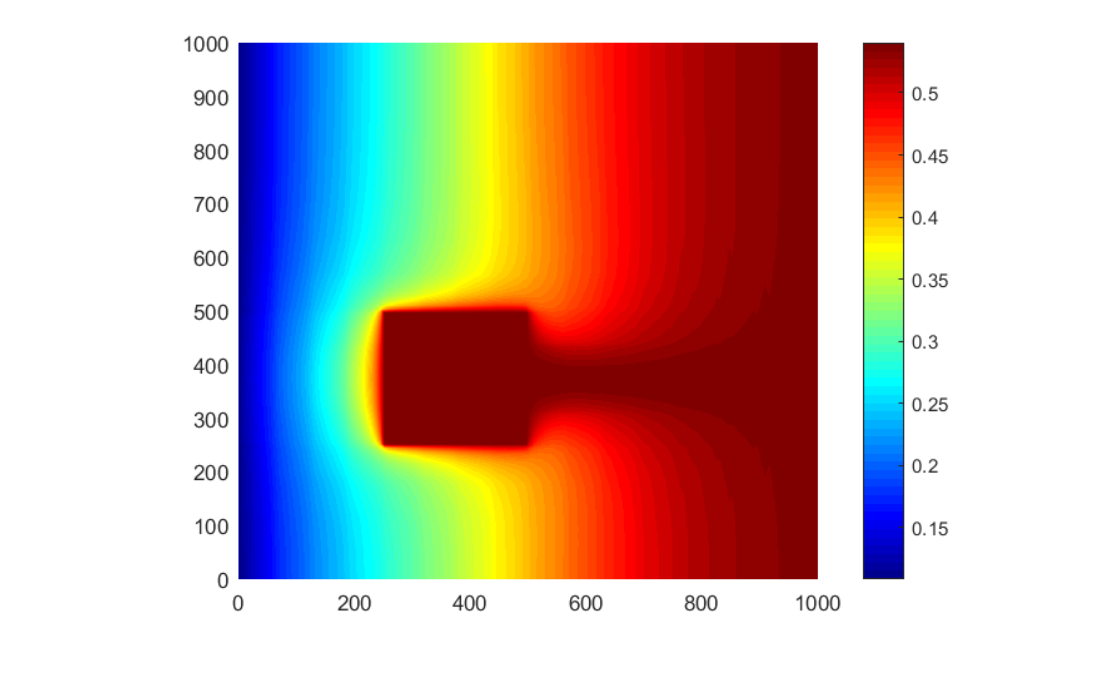}}
\newline
\hspace*{-10ex}
\subfigure[Wetting saturation, $k=16$]{\includegraphics[trim = 10mm 80mm 20mm 85mm, clip, scale = 0.4]{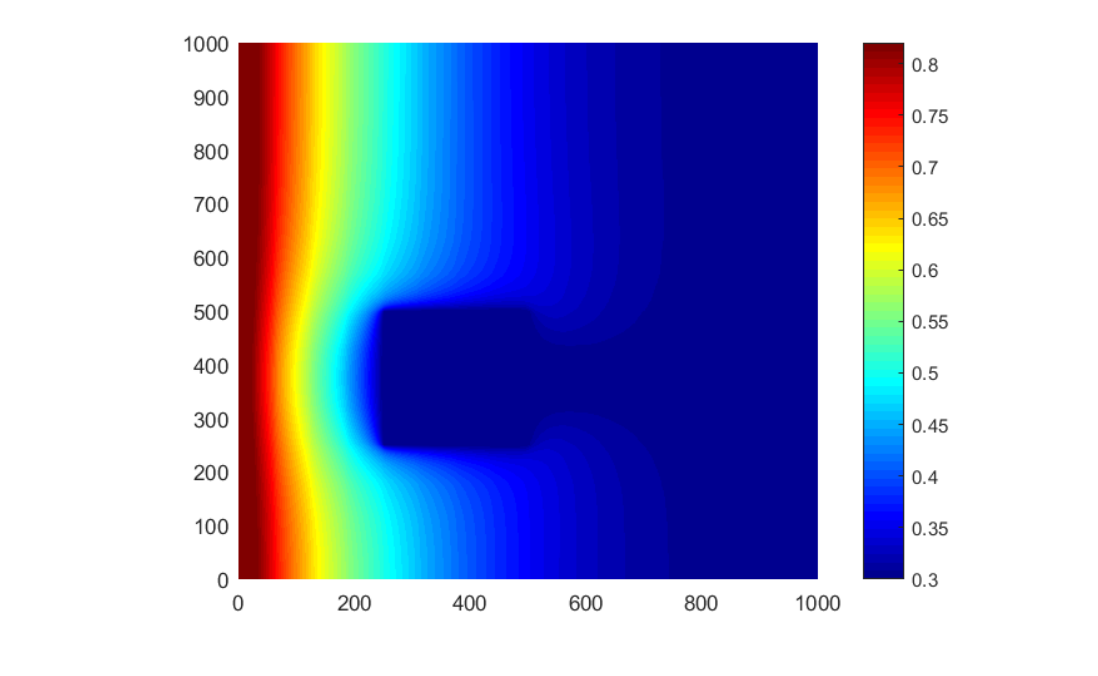}}
\hspace*{-15ex}
\subfigure[Light oil saturation, $k=16$]{\includegraphics[trim = 10mm 80mm 20mm 85mm, clip, scale = 0.4]{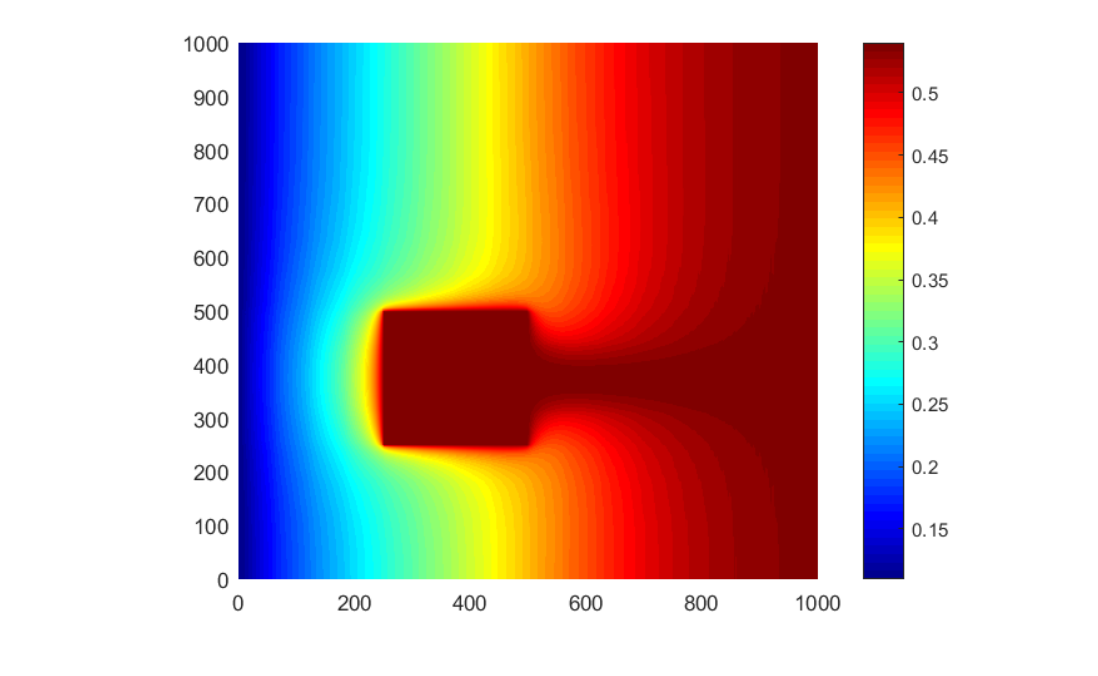}}
\caption{Wetting and light oil phase saturations for various polynomial orders.  Simulation is terminated at $t=100$ days.  Notice the use of polynomial orders that are rarely (never?) seen in approximations for flows in porous media.  These results challenge the standard rule of thumb which states that high-order polynomials }
\label{fig_het0}
\end{figure}

 Figs.~\ref{fig_het000} and \ref{fig_het0} plot the saturation approximations for different polynomial orders.  For piecewise linears, the mesh of 512 elements provides poor resolution.  By increasing the polynomial order (on the same mesh of 512 elements), we are able to capture the permeability lens with high accuracy.  Discontinuous piecewise octics significantly remove numerical artifacts.
 
\begin{figure}[ht!]   %5_1_sw
\hspace*{-10ex}
\subfigure[Wetting saturation, $k=1$, $|\mathcal{E}_h|=$ 2048 elements]{\includegraphics[trim = 10mm 80mm 20mm 85mm, clip, scale = 0.4]{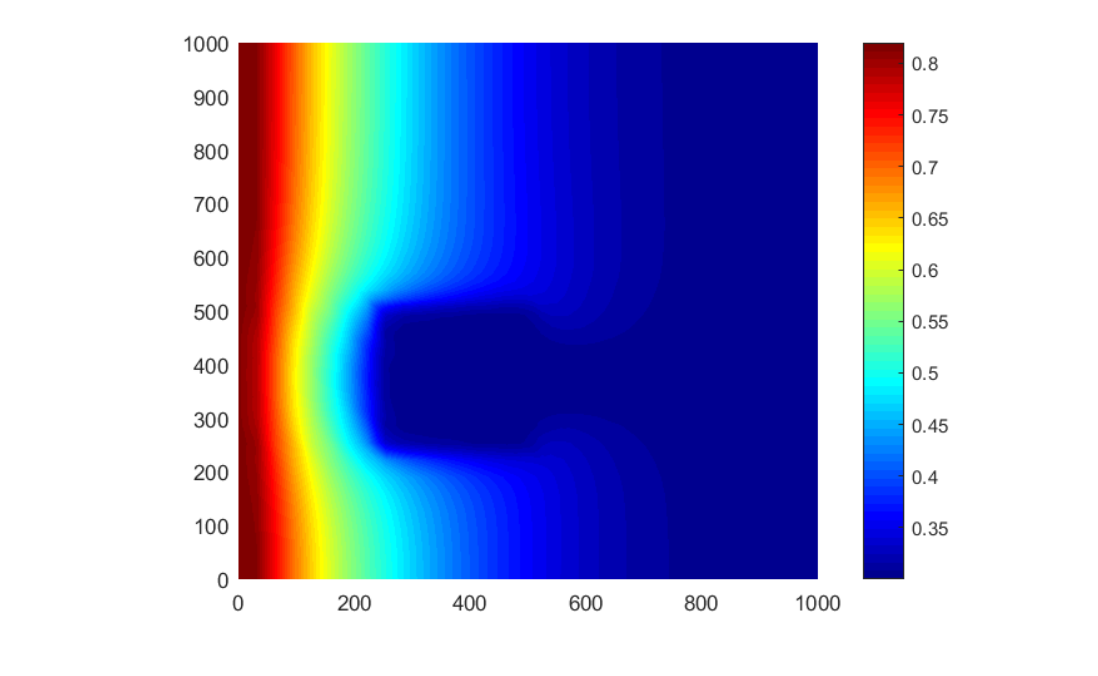}}
\hspace*{-15ex}
\subfigure[Light oil saturation, $k=1$, $|\mathcal{E}_h|=$ 2048 elements]{\includegraphics[trim = 10mm 80mm 20mm 85mm, clip, scale = 0.4]{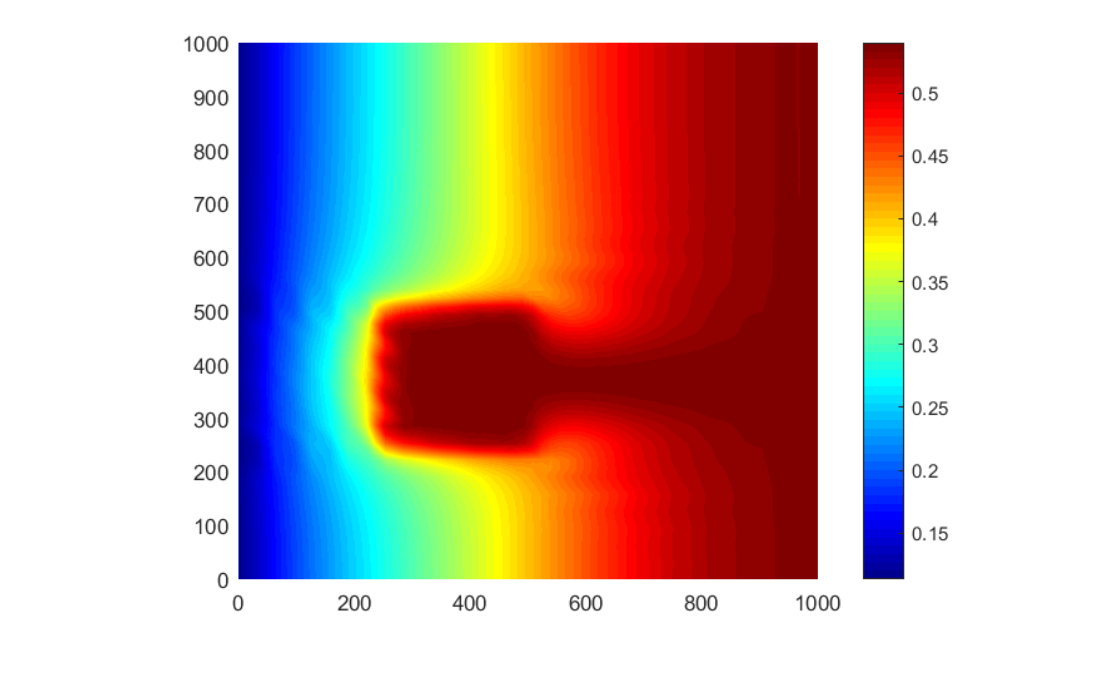}}
\newline
\hspace*{-10ex}
\subfigure[Wetting saturation, $k=1$, $|\mathcal{E}_h|=$ 8192 elements]{\includegraphics[trim = 10mm 80mm 20mm 85mm, clip, scale = 0.4]{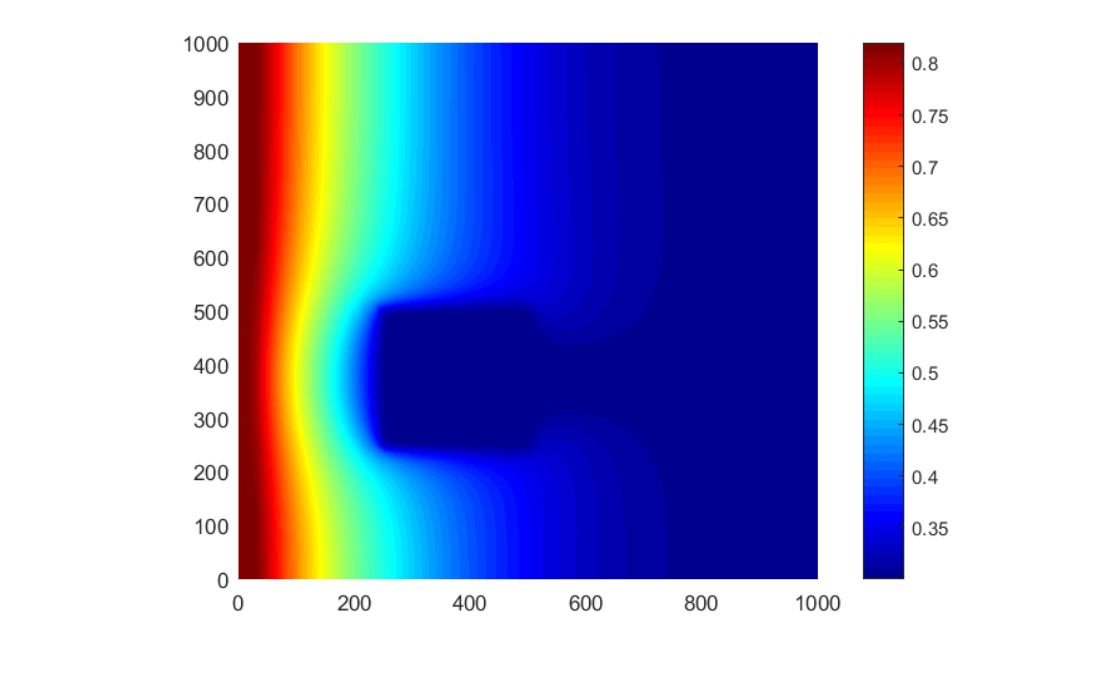}}
\hspace*{-15ex}
\subfigure[Light oil saturation, $k=1$, $|\mathcal{E}_h|=$ 8192 elements]{\includegraphics[trim = 10mm 80mm 20mm 85mm, clip, scale = 0.4]{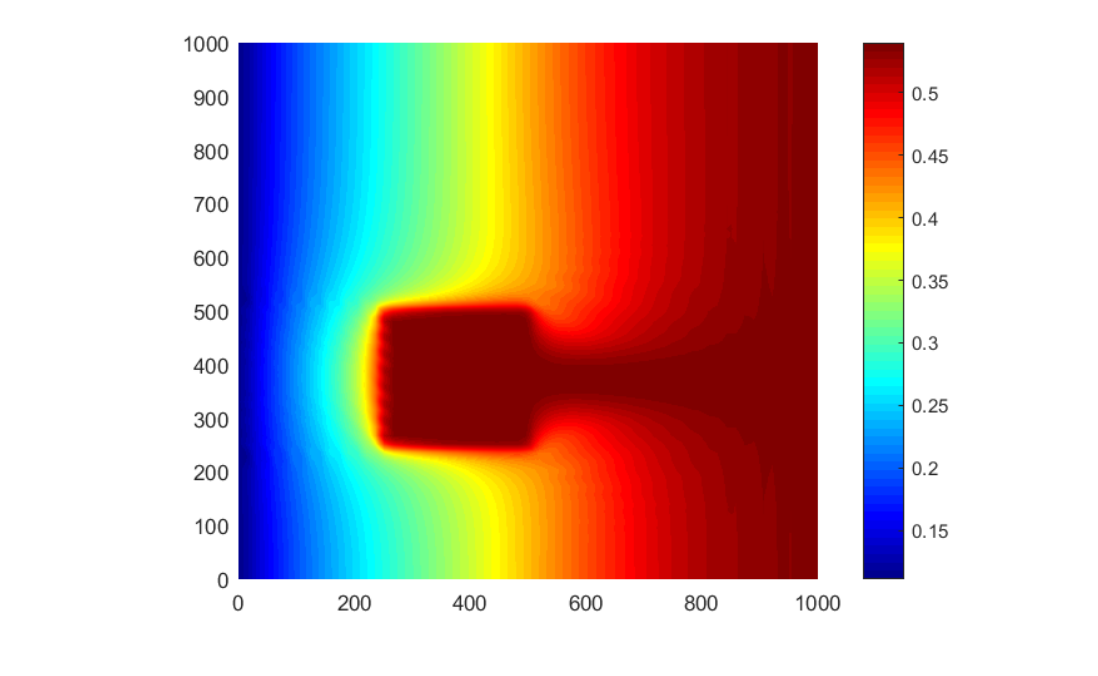}}
\newline
\hspace*{-10ex}
\subfigure[Wetting saturation, $k=1$, $|\mathcal{E}_h|=$ 32768 elements]{\includegraphics[trim = 10mm 80mm 20mm 85mm, clip, scale = 0.4]{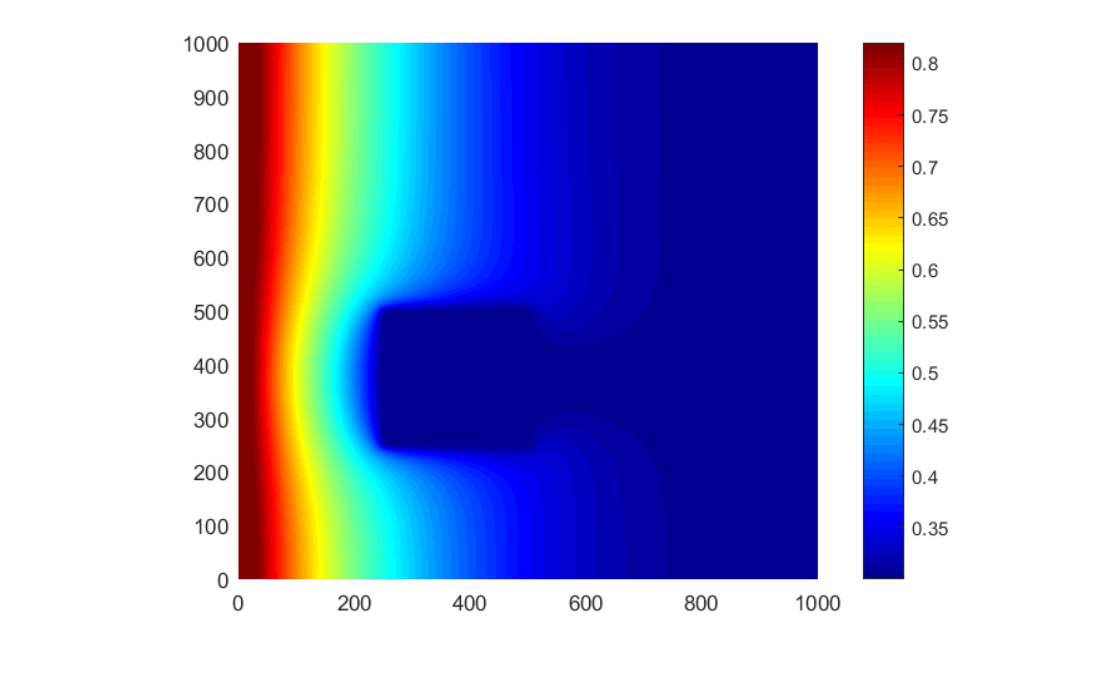}}
\hspace*{-15ex}
\subfigure[Light oil saturation, $k=1$, $|\mathcal{E}_h|=$ 32768 elements]{\includegraphics[trim = 10mm 80mm 20mm 85mm, clip, scale = 0.4]{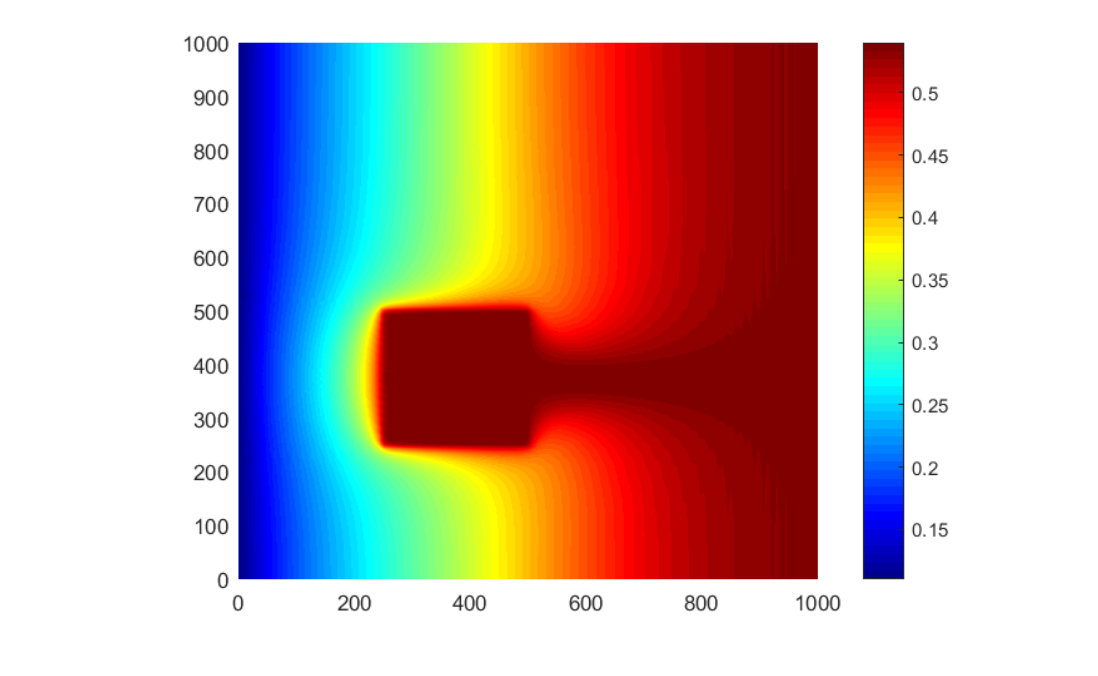}}
\caption{Wetting and light oil phase saturations for piecewise linears.  Simulation is terminated at $t=100$ days.  We use a sequence of uniform meshes with 2048, 8192, and 32768 elements.  Refining the mesh sufficiently reduces the oscillations and the saturations do not infiltrate the region of low permeability.}
\label{fig_het000r}
\end{figure}

It may appear that the HDG approximation is too diffusive for low polynomials orders (see Fig.~\ref{fig_het000}), as the wetting and light oil phases infiltrate the region of low permeability.  However, this is not the case, it simply is an issue of the mesh being too coarse to capture the multiphysics.  The effect of refining the mesh is shown in Fig.~\ref{fig_het000r}.  Further discussion of the beneifit of higher order polynomials is continued in Example 2.
 
\begin{figure}[h]
\hspace*{-8ex}
\subfigure[Wetting saturation]{\includegraphics[trim = 10mm 80mm 20mm 85mm, clip, scale = 0.4]{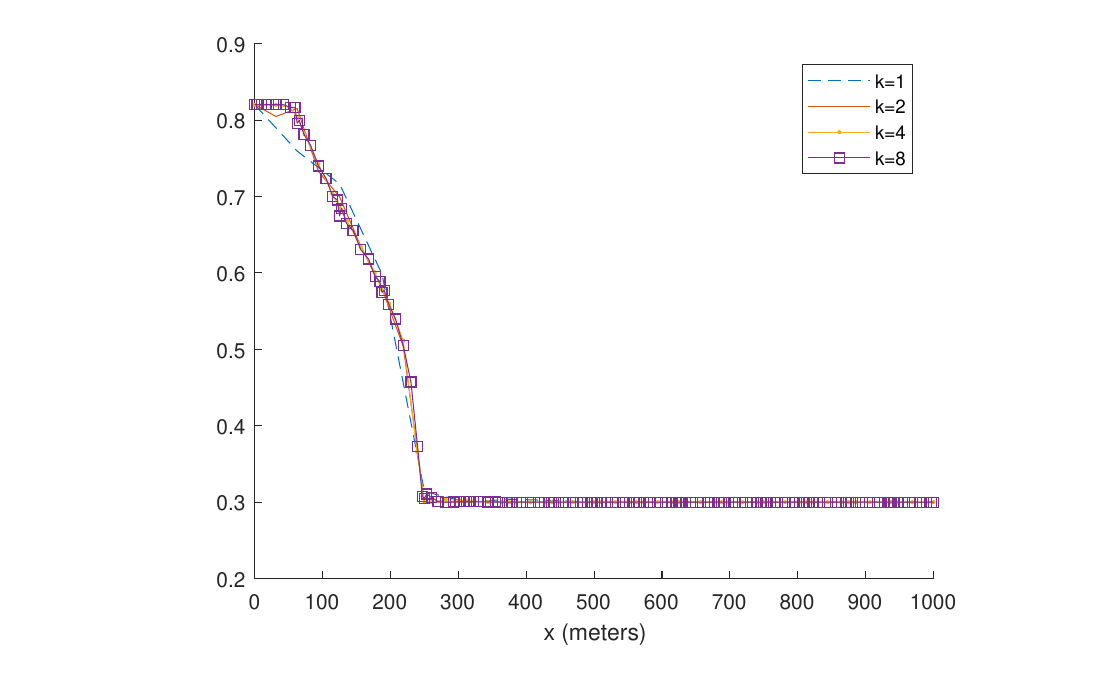}}
\hspace*{-8ex}
\subfigure[Light oil saturation]{\includegraphics[trim = 10mm 80mm 20mm 85mm, clip, scale = 0.4]{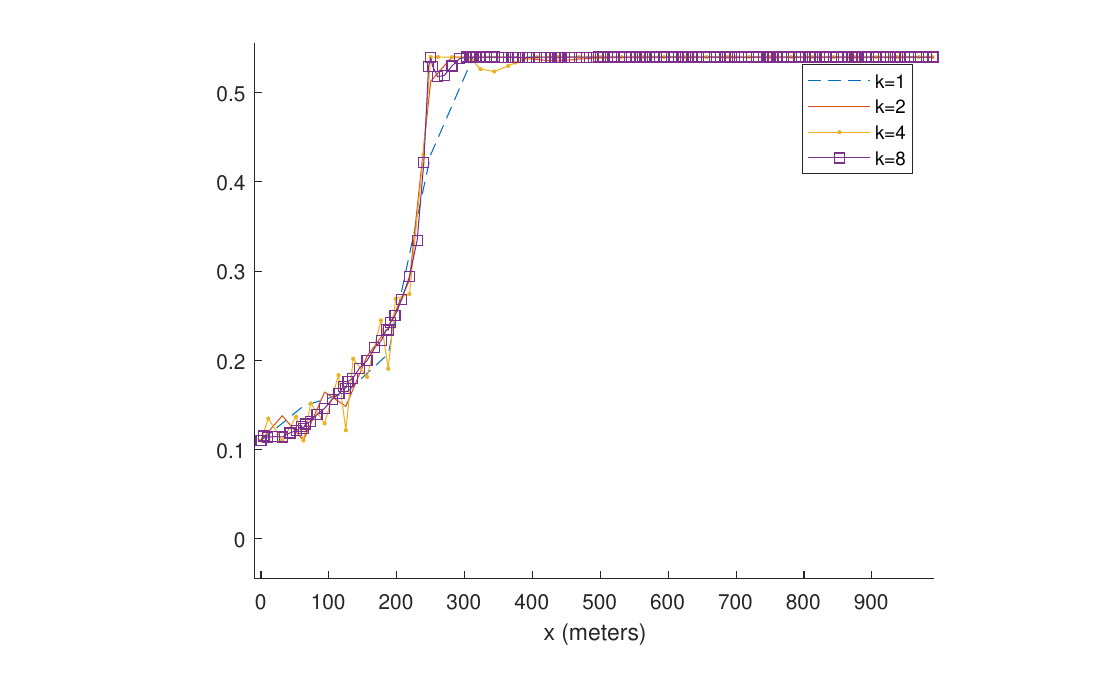}}
\caption{Wetting and light oil phase saturation profile along $y=375$ for various polynomial orders.  Simulation is terminated at $t=100$ days.  Under $p$-refinement spurious oscillations are reduced, and the approximation converges.}
\label{fig_het00}
\end{figure}

\begin{figure}[h]
\hspace*{-8ex}
\subfigure[Wetting saturation]{\includegraphics[trim = 10mm 80mm 20mm 85mm, clip, scale = 0.4]{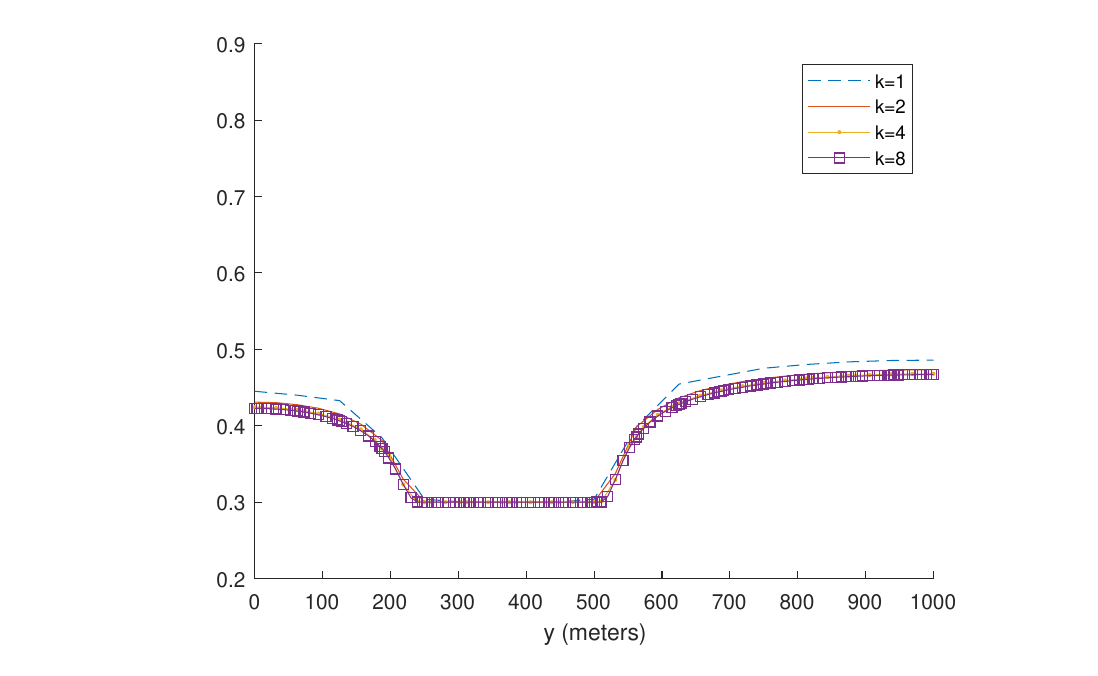}}
\hspace*{-8ex}
\subfigure[Light oil saturation]{\includegraphics[trim = 10mm 80mm 20mm 85mm, clip, scale = 0.4]{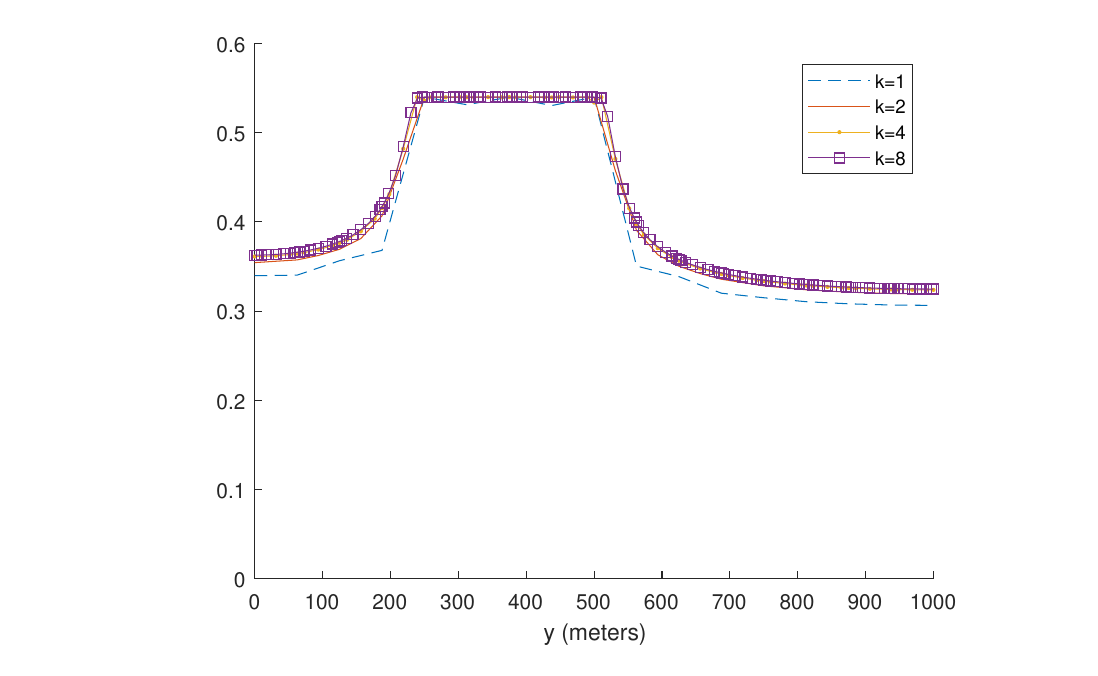}}
\caption{Wetting and light oil phase saturation profile along $x=375$ for various polynomial orders.  Simulation is terminated at $t=100$ days.  According to these profiles the approximation converges as $k$ increases.}
\label{fig_het01}
\end{figure}

\begin{figure}[ht!]
\hspace*{-10ex}
\subfigure[Full profile]{\includegraphics[trim = 10mm 80mm 20mm 85mm, clip, scale = 0.4]{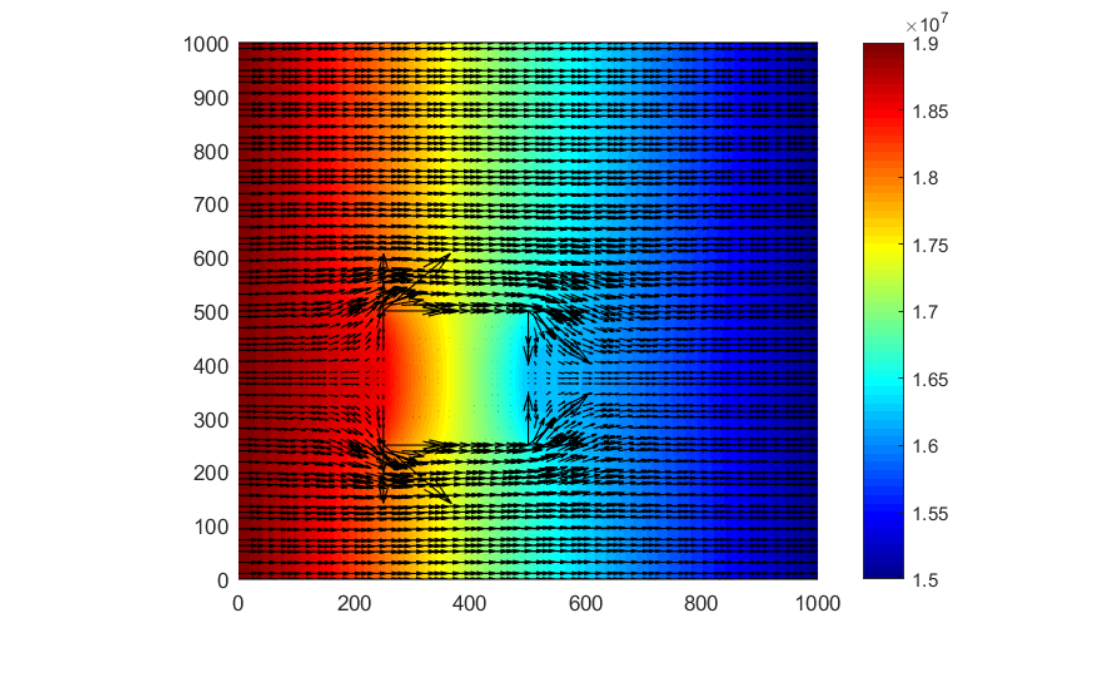}}
\hspace*{-15ex}
\subfigure[Zoom in near lens]{\includegraphics[trim = 10mm 80mm 20mm 85mm, clip, scale = 0.4]{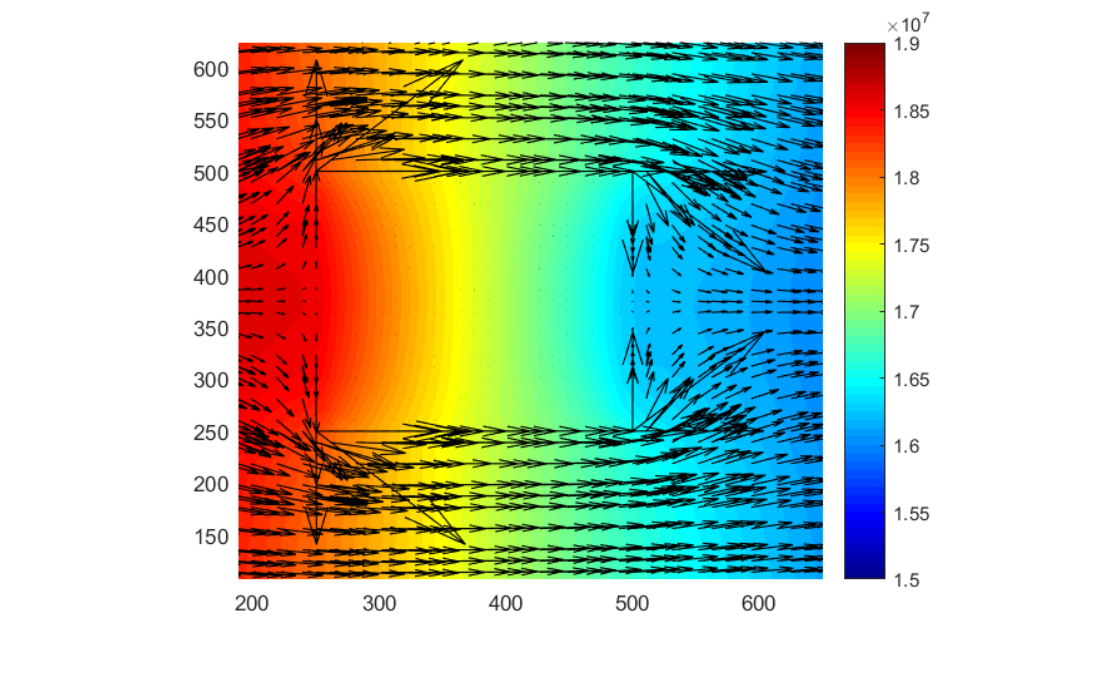}}
\caption{Heavy oil pressure with velocity field at $t=100$ days, $k=4$.  Arrow lengths are not to scale.}
\label{fig_het02}
\end{figure}

  Another visualization of convergence is provided in Fig.~\ref{fig_het00}; where we plot the approximate trace variables along the line $y=375$.  Convergence is more noticeable for $s_g$, as spurious oscillations are removed.  Horizontal profiles along $x=375$ can be observed in Fig.~\ref{fig_het01}.  The heavy oil pressure with an overlaid total velocity field is plotted in Fig.~\ref{fig_het02}.  The total velocity is uniform away from the lens, indicating flow from left to right.  Near the lens, the total velocity illustrates that flow must avoid the region of low permeability.
%\clearpage
\vspace*{-10ex}
\newpage
\paragraph{Example 2}
The domain $\Omega$ is a rectangle of dimensions $[0,300]\times [0,200]$ (meters).  The permeability $\bm K$ is equal to $10^{-9}$ everywhere in $\Omega$ except the disk centered at $(100,100)$ with radius 50 meters, where the permeability is 10,000 times smaller.  An unstructured mesh of 640 triangular elements is used, and the region of low permeability is grid-aligned.  The viscosities are $\mu_w=\mu_o=\mu_g = 1 $ cp.  The semi-implicit nature of our algorithm permits us to use large time steps.  In this case we set $\Delta t = 1$ day.  Porosity is held constant, $\phi\equiv 0.2$.  The left boundary has $s_w=0.82$, $s_g=0.11$, and $p_o= 3 \cdot 10^6$ Pa.  On the right boundary we set $s_w=0.3$, $s_g=0.54$, and $p_o= 1 \cdot 10^6$ Pa.  No flow conditions are imposed on the remaining boundaries.

\begin{figure}[ht!]
\hspace*{-5ex}
\subfigure[$p_o$ with velocity field at $t=50$ days, $k=4$]{\includegraphics[trim = 10mm 80mm 20mm 85mm, clip, scale = 0.4]{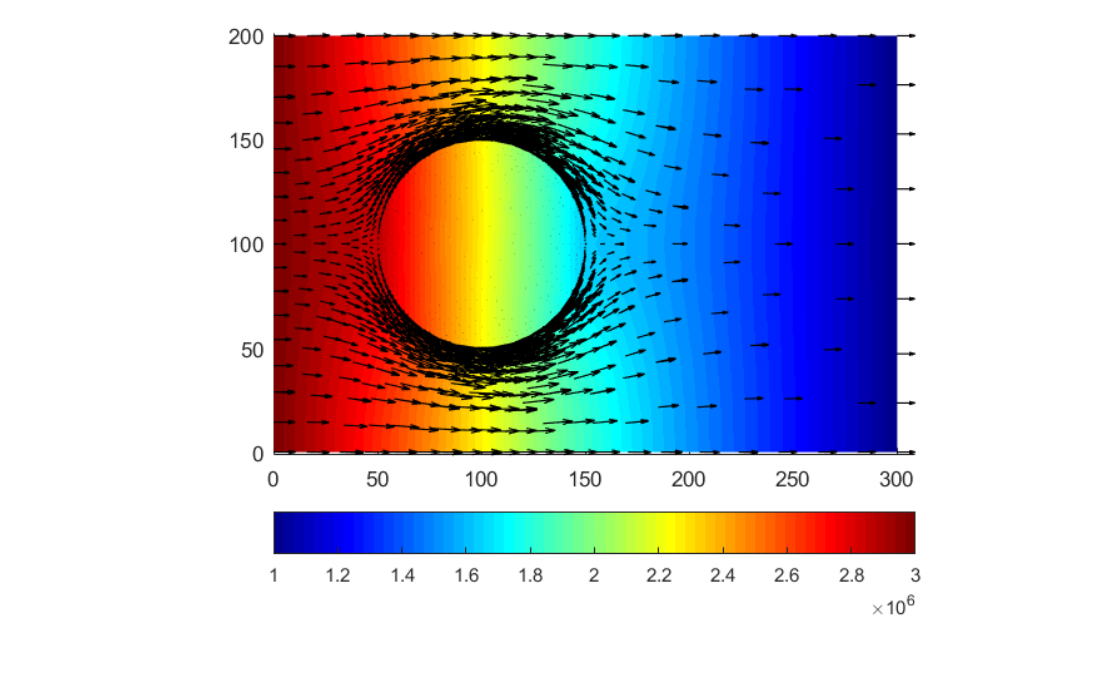}}
\hspace*{-5ex}
\subfigure[$p_o$ with velocity field at $t=50$ days, $k=4$]{\includegraphics[trim = 10mm 80mm 20mm 85mm, clip, scale = 0.4]{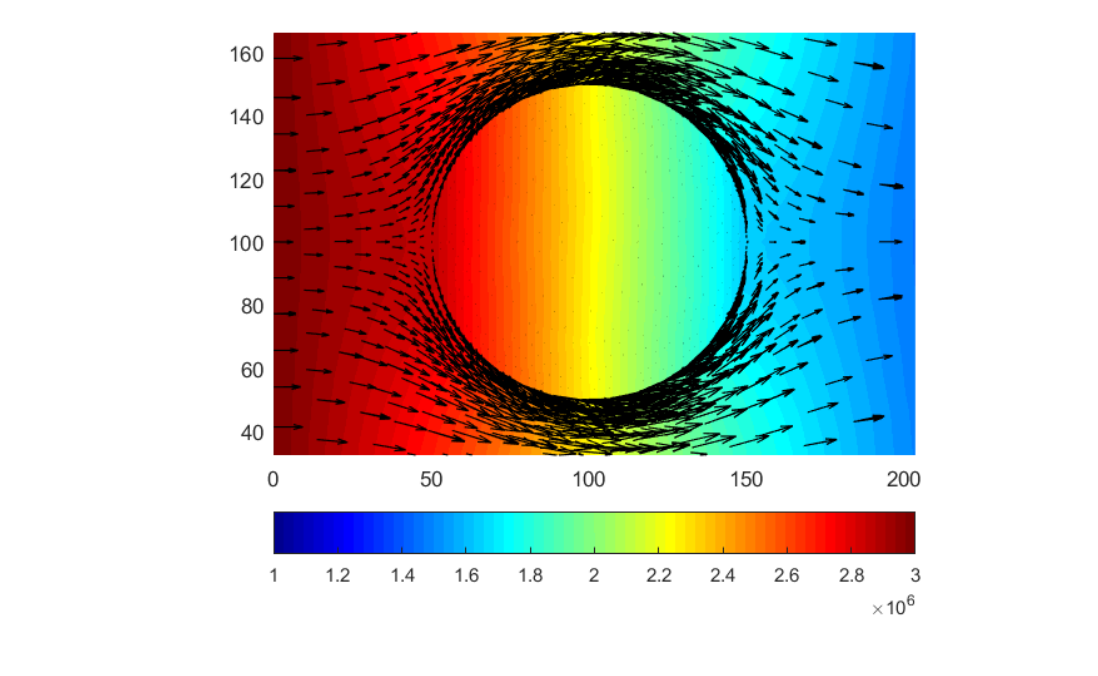}}
\newline
\hspace*{-5ex}
\subfigure[$s_w$ at $t=12.5$ days, $k=4$]{\includegraphics[trim = 10mm 80mm 20mm 85mm, clip, scale = 0.4]{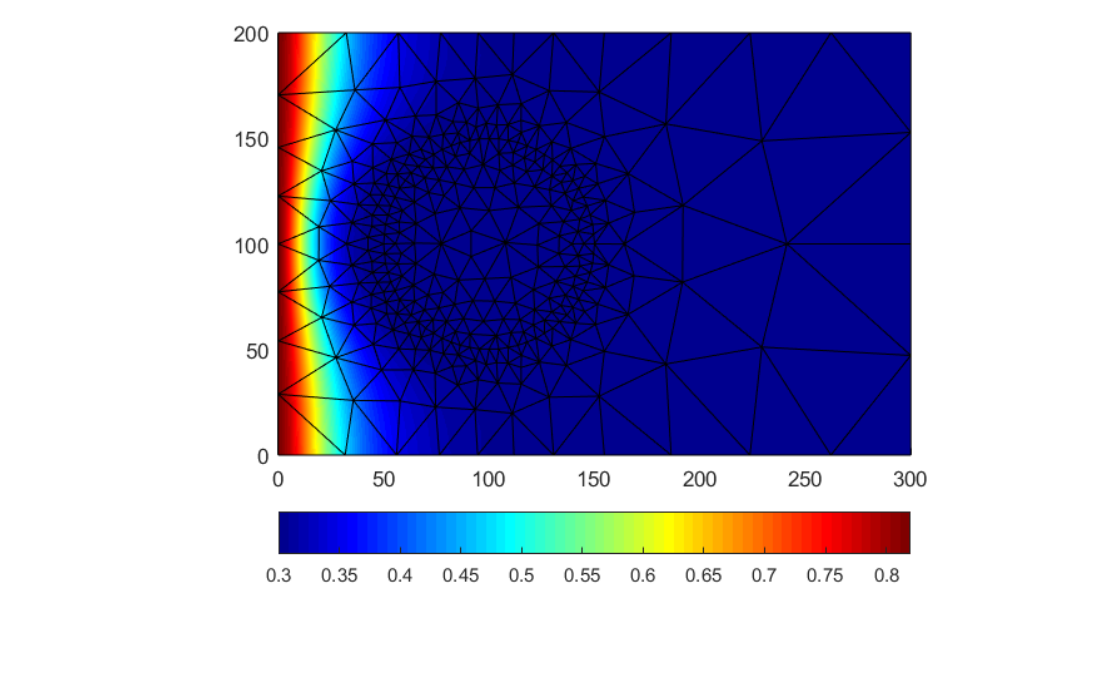}}
\hspace*{-5ex}
\subfigure[$s_w$ at $t=25$ days, $k=4$]{\includegraphics[trim = 10mm 80mm 20mm 85mm, clip, scale = 0.4]{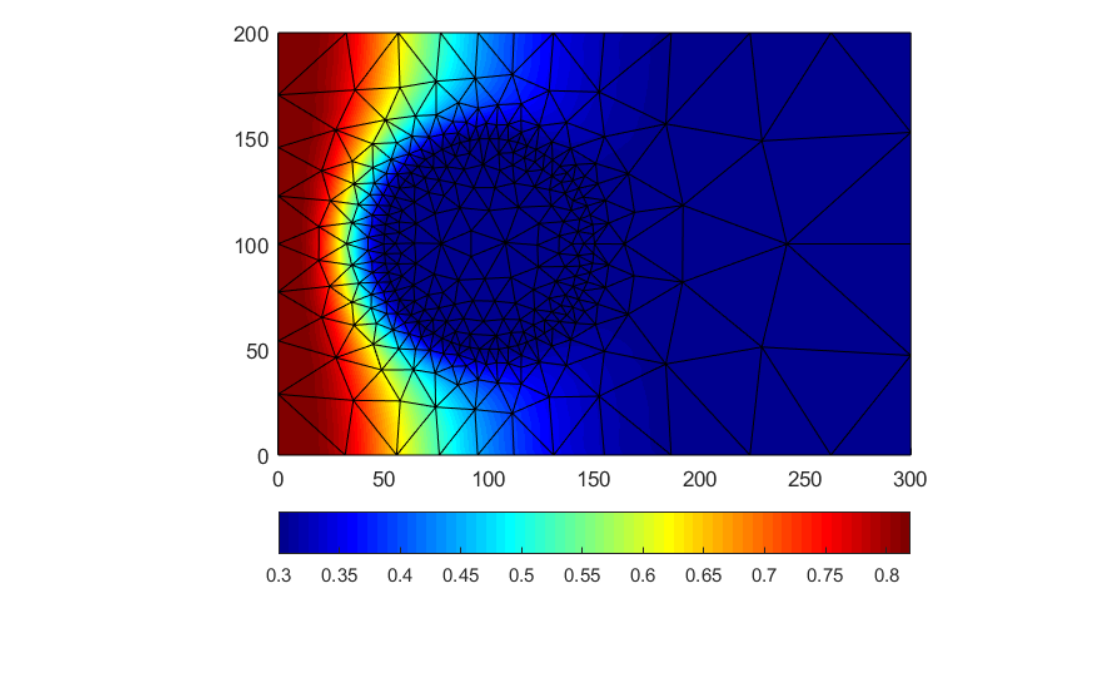}}
\newline
\hspace*{-5ex}
\subfigure[$s_w$ at $t=37.5$ days, $k=4$]{\includegraphics[trim = 10mm 80mm 20mm 85mm, clip, scale = 0.4]{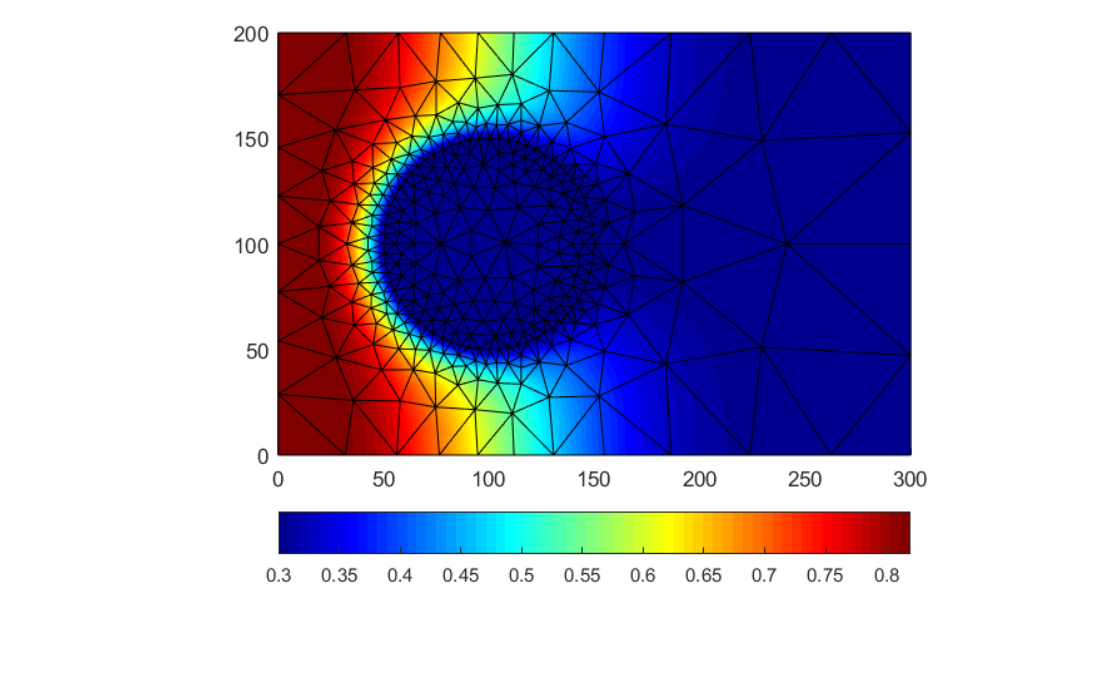}}
\hspace*{-5ex}
\subfigure[$s_w$ at $t=50$ days, $k=4$]{\includegraphics[trim = 10mm 80mm 20mm 85mm, clip, scale = 0.4]{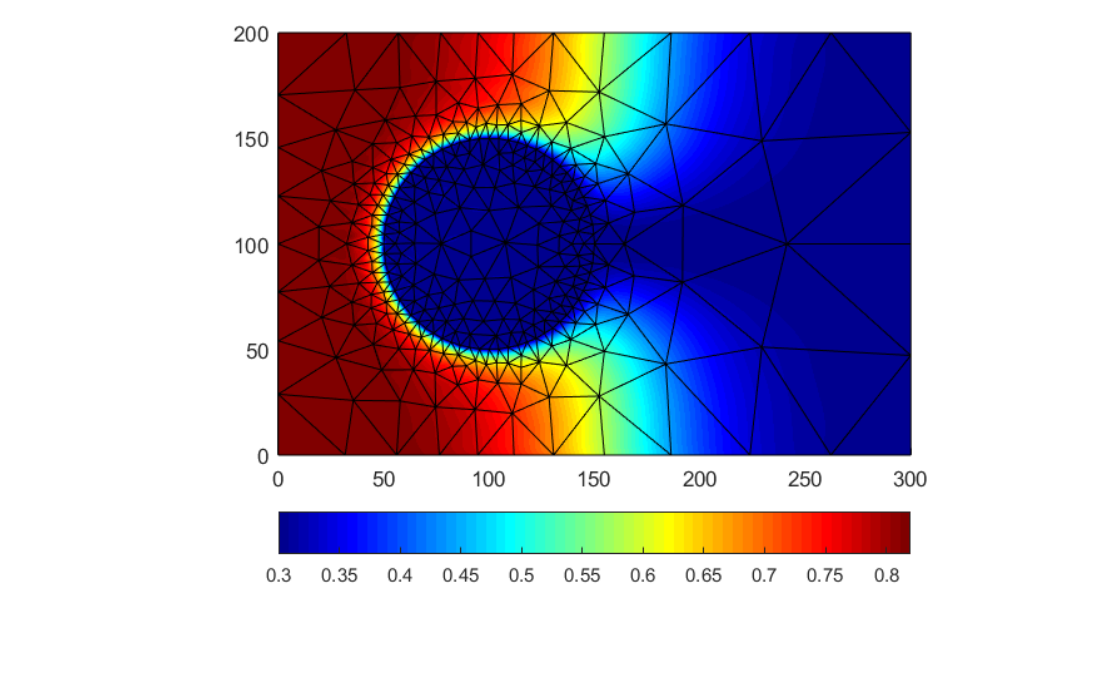}}
\caption{Heavy oil pressure at 50 days (see (a) and (b)).  Wetting saturation at $t=12.5$, 25, 37.5, and 50 days(see (c), (d), (e), and (f)).}
\label{het_2}
 
\end{figure}

\begin{figure}[htb!]
\hspace*{-5ex}
\subfigure[Wetting saturation, $k=1$]{\includegraphics[trim = 10mm 80mm 20mm 85mm, clip, scale = 0.4]{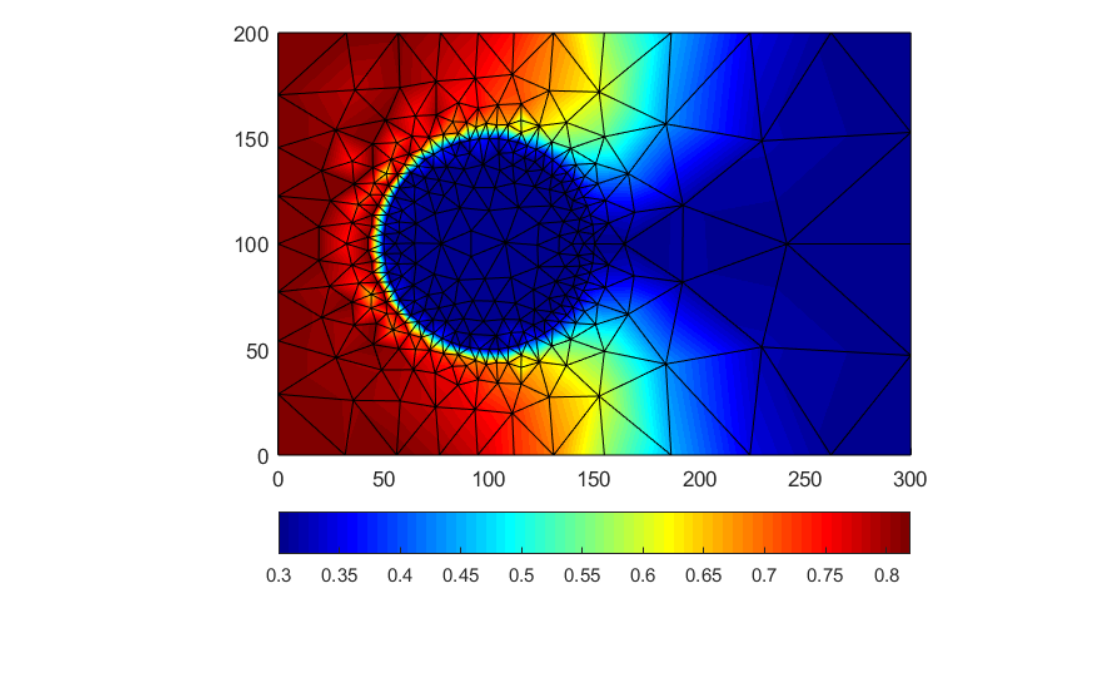}}
\hspace*{-22ex}
\subfigure[Light oil saturation, $k=1$]{\includegraphics[trim = 10mm 80mm 20mm 85mm, clip, scale = 0.4]{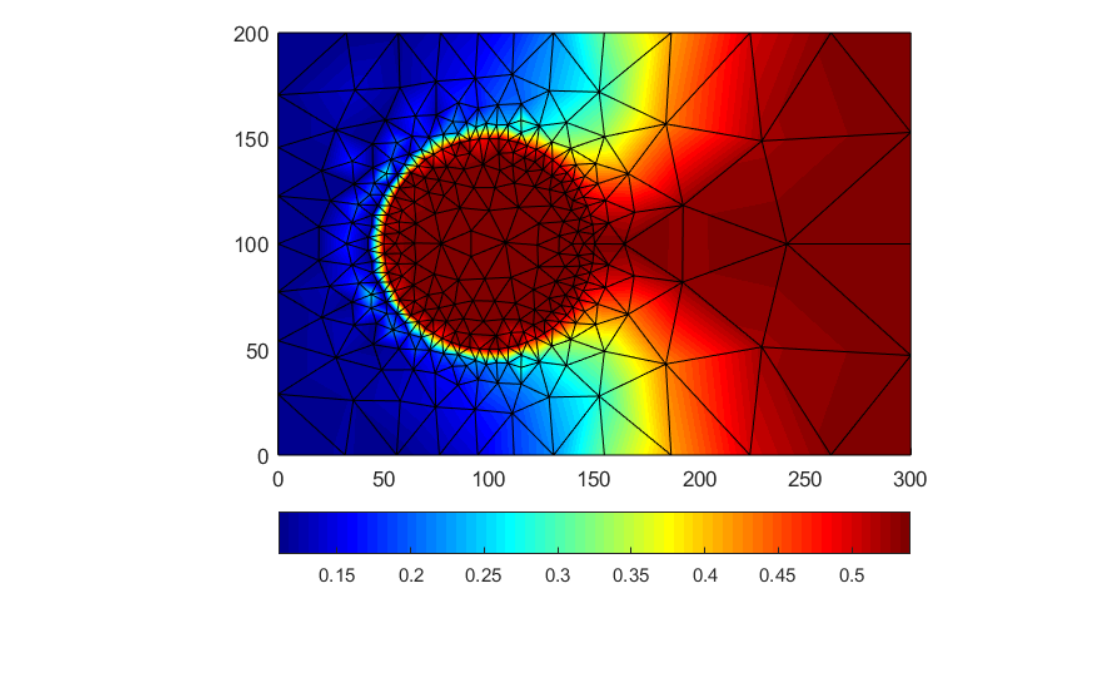}}
\newline
\hspace*{-5ex}
\subfigure[Wetting saturation, $k=2$]{\includegraphics[trim = 10mm 80mm 20mm 85mm, clip, scale = 0.4]{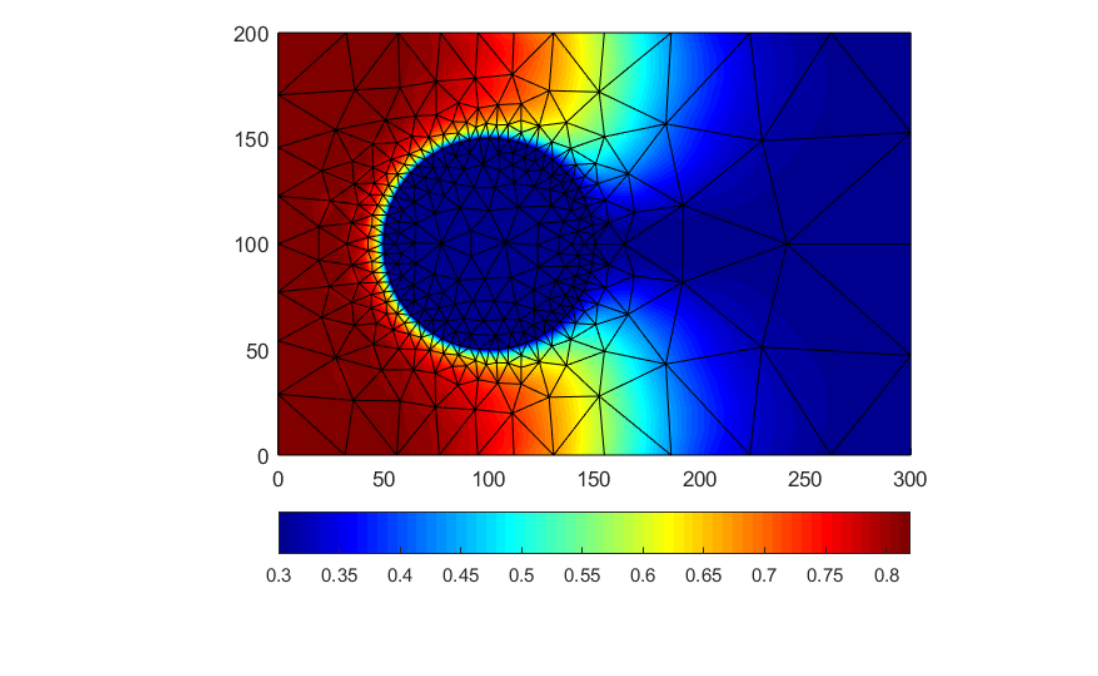}}
\hspace*{-22ex}
\subfigure[Light oil saturation, $k=2$]{\includegraphics[trim = 10mm 80mm 20mm 85mm, clip, scale = 0.4]{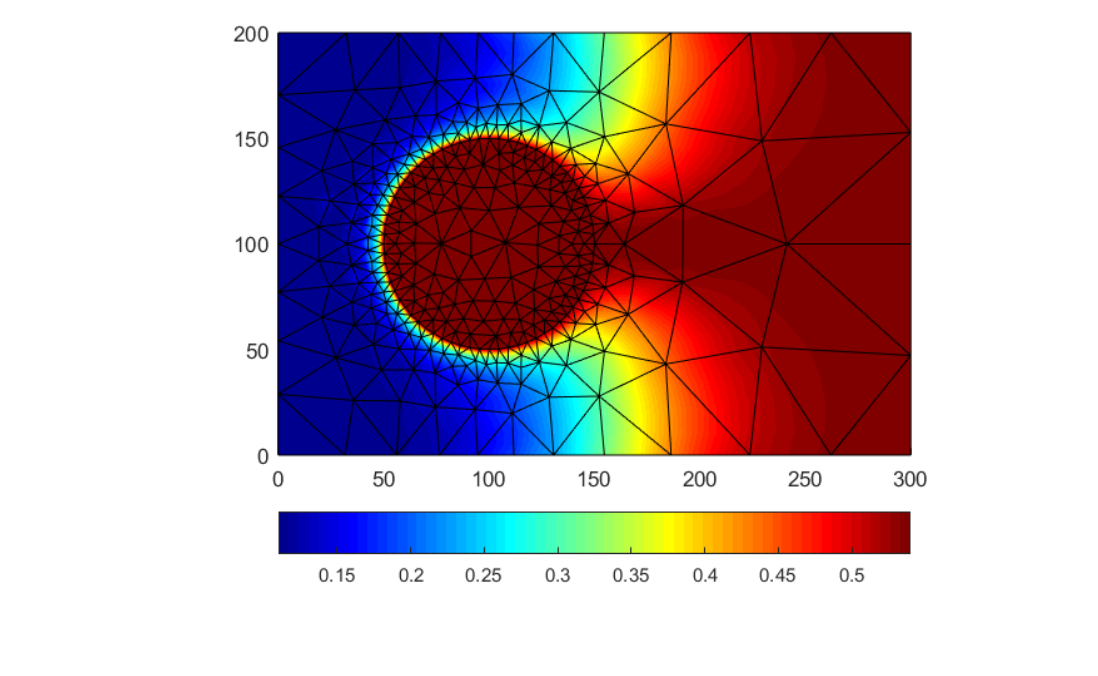}}
\newline
\hspace*{-5ex}
\subfigure[Wetting saturation, $k=4$]{\includegraphics[trim = 10mm 80mm 20mm 85mm, clip, scale = 0.4]{het_1_p4_wett.pdf}}
\hspace*{-22ex}
\subfigure[Light oil saturation, $k=4$]{\includegraphics[trim = 10mm 80mm 20mm 85mm, clip, scale = 0.4]{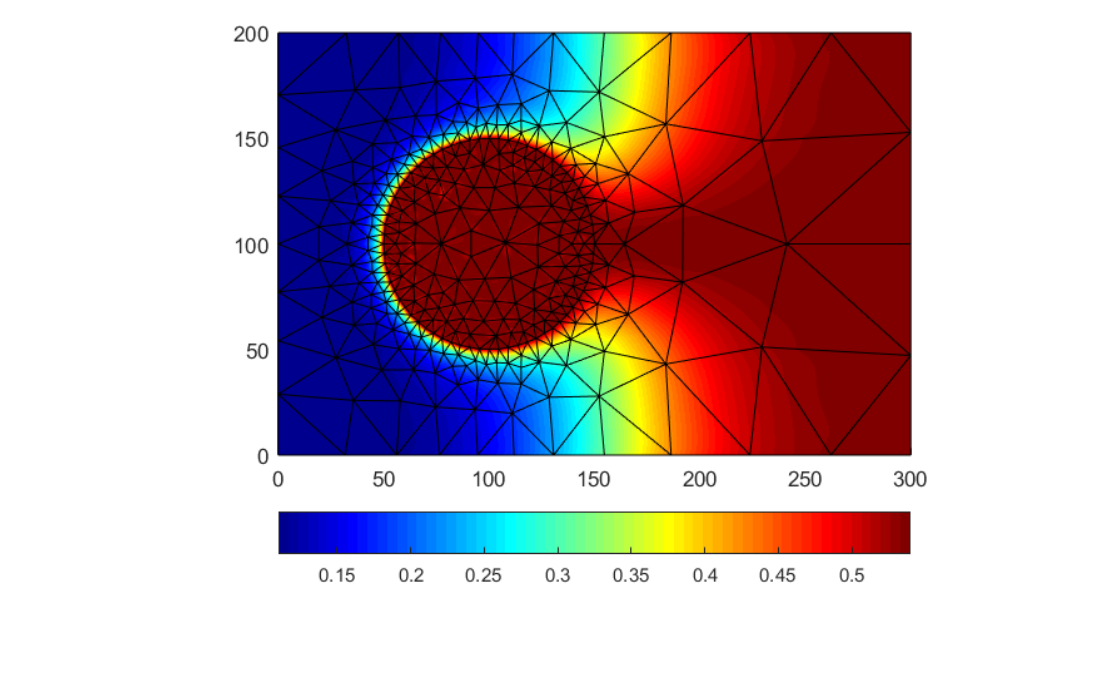}}

\caption{Three-phase flow in a heterogeneous medium.  Simulation is terminated at $t=50$ days.  Spurious oscillations are reduced for higher order polynomials.}
\label{het_1}
\end{figure}

Snapshots of the simulation at $t\in \{12.5,25,37.5,50\}$ (days), with piecewise quartics are provided in Fig.~\ref{het_2} (c), (d), (e) and (f).  Fig.~\ref{het_1} has plots of the wetting and light oil saturations at $t=50$ days for different polynomial orders, $k\in \{1,2,4\}$.  It is visible that the piecewise linear approximation has some numerical artifacts, particularly near the boundary of the permeability disk.  Increasing the polynomial order to piecewise quadratics and quartics allows us to generate simulations that more accurately capture the permeability disk.  The heavy oil pressure with the total velocity overlaid is displayed in Fig.~\ref{het_2} (a) and (b).  As we would expect, the velocity indicates that flows must move around the region of low permeability.

\begin{figure}[ht!] 
\centering
\includegraphics[trim = 10mm 80mm 20mm 85mm, clip, scale = 0.4]{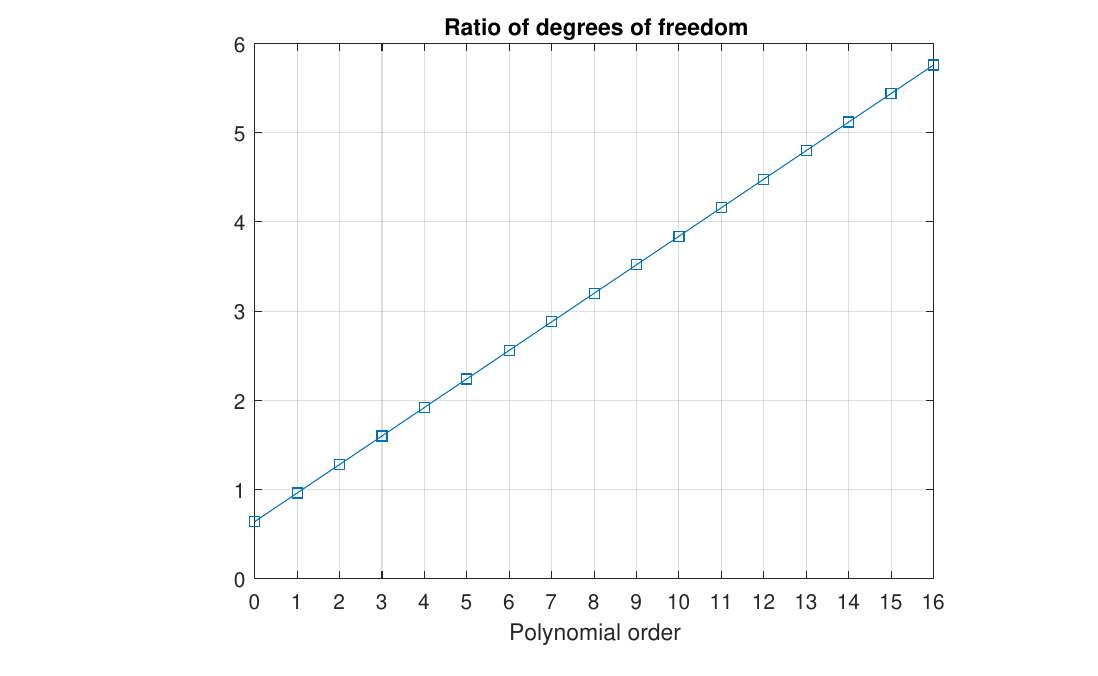}
\caption{Comparison of degrees of freedom for DG and HDG.   The reduction in unknowns for the HDG method is more drastic for higher orders.}
\label{ratio_1}
\end{figure}

High order HDG approximations are viable even in the heterogeneous case.  This is substantiated by Examples 1 and 2.  Compared to classical DG methods HDG has the advantage of static condensation.  In 2D, for triangular mesh with $|\mathcal{E}_h|$ elements, classical DG methods have $|\mathcal{E}_h|(k+2)(k+1)/2$ degrees of freedom.  Whereas, for HDG methods, through static condensation have $| \Gamma_h |(k+1)$ degrees of freedom.  To illustrate this fact, suppose we have a uniform triangular mesh of a square with $2N^2$ elements.  It can be shown that $\lim_{N\to \infty}\frac{|\mathcal{E}_h|(k+2)(k+1)/2}{| \Gamma_h |(k+1)}=(k+2)/3$.  In particular,  the ratio approaches 2.0, 3.0, and 6.0 for polynomial orders $k=4$, $k=8$, and $k=16$, respectively.  A visualization of this is given in Fig.~\ref{ratio_1}.

In Example 1, we successfully use polynomial orders up to degree 16.  The maximum polynomial order used in Example 2 is degree 4.  Example 3 uses piecewise quintics, and Example 4 has $k=8$ as the largest polynomial order.  We find that high-order approximations in our semi-implicit framework generate the best simulations.  A consequence of this is that we may use meshes that are much coarser than those needed for lower order approximations.

As evidence for this, we return to Example 1.  Comparing Figs~\ref{fig_het0} and~\ref{fig_het000r}, it is clear that the low order approximation needs significantly more elements than the higher order approximations to generate good quality data.  The $k=1$ approximation needs a uniform mesh with 32768 elements, whereas the $k=4,$ $8,$ or $16$ approximations need only 512 elements.  In terms of degrees of freedom, $k=1$ with 32768 elements yields 98816 unknowns, and $k=4,$ $8,$ and $16$ with 512 elements have 4000, 7200, and 13600 unknowns, respectively.  %An area of ongoing research is $hp$-adaptivity for DG method, which certainly would be advantageous with respect to controlling the total number of degrees of freedom.

Due to the local nature of the HDG method, it naturally exploits parallelism.  Especially the element by element local solvers and postprocessing, recovery of the volume unknowns, and the pairwise connectivity stencil, as the HDG globally coupled degrees of freedom reside on element faces.  High order DG methods are able to provide a sufficient amount of work at the element level, which is appropriate for emerging computer architectures such as accelerators \cite{fabien2019manycore}.

%For the phase viscosities we make the physically sound assumption $\mu_w<\mu_g<\mu_o$.  In this section we set $\mu_w= 0.890 $ cP, $\mu_g = 1.0$ cP, $\mu_o = 50.0$ cP
%\twocolumn
 
\subsubsection{Highly heterogeneous porous media}
\label{sec:het2}
To test the robustness of our method, we next consider domains that are highly heterogeneous.

\begin{figure}[htb!]
\hspace*{-5ex}
\subfigure[Wetting saturation, $k=5$]{\includegraphics[trim = 10mm 80mm 20mm 85mm, clip, scale = 0.4]{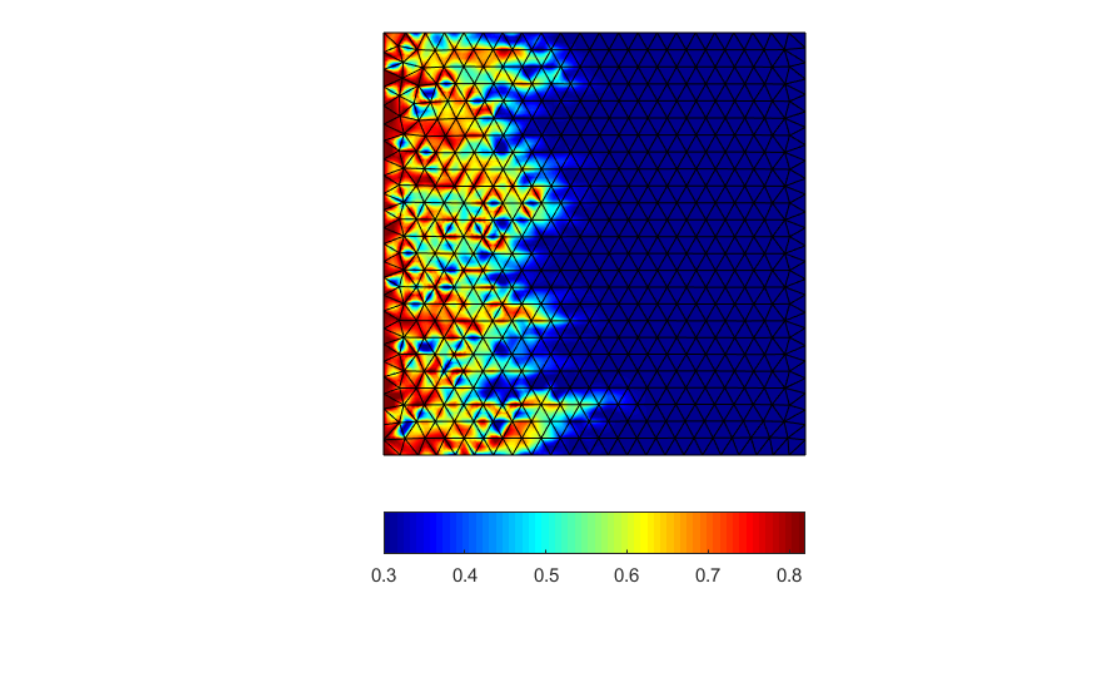}}
\hspace*{-25ex}
\subfigure[Light oil saturation, $k=5$]{\includegraphics[trim = 10mm 80mm 20mm 85mm, clip, scale = 0.4]{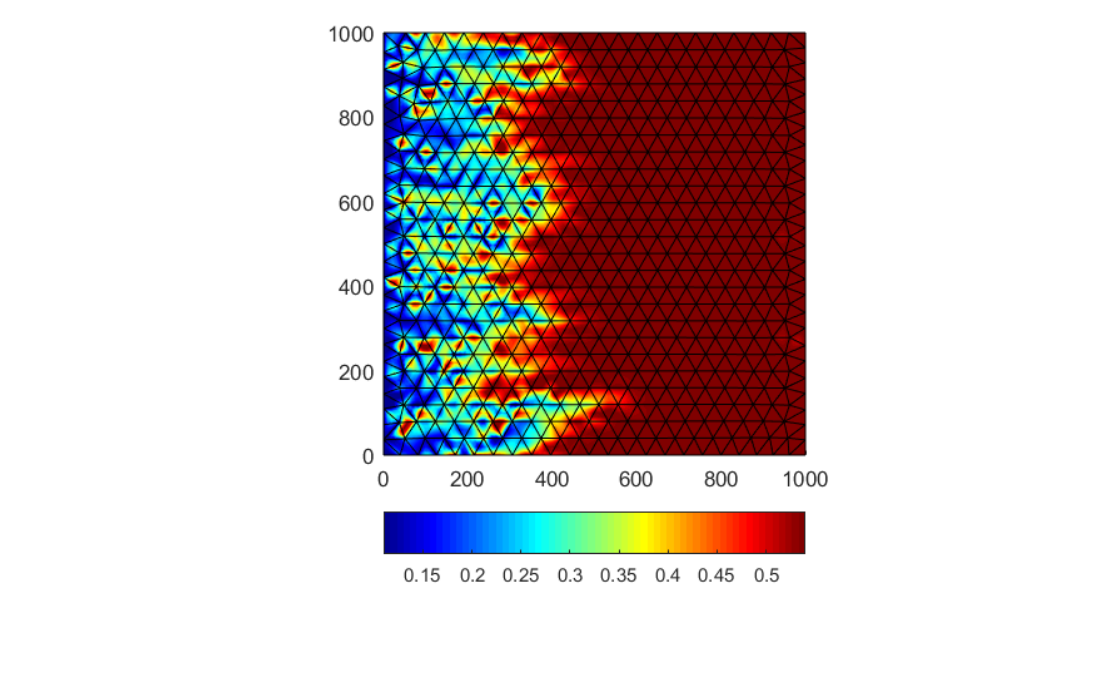}}
\caption{Three-phase flow in a heterogeneous medium with random permeability.  Simulation is terminated at $t=64$ days.}
\label{fig_rando}
\end{figure}

\paragraph{Example 3}  We set $\Omega=[0,1000]\times[0,1000]$ (meters), and the permeability $\bf K$ randomly takes values in the closed interval $[10^{-7},10^{-16}].$  For the viscosities we put $\mu_w= 1$ cp, $\mu_o=0.8$ cp, and $\mu_g = 0.9 $ cp.  A mesh with 1104 triangular elements is used, with piecewise quintic polynomials.  The remaining parameters and functions are the same as in section~\ref{section_het1}.  The simulation is run for 65 days.  From Fig.~\ref{fig_rando}, it is evident that the saturation wetting and light oil phase saturations are avoiding pockets of low permeability.  The calculated saturation profiles are bounded, and any overshoot or undershoot is negligible.  

\begin{figure}[htb!]
\centering
 \includegraphics[trim = 10mm 80mm 20mm 85mm, clip, scale = 0.65]{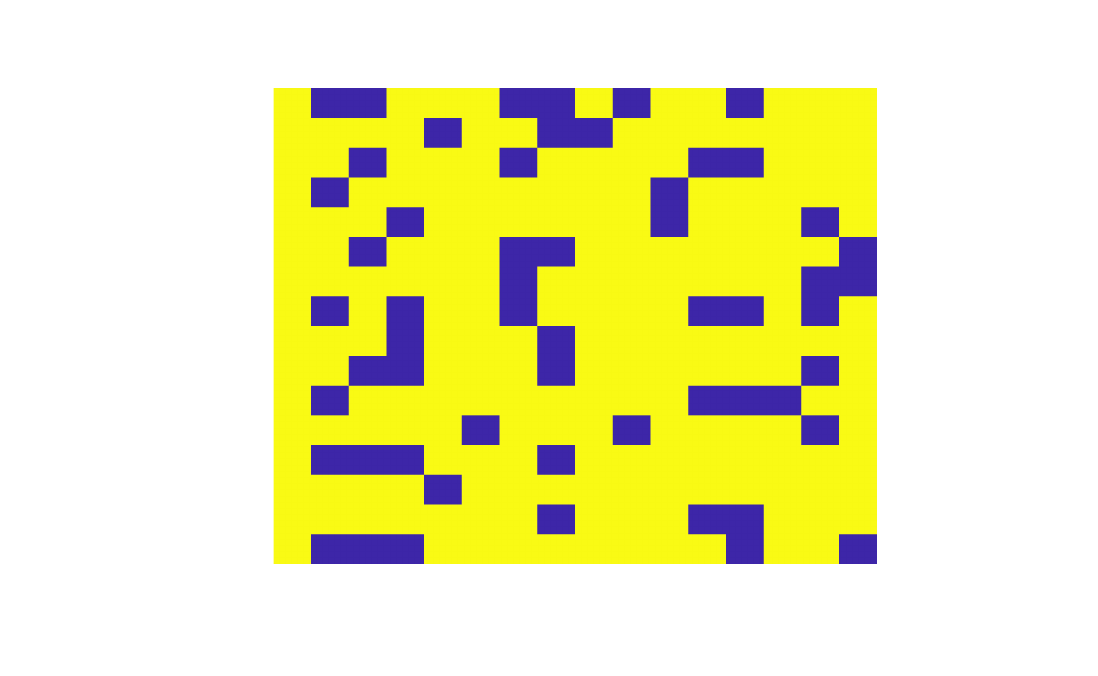}
\caption{Permeability field.  Shaded regions indicate lower permeability}
\label{hi_00}
\end{figure}

\paragraph{Example 4}
For this example we take all parameters and boundary conditions as the same as subsection~\ref{subsub_hom}.  The permeability is replaced with a highly varying field, as seen in Fig.~\ref{hi_00}.  Shaded squares mean a permeability of $10^{-15}$, and the permeability is $10^{-10}$ otherwise.  The simulation is run for 500 days, and we vary the polynomial order from piecewise linears to pieceiwise octics.

\begin{figure}[htb!]
\hspace*{-12ex}
\subfigure[Wetting saturation, $k=1$, $t=500$ days]{\includegraphics[trim = 10mm 80mm 20mm 85mm, clip, scale = 0.4]{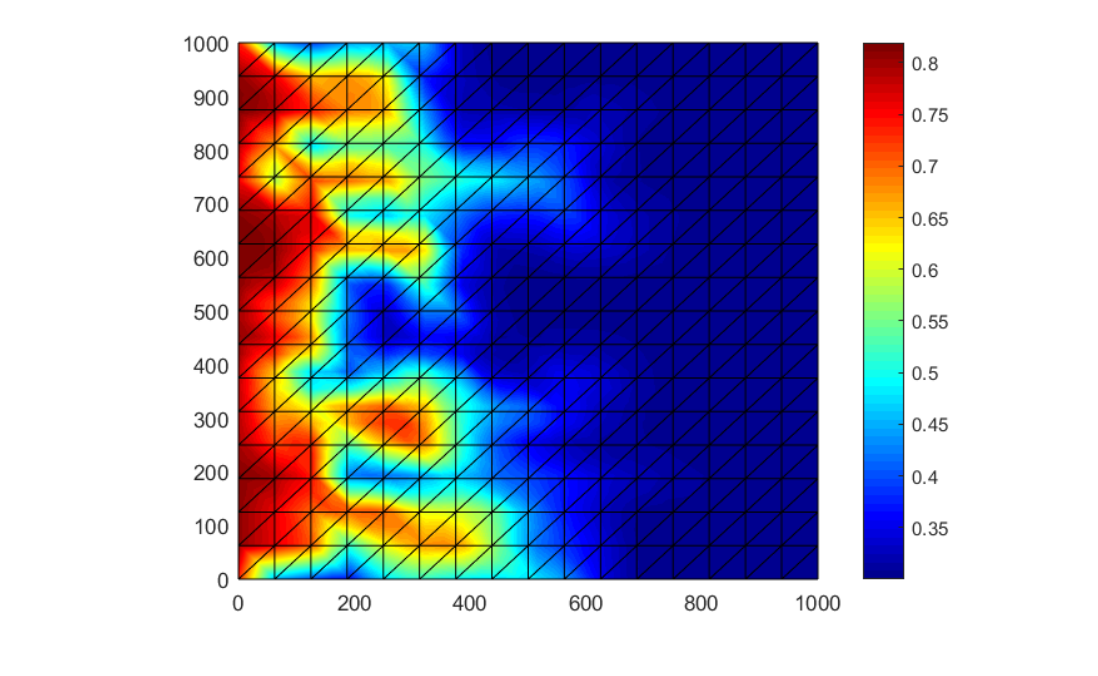}}
\hspace*{-18ex}
\subfigure[Light oil saturation, $k=1$, $t=500$ days]{\includegraphics[trim = 10mm 80mm 20mm 85mm, clip, scale = 0.4]{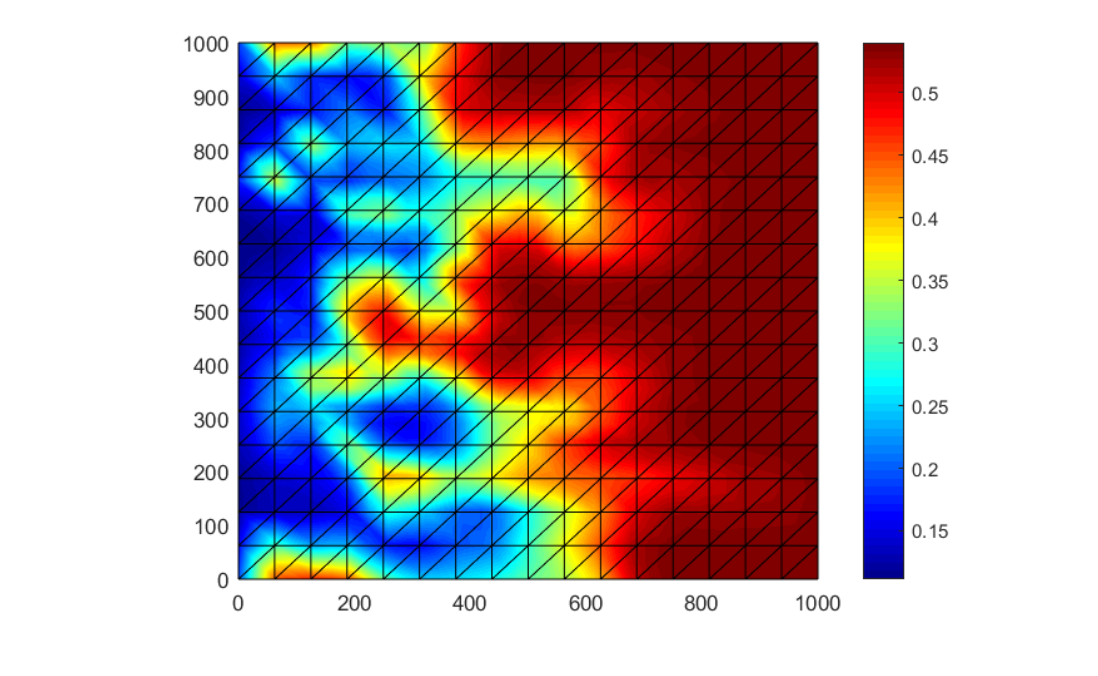}}
\caption{Piecewise linears on a coarse mesh create approximations that are too diffusive.}
\label{hi_0}
\end{figure}

\begin{figure}[ht!]
\hspace*{-12ex}
\subfigure[Wetting saturation at $t=150$ days, $k=8$]{\includegraphics[trim = 10mm 80mm 20mm 85mm, clip, scale = 0.4]{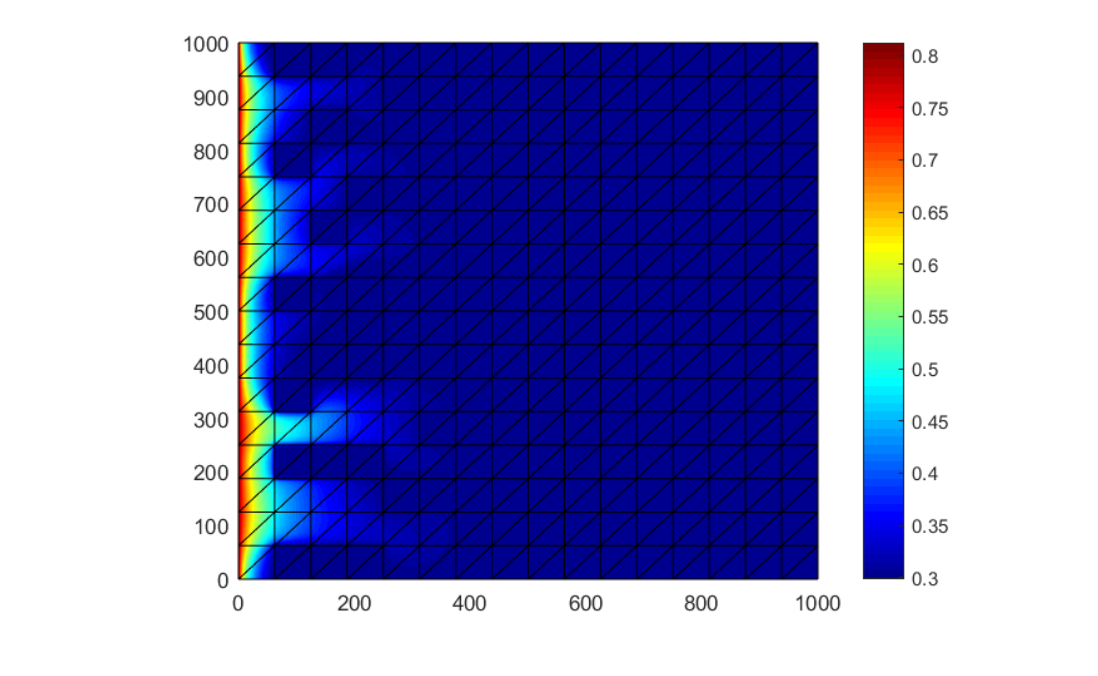}}
\hspace*{-18ex}
\subfigure[Wetting saturation at $t=300$ days, $k=8$]{\includegraphics[trim = 10mm 80mm 20mm 85mm, clip, scale = 0.4]{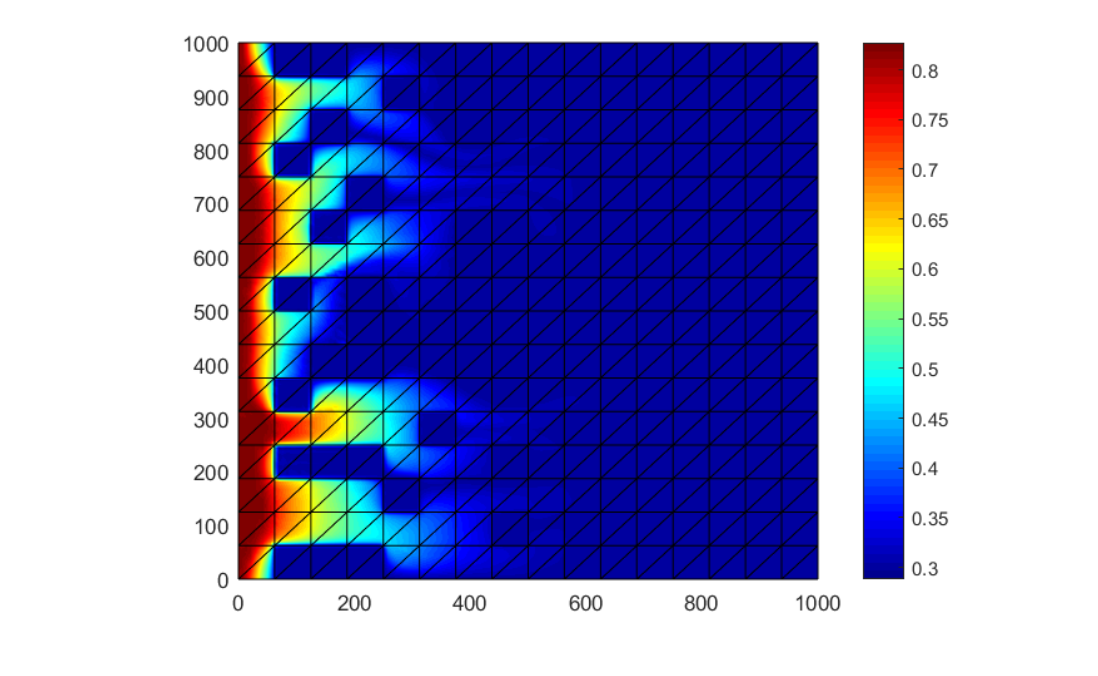}}
\newline
\hspace*{-12ex}
\subfigure[Wetting saturation at $t=450$ days, $k=8$]{\includegraphics[trim = 10mm 80mm 20mm 85mm, clip, scale = 0.4]{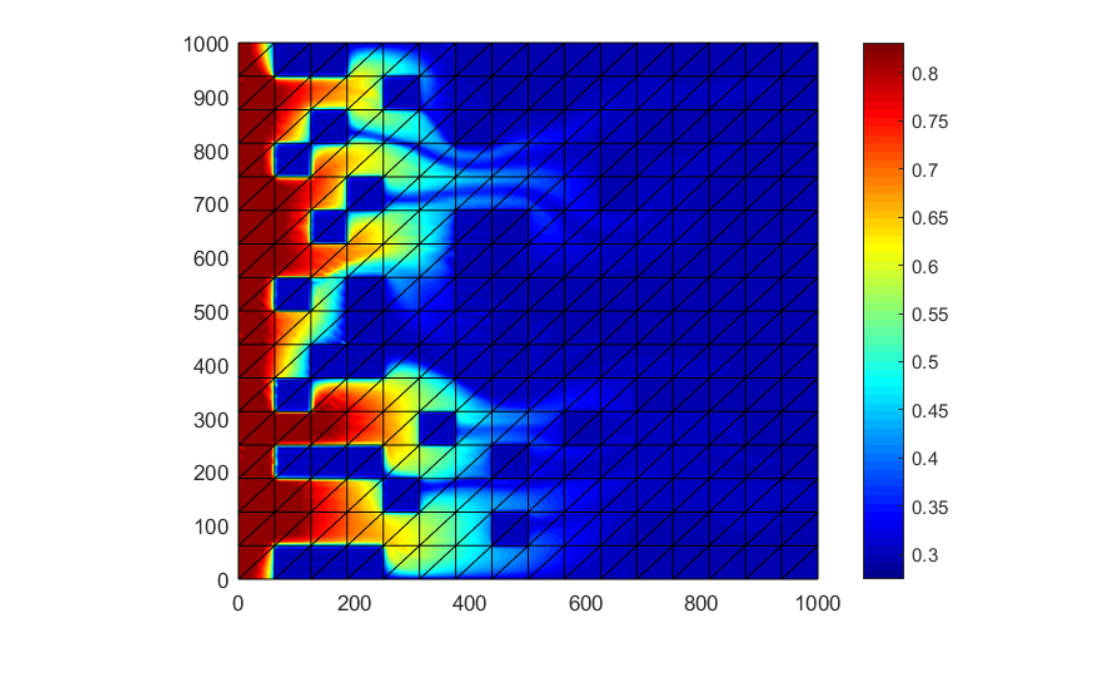}}
\hspace*{-18ex}
\subfigure[Wetting saturation at $t=500$ days, $k=8$]{\includegraphics[trim = 10mm 80mm 20mm 85mm, clip, scale = 0.4]{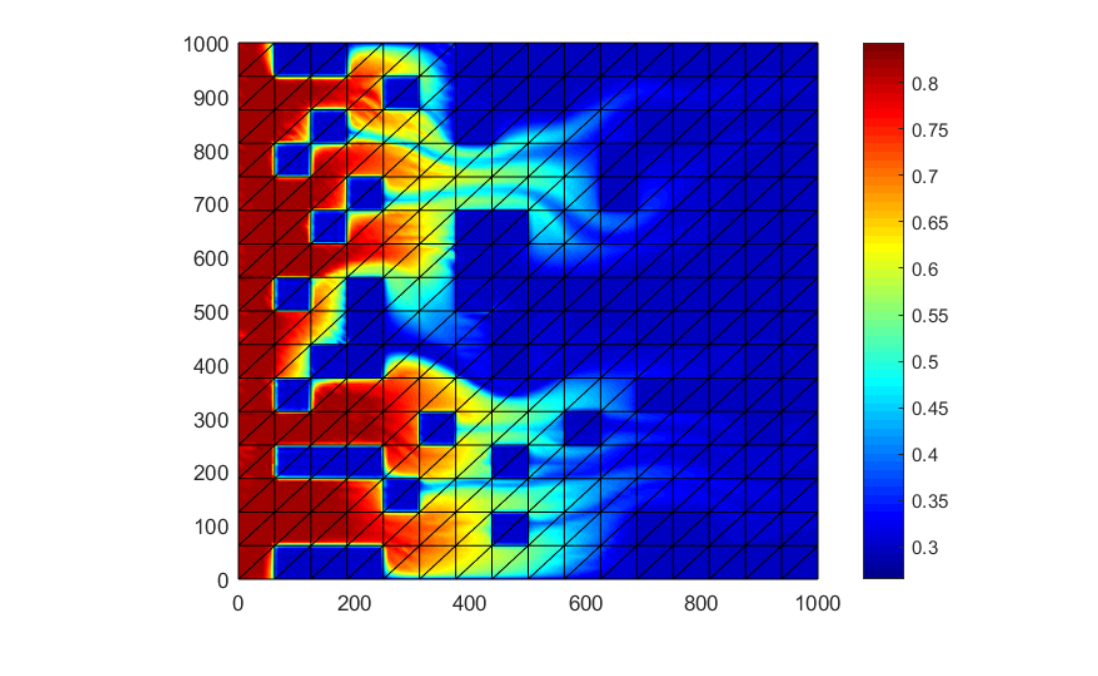}} 
\caption{Evolution of the wetting phase saturation at $t=150,$ 300, 450, and 500 days.}
\label{hi_000}
\end{figure}

\begin{figure}[ht!]
\hspace*{-16ex}
\subfigure[Wetting saturation, $k=2$]{\includegraphics[trim = 10mm 80mm 20mm 85mm, clip, scale = 0.4]{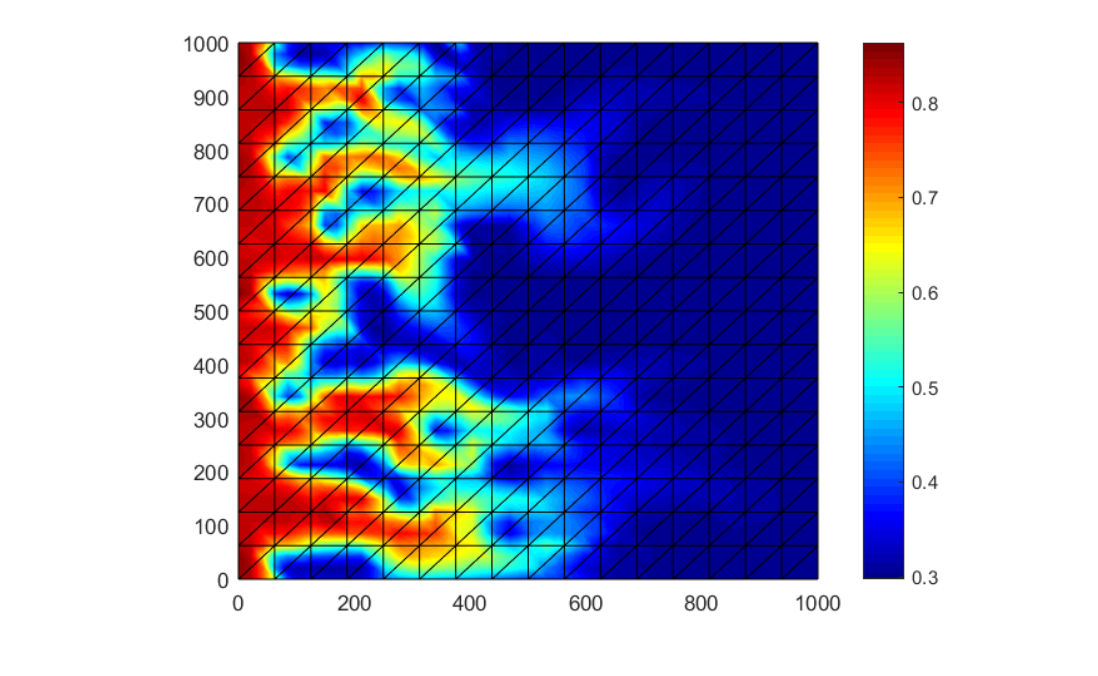}}
\hspace*{-20ex}
\subfigure[Light oil saturation, $k=2$]{\includegraphics[trim = 10mm 80mm 20mm 85mm, clip, scale = 0.4]{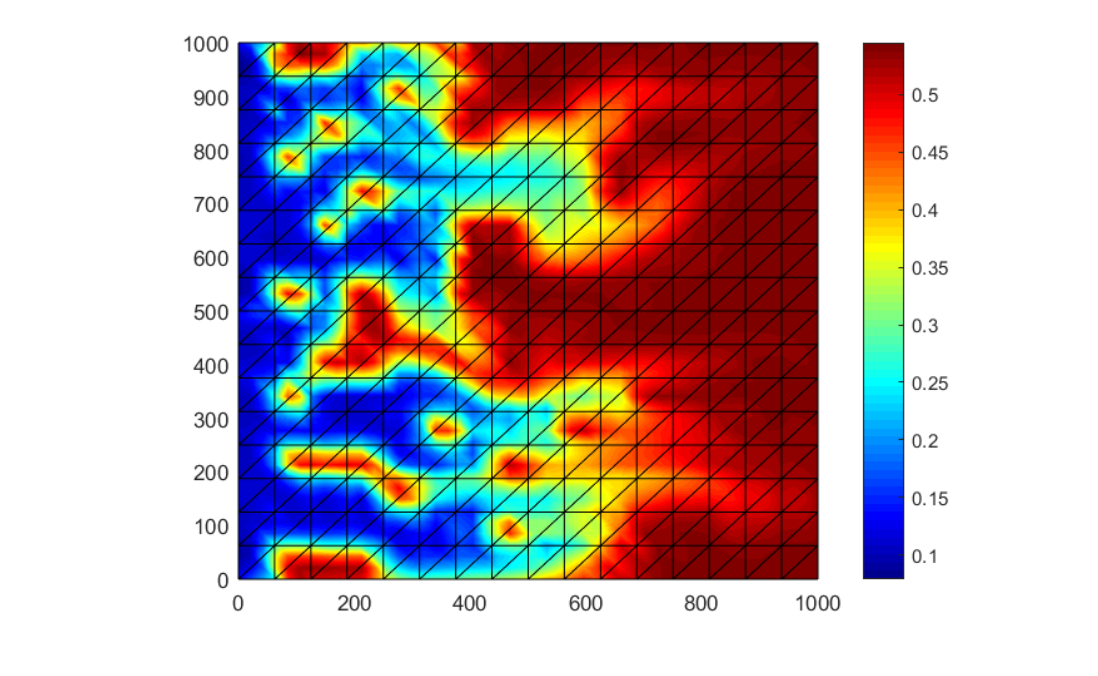}}
\newline
\hspace*{-16ex}
\subfigure[Wetting saturation, $k=4$]{\includegraphics[trim = 10mm 80mm 20mm 85mm, clip, scale = 0.4]{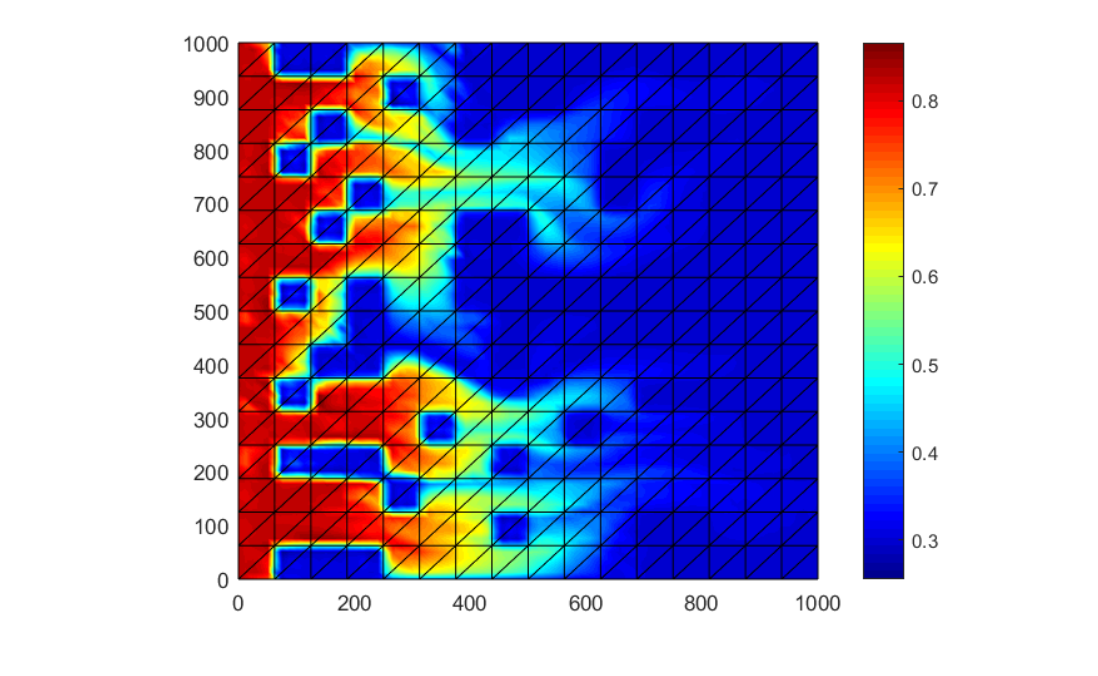}}
\hspace*{-20ex}
\subfigure[Light oil saturation, $k=4$]{\includegraphics[trim = 10mm 80mm 20mm 85mm, clip, scale = 0.4]{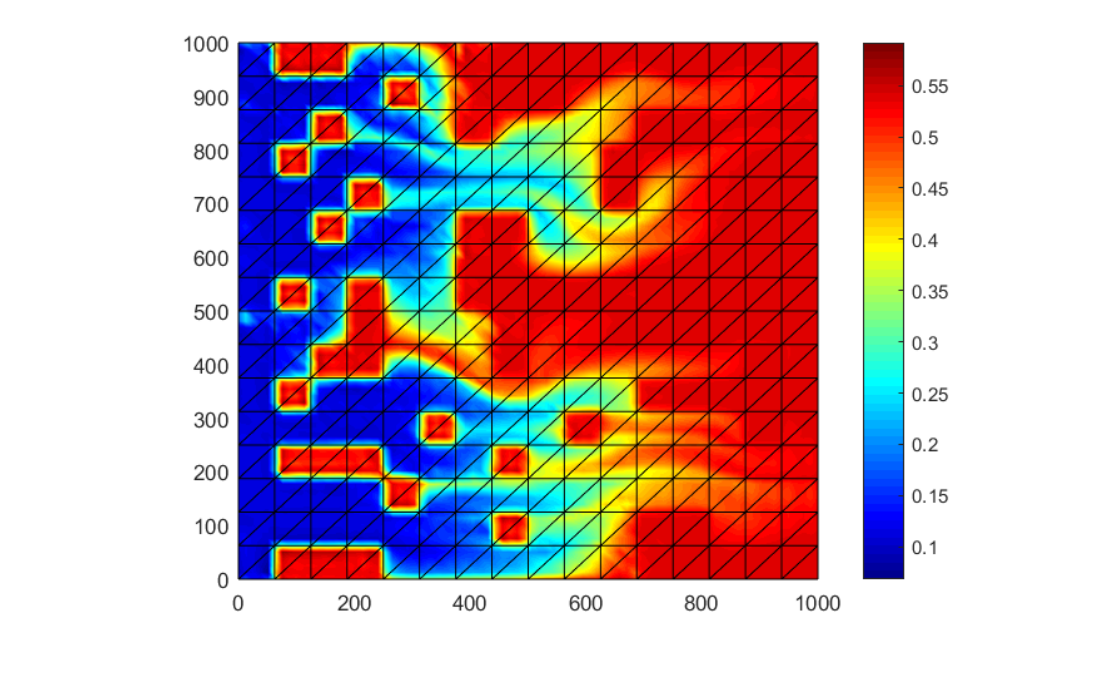}}
\newline
\hspace*{-16ex}
\subfigure[Wetting saturation, $k=8$]{\includegraphics[trim = 10mm 80mm 20mm 85mm, clip, scale = 0.4]{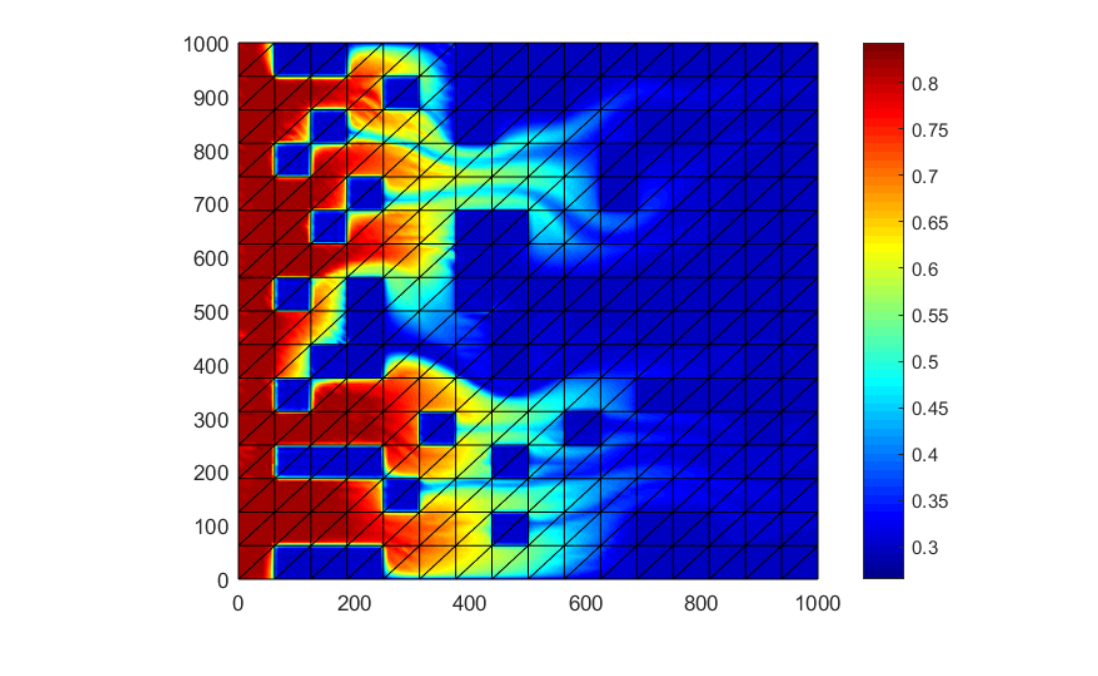}}
\hspace*{-20ex}
\subfigure[Light oil saturation, $k=8$]{\includegraphics[trim = 10mm 80mm 20mm 85mm, clip, scale = 0.4]{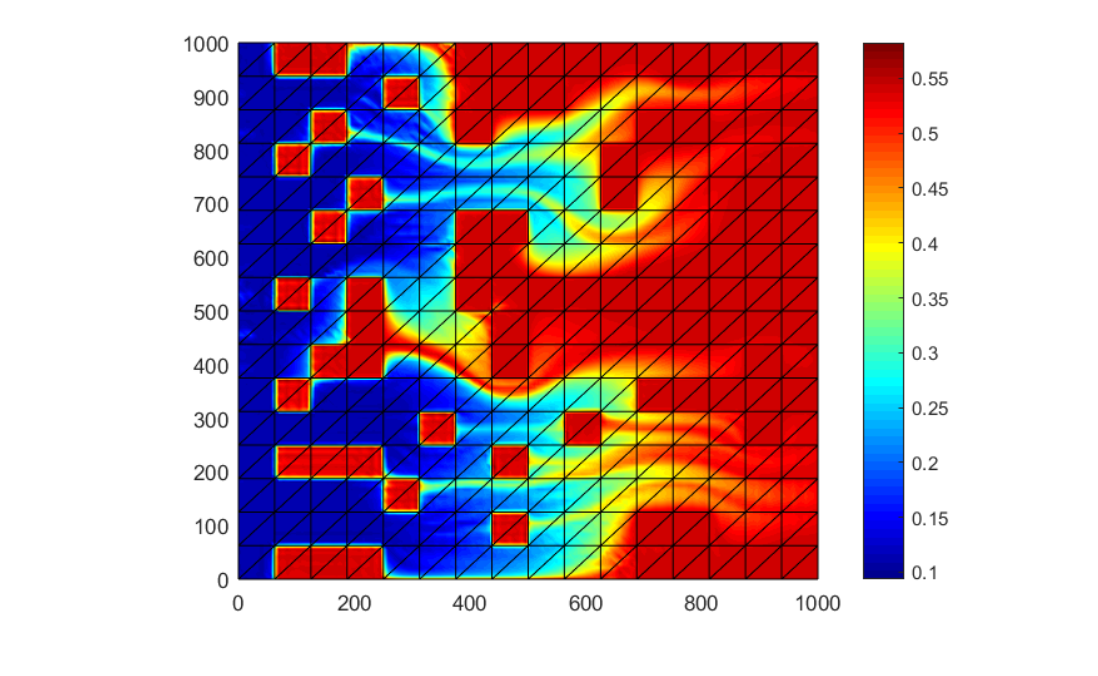}}

\caption{Three-phase flow in a highly heterogeneous medium.  Simulation is terminated at $t=500$ days.  Higher order approximations properly capture the expected behavior.}
\label{hi_1}
\end{figure}

  We gather from Fig.~\ref{hi_0} and Figs.~\ref{hi_1} (a) \& (b) that lower order approximations fail to realize the regions of low permeability.  Enriching the approximation space reveals that our numerical method can capture the expected behavior; the saturations evade the low permeability zones.  Additionally, the low permeability zones are captured with a sharp resolution, even on a coarse mesh.  The wetting phase saturation is plotted at $t\in\{150,300,450,500\}$ days (see Fig.~\ref{hi_000}).


\section{Conclusions} % The \section*{} command stops section numbering
This work demonstrates the effectiveness of a new hybridizable discontinuous Galerkin method for incompressible immiscible three-phase flow problems in porous media.  A splitting algorithm relaxes the nonlinearity present in the governing equations.  To the best of our knowledge, applying HDG to flows in porous media has not been done before.  In addition, piecewise quadratics is often considered high-order for flow-transport systems, but in this paper we regularly use polynomial orders up to and including degree 16.  The method is robust even when highly discontinuous heterogeneous properties of the porous medium are present.  Several numerical examples confirm this.  The high-order accuracy is verified by the method of manufactured solutions.

The HDG method simultaneously solves for both a scalar quantity and its gradient; which both converge at the optimal rate of $k+1$ in the $L^2$ norm for smooth solutions.  This observation is relevant as the Darcy system requires $\nabla s_w$ and $\nabla s_g$. Moreover, the saturation equations require the total velocity. For classical DG methods, one needs to locally differentiate the primary scalar quantities to obtain their vector counterparts, which leads to a loss in accuracy.  To upgrade the accuracy of our approximations even further, a local post processing exists for the HDG method for the scalar variable.  For smooth solutions, the resulting post processed approximation is superconvergent in the $L^2$ norm, with a rate of $k+2$.

	Another important feature of the HDG method is that it is a compatible discretization for flow-transport systems.  Not only is it locally conservative, but its approximations to gradient quantities do not lead to instability and oscillations.  As the HDG method falls under the umbrella of stabilized DG methods, it has no issue with handling unstructured meshes.  The HDG stablilization parameter is $\mathcal{O}(1)$, and is independent of the polynomial order as well as the mesh spacing; a convenient feature as simulations already have an abundance of parameters to tune.

Through several numerical examples we demonstrated that high-order approximations are beneficial, as low order approximations may not appropriately capture important phenomena in simulations \cite{li2015high}.  Further, in the low order regime, the mesh spacing may have to be taken prohibitively small so that simulations generate accurate data.  Notwithstanding, using high-order DG methods for flow-transport problem is challenging, but shows great promise. 

One issue is that high-order DG methods give rise to a proliferation of degrees of freedom.  A second issue is that semi-implicit algorithms require one to resolve nonlinearities through Newton or Picard type techniques.  So, at each time step, multiple linear solves must be performed.  This situation is exacerbated by the fact that three-phase flow calls for the solution of three equations, with at least two of them being nonlinear.  By the same token, high-order time stepping may also require additional linear solves on top everything else.

  The HDG method has the ability to call upon static condensation, which significantly reduces the number of degrees of freedom compared to traditional DG methods.  Thus, the HDG method is well suited for implicit problems.

%\phantomsection
%\section*{Acknowledgments} % The \section*{} command stops section numbering

%\addcontentsline{toc}{section}{Acknowledgments} % Adds this section to the table of contents

%So long and thanks for all the fish \cite{Figueredo:2009dg}.

%----------------------------------------------------------------------------------------
%	REFERENCE LIST
%----------------------------------------------------------------------------------------
%\newpage
%\phantomsection
%\bibliographystyle{unsrt}
%\bibliographystyle{spmpsci}
%\bibliography{sample}
\bibliographystyle{spmpsci} 
\bibliography{bibliog}

\begin{thebibliography}{10}
\providecommand{\url}[1]{{#1}}
\providecommand{\urlprefix}{URL }
\expandafter\ifx\csname urlstyle\endcsname\relax
  \providecommand{\doi}[1]{DOI~\discretionary{}{}{}#1}\else
  \providecommand{\doi}{DOI~\discretionary{}{}{}\begingroup
  \urlstyle{rm}\Url}\fi

\bibitem{anderson1965iterative}
Anderson, D.G.: Iterative procedures for nonlinear integral equations.
\newblock Journal of the ACM (JACM) \textbf{12}(4), 547--560 (1965)

\bibitem{arbogast2013discontinuous}
Arbogast, T., Juntunen, M., Pool, J., Wheeler, M.F.: A discontinuous {G}alerkin
  method for two-phase flow in a porous medium enforcing h (div) velocityand
  continuous capillary pressure.
\newblock Computational Geosciences \textbf{17}(6), 1055--1078 (2013)

\bibitem{ArnoldBCM02}
Arnold, D.N., F.~Brezzi, B.C., Marini, L.D.: Unified analysis of discontinuous
  {G}alerkin methods for elliptic problems.
\newblock {SIAM} J. Numerical Analysis \textbf{39}, 1749--1779 (2002)

\bibitem{babuvska1988p}
Babu{\v{s}}ka, I.: The p and hp versions of the finite element method: {T}he
  state of the art.
\newblock In: Finite Elements: Theory and Application Proceedings of the ICASE
  Finite Element Theory and Application Workshop Held July 28--30, 1986, in
  Hampton, Virginia, pp. 199--239. Springer (1988)

\bibitem{babuvska1992h}
Babu{\v{s}}ka, I., Guo, B.: The h, p and hp version of the finite element
  method; basis theory and applications.
\newblock Advances in Engineering Software \textbf{15}(3-4), 159--174 (1992)

\bibitem{bastian2014fully}
Bastian, P.: A fully-coupled discontinuous {G}alerkin method for two-phase flow
  in porous media with discontinuous capillary pressure.
\newblock Computational Geosciences \textbf{18}(5), 779--796 (2014)

\bibitem{blyth2006comparison}
Blyth, M., Luo, H., Pozrikidis, C.: A comparison of interpolation grids over
  the triangle or the tetrahedron.
\newblock Journal of engineering mathematics \textbf{56}(3), 263--272 (2006)

\bibitem{bui2015godunov}
Bui-Thanh, T.: From {G}odunov to a unified hybridized discontinuous {G}alerkin
  framework for partial differential equations.
\newblock Journal of Computational Physics \textbf{295}, 114--146 (2015)

\bibitem{chen1997comparison}
Chen, Z., Ewing, R.E.: Comparison of various formulations of three-phase flow
  in porous media.
\newblock Journal of Computational Physics \textbf{132}(2), 362--373 (1997)

\bibitem{chen2006computational}
Chen, Z., Huan, G., Ma, Y.: Computational methods for multiphase flows in
  porous media.
\newblock SIAM (2006)

\bibitem{class2009benchmark}
Class, H., Ebigbo, A., Helmig, R., Dahle, H.K., Nordbotten, J.M., Celia, M.A.,
  Audigane, P., Darcis, M., Ennis-King, J., Fan, Y., et~al.: A benchmark study
  on problems related to {CO2} storage in geologic formations.
\newblock Computational Geosciences \textbf{13}(4), 409 (2009)

\bibitem{CockburnDG08}
Cockburn, B., Dong, B., Guzm{\'{a}}n, J.: A superconvergent {LDG}-hybridizable
  {G}alerkin method for second-order elliptic problems.
\newblock Math. Comput. \textbf{77}(264), 1887--1916 (2008)

\bibitem{flux_cg}
Cockburn, B., Gopalakrishnan, J., Wang, H.: Locally conservative fluxes for the
  continuous {G}alerkin method.
\newblock SIAM Journal on Numerical Analysis \textbf{45}(4), 1742--1776 (2007).
\newblock \doi{10.1137/060666305}.
\newblock \urlprefix\url{https://doi.org/10.1137/060666305}

\bibitem{cockburn1998runge}
Cockburn, B., Shu, C.W.: The {R}unge--{K}utta discontinuous {G}alerkin method
  for conservation laws {V}: multidimensional systems.
\newblock Journal of Computational Physics \textbf{141}(2), 199--224 (1998)

\bibitem{costa2021high1}
Costa-Sol{\'e}, A., Ruiz-Giron{\'e}s, E., Sarrate, J.: High-order hybridizable
  discontinuous {G}alerkin formulation for one-phase flow through porous media.
\newblock Journal of Scientific Computing \textbf{87}, 1--31 (2021)

\bibitem{costa2021one}
Costa-Sol{\'e}, A., Ruiz-Giron{\'e}s, E., Sarrate, J.: One-phase and two-phase
  flow simulation using high-order hdg and high-order diagonally implicit time
  integration schemes.
\newblock In: Applied Mathematics for Environmental Problems, pp. 53--84.
  Springer (2021)

\bibitem{costa2021high2}
Costa~Sol{\'e}, A., Ruiz~Giron{\`e}s, E., Sarrate~Ramos, J.: High-order hdg
  formulation with fully implicit temporal schemes for the simulation of
  two-phase flow through porous media.
\newblock International journal for numerical methods in engineering
  \textbf{122}(14), 3583--3612 (2021)

\bibitem{dawson2004compatible}
Dawson, C., Sun, S., Wheeler, M.F.: Compatible algorithms for coupled flow and
  transport.
\newblock Computer Methods in Applied Mechanics and Engineering
  \textbf{193}(23), 2565--2580 (2004)

\bibitem{dong2014high}
Dong, J.: A high order method for three phase flow in homogeneous porous media.
\newblock Technical report  (2014)

\bibitem{dong2016semi}
Dong, J., Rivi{\`e}re, B.: A semi-implicit method for incompressible
  three-phase flow in porous media.
\newblock Computational Geosciences \textbf{20}(6), 1169--1184 (2016)

\bibitem{dubiner1991spectral}
Dubiner, M.: Spectral methods on triangles and other domains.
\newblock Journal of Scientific Computing \textbf{6}(4), 345--390 (1991)

\bibitem{epshteyn2006solution}
Epshteyn, Y., Riviere, B.: On the solution of incompressible two-phase flow by
  ap-version discontinuous {G}alerkin method.
\newblock International Journal for Numerical Methods in Biomedical Engineering
  \textbf{22}(7), 741--751 (2006)

\bibitem{epshteyn2007fully}
Epshteyn, Y., Rivi{\`e}re, B.: Fully implicit discontinuous finite element
  methods for two-phase flow.
\newblock Applied Numerical Mathematics \textbf{57}(4), 383--401 (2007)

\bibitem{ern2009accurate}
Ern, A., Mozolevski, I., Schuh, L.: Accurate velocity reconstruction for
  discontinuous {G}alerkin approximations of two-phase porous media flows.
\newblock Comptes Rendus Mathematique \textbf{347}(9-10), 551--554 (2009)

\bibitem{ern2010discontinuous}
Ern, A., Mozolevski, I., Schuh, L.: Discontinuous {G}alerkin approximation of
  two-phase flows in heterogeneous porous media with discontinuous capillary
  pressures.
\newblock Computer methods in applied mechanics and engineering
  \textbf{199}(23), 1491--1501 (2010)

\bibitem{eslinger2005discontinuous}
Eslinger, O.J.: Discontinuous {G}alerkin finite element methods applied to
  two-phase, air-water flow problems.
\newblock Ph.D. thesis, University of Texas (2005)

\bibitem{fabien2}
Fabien, M.S., Knepley, M., Riviere, B.: A high order hybridizable discontinuous
  {G}alerkin method for incompressible miscible displacement in heterogeneous
  media.
\newblock Results in Applied Mathematics \textbf{8}, 100089 (2020).
\newblock \doi{https://doi.org/10.1016/j.rinam.2019.100089}.
\newblock
  \urlprefix\url{https://www.sciencedirect.com/science/article/pii/S2590037419300895}.
\newblock Special Issue on Recent Advances in Computational Mathematics and
  Applications

\bibitem{fabien2019manycore}
Fabien, M.S., Knepley, M.G., Mills, R.T., Rivi{\`e}re, B.M.: Manycore parallel
  computing for a hybridizable discontinuous {G}alerkin nested multigrid
  method.
\newblock SIAM Journal on Scientific Computing \textbf{41}(2), C73--C96 (2019)

\bibitem{fabien1}
Fabien, M.S., Knepley, M.G., Rivière, B.M.: A hybridizable discontinuous
  {G}alerkin method for two-phase flow in heterogeneous porous media.
\newblock International Journal for Numerical Methods in Engineering
  \textbf{116}(3), 161--177 (2018).
\newblock \doi{https://doi.org/10.1002/nme.5919}.
\newblock
  \urlprefix\url{https://onlinelibrary.wiley.com/doi/abs/10.1002/nme.5919}

\bibitem{jamei2016novel}
Jamei, M., Ghafouri, H.: A novel discontinuous {G}alerkin model for two-phase
  flow in porous media using an improved impes method.
\newblock International Journal of Numerical Methods for Heat \& Fluid Flow
  \textbf{26}(1), 284--306 (2016)

\bibitem{klieber2006adaptive}
Klieber, W., Riviere, B.: Adaptive simulations of two-phase flow by
  discontinuous {G}alerkin methods.
\newblock Computer methods in applied mechanics and engineering
  \textbf{196}(1), 404--419 (2006)

\bibitem{koornwinder1975two}
Koornwinder, T.: Two-variable analogues of the classical orthogonal
  polynomials.
\newblock In: Theory and application of special functions (Proc. Advanced Sem.,
  Math. Res. Center, Univ. Wisconsin, Madison, Wis., 1975), pp. 435--495.
  Academic Press New York (1975)

\bibitem{lacroix2000iterative}
Lacroix, S., Vassilevski, Y.V., Wheeler, M.F.: Iterative solvers of the
  implicit parallel accurate reservoir simulator (ipars), i: single processor
  case.
\newblock TICAM report 00-28, The University of Texas at Austin, Austin  (2000)

\bibitem{drad054}
Leng, H., Chen, H.: {Adaptive interior penalty hybridized discontinuous
  {G}alerkin methods for {D}arcy flow in fractured porous media}.
\newblock IMA Journal of Numerical Analysis p. drad054 (2023).
\newblock \doi{10.1093/imanum/drad054}.
\newblock \urlprefix\url{https://doi.org/10.1093/imanum/drad054}

\bibitem{leverett1941steady}
Leverett, M., Lewis, W., et~al.: Steady flow of gas-oil-water mixtures through
  unconsolidated sands.
\newblock Transactions of the AIME \textbf{142}(01), 107--116 (1941)

\bibitem{li2015high}
Li, J., Riviere, B.: High order discontinuous {G}alerkin method for simulating
  miscible flooding in porous media.
\newblock Computational Geosciences \textbf{19}(6), 1251 (2015)

\bibitem{lu2008iteratively}
Lu, B.: Iteratively coupled reservoir simulation for multiphase flow in porous
  media.
\newblock The University of Texas at Austin (2008)

\bibitem{moon2022}
Moon, M.: Generalized multiscale hybridizable discontinuous {G}alerkin
  ({GMsHDG}) method for flows in nonlinear porous media.
\newblock J. Comput. Appl. Math. \textbf{415}(C) (2022).
\newblock \doi{10.1016/j.cam.2022.114441}.
\newblock \urlprefix\url{https://doi.org/10.1016/j.cam.2022.114441}

\bibitem{moortgat2013higher}
Moortgat, J., Firoozabadi, A.: Higher-order compositional modeling of
  three-phase flow in 3d fractured porous media based on cross-flow
  equilibrium.
\newblock Journal of Computational Physics \textbf{250}, 425--445 (2013)

\bibitem{Owens857}
Owens, R.G.: Spectral approximations on the triangle.
\newblock Proceedings of the Royal Society of London A: Mathematical, Physical
  and Engineering Sciences \textbf{454}(1971), 857--872 (1998)

\bibitem{peraire2008compact}
Peraire, J., Persson, P.O.: The compact discontinuous {G}alerkin ({CDG}) method
  for elliptic problems.
\newblock SIAM Journal on Scientific Computing \textbf{30}(4), 1806--1824
  (2008)

\bibitem{proriol1957famille}
Proriol, J.: Sur une famille de polynomes {\'a} deux variables orthogonaux dans
  un triangle.
\newblock Comptes rendus hebdomadaires des s{\'e}ances de l'Acad{\'e}mie des
  sciences \textbf{245}(26), 2459--2461 (1957)

\bibitem{flux_eg}
R.~Becker E.~Burman, P.H., Larson, M.G.: A reduced p1–discontinuous
  {G}alerkin method.
\newblock Chalmers Finite Element Center Preprint  (2003)

\bibitem{reed1973triangularmesh}
Reed, W.H., Hill, T.: Triangularmesh methodsfor the neutrontransportequation.
\newblock Los Alamos Report LA-UR-73-479  (1973)

\bibitem{ricardo2012single}
Ricardo~Dias, A.: Single and Two-Phase Flows on Chemical and Biomedical
  Engineering:.
\newblock Bentham Science Publishers (2012).
\newblock \urlprefix\url{https://books.google.com/books?id=jXszDgAAQBAJ}

\bibitem{stenberg1991postprocessing}
Stenberg, R.: Postprocessing schemes for some mixed finite elements.
\newblock ESAIM: Mathematical Modelling and Numerical Analysis \textbf{25}(1),
  151--167 (1991)

\bibitem{TEZDUYAR19911}
Tezduyar, T.: Stabilized finite element formulations for incompressible flow
  computations††this research was sponsored by nasa-johnson space center
  (under grant nag 9-449), nsf (under grant msm-8796352), u.s. army (under
  contract daal03-89-c-0038), and the university of paris vi.
\newblock pp. 1--44. Elsevier (1991).
\newblock \doi{https://doi.org/10.1016/S0065-2156(08)70153-4}.
\newblock
  \urlprefix\url{https://www.sciencedirect.com/science/article/pii/S0065215608701534}

\bibitem{trefethen2013approximation}
Trefethen, L.N.: Approximation theory and approximation practice.
\newblock Siam (2013)

\bibitem{troescher2023fully}
Troescher, N., Higdon, J.: A fully-implicit hybridized discontinuous {G}alerkin
  spectral element method for two phase flow in petroleum reservoirs.
\newblock Journal of Computational Physics \textbf{474}, 111824 (2023)

\bibitem{wanner1996solving}
Wanner, G., Hairer, E.: Solving ordinary differential equations {II}, vol. 375.
\newblock Springer Berlin Heidelberg New York (1996)

\bibitem{woopen2014hybridized}
Woopen, M., Ludescher, T., May, G.: A hybridized discontinuous {G}alerkin
  method for turbulent compressible flow.
\newblock In: 44th AIAA Fluid Dynamics Conference, p. 2783 (2014)

\end{thebibliography}

\end{document}